\documentclass{aa}  
\usepackage{natbib}
\usepackage{graphicx}
\usepackage{txfonts}

\begin{document}
   \title{Testing grain-surface chemistry in massive hot-core regions}


   \author{S.~E. Bisschop\inst{1} \and J.~K. J{\o}rgensen\inst{2}
   \and E.~F. van Dishoeck\inst{1} \and E.~B.~M. de Wachter\inst{1}}

   \offprints{S. Bisschop, bisschop@strw.leidenuniv.nl }
 
  \institute{Leiden Observatory, P.O. Box 9513, 2300 RA Leiden, Netherlands
\and
{Harvard-Smithsonian Center for Astrophysics, 60 Garden Street MS42, MA 02138, U.S.A.}
 }

   \date{Received ; accepted}

 
\abstract 
{} 
{To establish the chemical origin of a set of complex organic
molecules thought to be produced by grain surface chemistry in high
mass young stellar objects (YSOs).}
{A partial submillimeter line-survey was performed toward 7 high-mass
YSOs aimed at detecting H$_2$CO, CH$_3$OH, CH$_2$CO, CH$_3$CHO,
C$_2$H$_5$OH, HCOOH, HNCO and NH$_2$CHO. In addition, lines of
CH$_3$CN, C$_2$H$_5$CN, CH$_3$CCH, HCOOCH$_3$, and CH$_3$OCH$_3$ were
observed. Rotation temperatures and beam-averaged column densities are
determined. To correct for beam dilution and determine abundances for
hot gas, the radius and H$_2$ column densities of gas at temperatures
$>$100~K are computed using 850 $\mu$m dust continuum data and source
luminosity.}
{Based on their rotation diagrams, molecules can be classified as
either cold ($<$100~K) or hot ($>$100~K). This implies that complex
organics are present in at least two distinct regions. Furthermore,
the abundances of the hot oxygen-bearing species are correlated, as
are those of HNCO and NH$_2$CHO. This is suggestive of chemical
relationships within, but not between, those two groups of molecules.}
{The most likely explanation for the observed correlations of the
various hot molecules is that they are ``first generation'' species
that originate from solid-state chemistry. This includes H$_2$CO,
CH$_3$OH, C$_2$H$_5$OH, HCOOCH$_3$, CH$_3$OCH$_3$, HNCO, NH$_2$CHO,
and possibly CH$_3$CN, and C$_2$H$_5$CN. The correlations between
sources implies very similar conditions during their formation or very
similar doses of energetic processing. Cold species such as CH$_2$CO,
CH$_3$CHO, and HCOOH, some of which are seen as ices along the same
lines of sight, are probably formed in the solid state as well, but
appear to be destroyed at higher temperatures. A low level of
non-thermal desorption by cosmic rays can explain their low rotation
temperatures and relatively low abundances in the gas phase compared
to the solid state. The CH$_3$CCH abundances can be fully explained by
low temperature gas phase chemistry. No cold N-containing molecules
are found.}


   \maketitle
%

\section{Introduction}
\label{intro}

A characteristic stage of high-mass star formation is the hot core
phase where high abundances of complex molecules are found in the
gas-phase and in the ice mantles around the dust grains in the cold
extended envelope \citep[e.g.,][]{millar1996,ikeda2001,boogert2004a}.
The existence of both ices and gas-phase molecules around the same
high-mass protostars suggests that the envelopes have temperatures
that range from $>$200 K in the inner region to as low as 10~K in the
cold outer part
\citep[e.g.,][]{vdtak2000,gibb2000b,ikeda2001}. Together the
submillimeter and infrared data indicate that there is a rich
chemistry which involves interactions between gas and grain-surface
species. The precise origin of complex molecules is still heavily
debated: are they ``first generation'' species formed on the grains
during the cold collapse phase \citep[e.g.,][]{tielens1997}, or
``second generation'' species produced by high-temperature gas phase
reactions following ice evaporation
\citep[e.g.,][]{millar1991,charnley1992,charnley1995}? The aim of this
work is to test the formation schemes of key complex organic molecules
by presenting submillimeter observations of a sample of high-mass
cores using the same instrument and frequency settings.

High column densities of the organic molecules CH$_3$OH, CH$_4$,
HCOOH, OCN$^{-}$ and potentially H$_2$CO have been detected in ice
mantles in the cold outer envelopes of high-mass young stellar objects
(YSOs) \citep{gibb2004}. Abundances for some ice species such as
CH$_3$OH and OCN$^-$ are large, up to 30\% with respect to H$_2$O ice
or 3$\times 10^{-5}$ with respect to H$_2$
\citep{brooke1999,dartois1999,pontoppidan2003,broekhuizen2005a}. Information
on minor species such as CH$_4$, HCOOH, H$_2$CO and CH$_3$CHO is more
limited, but they appear to have constant abundances at the level of a
few \% with respect to H$_2$O ice
\citep{schutte1999,keane2001a,gibb2004}. More complex species with
abundances lower than 0.1\% cannot be detected by infrared
observations because of their small optical depth.  Moreover, features
of the same functional groups such as C--H and O--H groups overlap, so
that they can only be assigned to a class of molecules rather than to
a specific species. Deep submillimeter surveys in the gas phase are
more sensitive to detect these minor species.

\begin{figure}
\centering
\includegraphics[width=6cm]{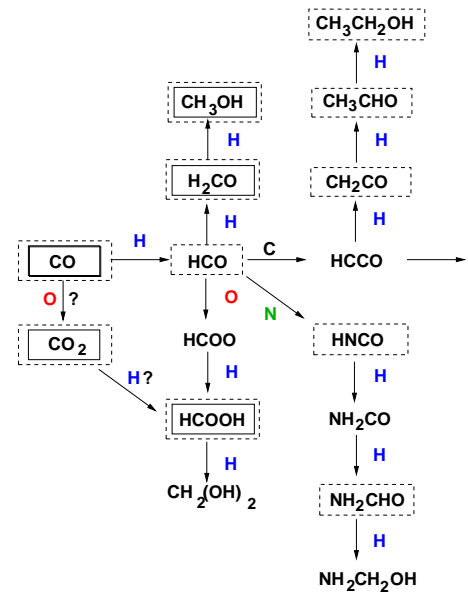}
\caption{Grain-surface chemistry routes involving hydrogenation of
CO. Solid rectangular boxes contain molecules which have been detected
in interstellar ices, whereas dashed boxes indicate molecules that
have been detected in the gas phase \citep[based
on][]{tielens1997}.}\label{grainchem}
\end{figure}

The abundances of the detected ice species are so large that they
cannot be produced by pure gas-phase processes but must be formed on
the grains. A proposed scheme for the formation of various ``first
generation'' molecules on grains \citep[based on][]{tielens1997} is
shown in Fig.~\ref{grainchem}. Through successive addition of H, C, O,
and N to CO, many complex organic species are formed. The outcome
depends on the relative gas-phase abundances of the atoms, their
mobility on the grains, and the CO abundance. If the formation routes
depicted in Fig.~\ref{grainchem} are indeed valid and proceed under
similar initial conditions, this would imply that the molecules in
this scheme are related and that they will have correlated abundance
ratios upon evaporation. For different initial conditions, the
relative abundances will vary from source to source; the
H$_2$CO/CH$_3$OH ratio depends for example on ice temperature, and the
C$_2$H$_5$OH/CH$_3$OH ratio on the atomic carbon abundance. In
contrast, gas phase formation of second generation species is expected
to cause time-dependent abundance variations where ``first'' and
``second generation'' as well as non-related ``first generation''
species are uncorrelated. A combination of rotation temperatures and
relative abundances between species can determine which species are
part of the same chemical group. To test the basic grain surface
chemistry scheme, our survey has targeted emission lines of all
species depicted in Fig.~\ref{grainchem}. Only CO$_2$, the unstable
species HCOO, HCCO, and NH$_2$CO, as well as CH$_2$(OH)$_2$ and
NH$_2$CH$_2$OH for which no laboratory frequencies are available, are
lacking.

Many line surveys have been performed to study the molecular
composition of high mass star forming regions
\citep[e.g.,][]{blake1987,sutton1995,helmich1997,schilke1997,nummelin2000,gibb2000a,schilke2001}.
However, they often cover a single atmospheric window, each at a
different telescope. These data thus probe a variety of excitation
conditions with different beam-sizes, which complicates the comparison
of column densities and abundances between different sources. Our
observing strategy employs deep, partial line surveys in multiple
atmospheric windows rather than complete, unbiased line surveys.
Partial line surveys exist, but are usually targeted at a few specific
molecules such as CH$_3$OH \citep{vdtak2000} and HNCO
\citep{zinchenko2000}. The advantage of our survey is that it
efficiently covers lines with a large range of excitation temperatures
and that the settings and telescopes are the same for all sources. The
lines of the molecules targeted in this work are sufficiently strong
so that no complete line survey is needed for their
identification. Compared with previous studies such as those by
\citet{ikeda2001} and \citet{zinchenko2000} our observations cover
higher frequencies, which implies smaller beams and, for some species,
lines from higher energy levels.

This paper is organized as follows: Sect.~\ref{sample} presents the
selection of sources plus frequencies, and the observational
techniques, Sect.~\ref{analysis} presents the data reduction
techniques, Sect.~\ref{results} the results on the derived rotation
temperatures plus column densities and compares those to other high
mass YSOs. In Sect.~\ref{sec_cor} the correlations of the abundances
between different molecular species are
determined. Sect.~\ref{discussion} discusses the origin of the
constant abundance ratios and compares these to astrochemical models,
and finally in Sect.~\ref{conclusions} the main conclusions are
summarized.


\section{Observations}
\label{sample}

\subsection{Source sample}
\label{source_sample}

\begin{table}
\caption{Positions, luminosities, distances and typical line width of
selected sources.\label{sources}}
\begin{center}
\begin{tabular}{l|lllll}
\hline
\hline
Sources & $\alpha$(2000) & $\delta$(2000) & $L$$^a$ & $d^b$ & $\Delta V$\\
        &                &                & [10$^5$ L$_\odot$] & [kpc] & km s$^{-1}$\\
\hline
AFGL~2591   & 20:29:24.6 & $+$40:11:19 & 0.2 & 1.0 & 3\\
G24.78      & 18:36:12.6 & $-$07:12:11 & 7.9 & 7.7 & 6\\
G75.78      & 20:21:44.1 & $+$37:26:40 & 1.9 & 4.1 & 4\\
NGC~6334~IRS1  & 17:20:53.0 & $-$35:47:02  & 1.1 & 1.7 & 5\\
NGC~7538~IRS1  & 23:13:45.4 & $+$61:28:12 & 1.3 & 2.8 & 4\\
W~3(H$_2$O) & 02:27:04.6 & $+$61:52:26 & 0.2 & 2.0 & 6\\
W~33A       & 18:14:38.9 & $-$17:52:04 & 1.0 & 4.0 & 5\\
\hline
\end{tabular}
\end{center}

$^a$ From \citet{vdtak2000}, except for
G24.78 and G75.78 where the luminosities are from \citet{zhang2005} and
\citet{churchwell1990} respectively.

$^b$ From \citet{vdtak2000}, except for G24.78 and G75.78 which are
based on \citet{forster1989} and \citet{churchwell1990} respectively.
\end{table}

The basis for this survey is formed by a sample of 7 high-mass young
stellar objects summarized in Table~\ref{sources}. The sources are
known to have hot gas in their inner envelope where ices have
evaporated and they either show complex organic molecules due to hot
core chemistry or are on the verge of reaching that stage
\citep[e.g.,][]{boonman2001}. Most of the sources were selected from
the sample of high mass YSOs studied in detail by \citet{vdtak2000}
based on the following criteria: (i) narrow line widths
($\lesssim$6~km~s$^{-1}$) to minimize line confusion; (ii) detection
of CH$_3$OH lines, where the strongest lines in the 7$_K$--6$_K$
338~GHz band should have main-beam temperatures of $\geq$1~K; (iii)
distances $<$5~kpc and (iv) visible from the James Clerk Maxwell
Telescope (JCMT)\footnote{The James Clerk Maxwell Telescope is
operated by the Joint Astronomy Centre, on behalf of the Particle
Physics and Astronomy Research Council of the United Kingdom, the
Netherlands Organization for Scientific Research and the National
Research Council of Canada. Project IDs are m04bn12, m05an05, and
m05bn11.}. Preference was also given to sources for which infrared
absorption line data of ices are available. The source list
additionally includes G75.78 and G24.78. G75.78 has been previously
observed by \citet{hatchell1998} and obeys all previously mentioned
selection criteria. G24.78 has a distance $>$5 kpc, but was selected
based on the large line intensities of CH$_3$CN detected by
\citet{beltran2005}. Trial JCMT observations of the 7$_K$--6$_K$
338~GHz CH$_3$OH band confirmed that the strongest lines are $>$1~K
and G24.78 was therefore added to the sample.

\subsection{Observed frequency settings}
\label{lines}

The observed frequencies are shown in Table~\ref{obs_lines} and were
selected based on whether they contained at least one strong line for
a target molecule as well as lines of other interesting
species. Strong lines of target molecules were chosen because of their
high main-beam temperatures and minimum line confusion in line surveys
of Orion~A by \citet{sutton1985} and Orion~KL by \citet{schilke1997}
at 230~GHz and 345~GHz respectively. The aim was to cover at least two
transitions for each molecule to determine rotation temperatures: one
with $E_{\rm u}>$100~K, and one with $E_{\rm
u}<$100~K. Table~\ref{obs_lines} shows the additional lines present in
the rest frequency side band. The molecular species detected in the
image band differed, because it varied per source. Finally, for
H$_2$CO, CH$_3$OH, and HNCO lines of the isotopic species were
observed to determine the optical depth of the main isotopologues.

An additional number of settings have been observed for CH$_2$CO,
CH$_3$CHO, and HCOOH with the Institut de Radioastronomie
Millimetrique\footnote{The IRAM 30~m telescope is operated by the
Institut de Radioastronomie Millim\'{e}trique on behalf of the Centre
National de la Recherche Scientifique of France, the Max Planck
Gesellschaft of Germany, and Instituto Geogr\'{a}fico Nacional of Spain.}
(IRAM) 30~m telescope for AFGL~2591, G75.78, NGC~7538~IRS1, and
W~3(H$_2$O). The settings were chosen because of their high
line strengths and low excitation temperatures. Combined with the JCMT
settings, they allow for the determination of any cold component of
these molecules.

\begin{table*}
\centering
\caption{List of observed frequency settings and molecular lines covered at the JCMT and IRAM.}
\begin{tabular}{llllll}
\hline
\hline
Molecule & Freq. & $E_{\rm u}$ & $\mu^2S$ & Transition & Additional molecules\\
         & [GHz]   & [K]           & [D$^2$]    &            &                     \\
\hline
\multicolumn{6}{c}{JCMT}\\
\hline
HCO       & 346.725 & 41.61  & 7.341 & 4$_{0,4,5,4}$-- 3$_{0,3,4,3}$ & C$_2$H$_5$CN/C$_2$H$_5$OH/H$_2$C$^{18}$O/HCOOCH$_3$\\
          & 348.778 & 172.36 & 5.313 & 4$_{2,3,4,4}$-- 3$_{2,2,3,3}$ & HCOOCH$_3$/CH$_3$$^{13}$CN/C$_2$H$_5$OH\\
H$_2$CO   & 218.222 & 20.96  & 16.2985 & 3$_{0,3}$-- 2$_{0,2}$       & CH$_3$OH/C$_2$H$_5$CN/HCOOCH$_3$\\
          & 364.275 & 158.42 & 52.1647 & 5$_{3,3}$-- 4$_{3,2}$       & C$_2$H$_5$OH\\
CH$_3$OH  & 338.344 & 5.3431 & 69.36 & 7$_{3,4,0}$-- 6$_{3,4,0}$     & HCOOCH$_3$\\
CH$_2$CO  & 240.187 & 88.01  & 71.489 & 12$_{1,12}$-- 11$_{1,11}$    & CH$_3$OH/H$_2$C$^{18}$O/C$_2$H$_5$CN/CH$_3$CN/HCOOCH$_3$\\
          & 220.178 & 76.48  & 65.434 & 11$_{1,11}$-- 10$_{1,10}$    & CH$_3$$^{13}$CN/$^{13}$CH$_3$OH/HCOOCH$_3$\\
          & 222.298 & 182.26 & 61.0830  & 11$_{3,9}$-- 10$_{3,8}$    & CH$_3$OCH$_3$/CH$_3$CCH/HCOOCH$_3$\\         
CH$_3$CHO & 226.590 & 71.30  & 77.779 & 12$_{0,12}$-- 11$_{0,11}$    & HCOOCH$_3$ \\
C$_2$H$_5$OH & 230.991 & 85.53 & 48.953 & 16$_{5,11}$-- 16$_{4,12}$  & CH$_3$OH/C$_2$H$)5$CN\\
             & 347.916 & 250.88 & 66.314 & 20$_{4,17,0}$-- 19$_{4,16,0}$ & C$_2$H$_5$CN/HCOOCH$_3$/H$_2$C$^{18}$O \\
             & 339.979 & 57.89 & 17.724 & 9$_{4,6}$-- 8$_{3,5}$ & C$_2$H$_5$CN\\
HNCO      & 219.798 & 58.02  & 28.112 & 10$_{0,10,11}$-- 9$_{0,9,10}$ & H$_2$$^{13}$CO/C$_2$H$_5$CN\\
          & 352.898 & 187.25 & 43.387 & 16$_{1,15,17}$-- 15$_{1,14,16}$ & C$_2$H$_5$CN/C$_2$H$_5$OH/HCOOCH$_3$ \\
HN$^{13}$CO & 240.881 & 112.53 & 30.4313 & 11$_{1,11,12}$-- 10$_{1,10,11}$ & CH$_3$OCH$_3$/CH$_3$OH/HNCO \\
NH$_2$CHO & 339.904 & 245.79 & 539.342 & 16$_{6,11}$-- 15$_{6,10}$ & same setting as C$_2$H$_5$OH\\
 & 340.491 & 165.59 & 605.520 & 16$_{3,14}$-- 15$_{3,13}$ & same setting as CH$_2$CO\\
 & 345.183 & 151.59 & 664.219 & 17$_{0,17}$-- 16$_{0,16}$ & HCOOCH$_3$/C$_2$H$_5$OH/$^{13}$CH$_3$OH\\
HCOOH & 225.238 & 88.02 & 17.734 & 10$_{3,8}$-- 9$_{3,7}$ & H$_2$CO/HCOOCH$_3$/CH$_3$OCH$_3$/$^{13}$CH$_3$OH \\
& 338.144 & 180.53 & 27.1548 & 15$_{4,12}$-- 14$_{4,11}$ & CH$_3$OH/C$_2$H$_5$OH\\
 & 346.719 & 143.92 & 28.726 & 15$_{2,13}$-- 14$_{2,12}$ & same setting as HCO\\ 
& 356.137 & 158.67 & 30.6650 & 16$_{2,15}$-- 15$_{2,14}$ & NH$_2$CHO/HCOOCH$_3$/C$_2$H$_5$CN/H$_2$$^{13}$CO/CH$_3$OH\\
\hline
\multicolumn{6}{c}{IRAM}\\
\hline
CH$_2$CO  & 81.586  & 22.86  & 22.4974 &  4$_{1,3}$-- 3$_{1,2}$      \\
          & 140.127 & 39.97  & 41.1295 & 7$_{1,7}$-- 6$_{1,6}$       \\
          & 244.712 & 89.42  & 71.4841 & 12$_{1,11}$-- 11$_{1,10}$   \\
          & 262.619 & 140.63 & 25.3788 & 13$_{2,12}$-- 12$_{2,11}$   \\
CH$_3$CHO & 98.901  & 16.51  & 31.2087 & 5$_{1,4}$-- 4$_{1,3}$       \\
          & 112.249 & 21.14  & 37.9261 & 6$_{1,6}$-- 5$_{1,5}$       \\
          & 149.507 & 34.59  & 51.1868 & 8$_{1,8}$-- 7$_{1,7}$       \\
          & 168.093 & 42.66  & 57.7848 & 9$_{1,9}$-- 8$_{1,8}$       \\
HCOOH     & 223.916 & 71.93  & 18.7011 & 10$_{2,9}$-- 9$_{2,8}$      \\
          & 247.514 & 150.71 & 17.0080 & 11$_{5,6/7}$-- 10$_{5,5/6}$ \\
          & 257.975 & 83.91  & 23.2080 & 12$_{1,12}$-- 11$_{1,11}$   \\
          & 262.103 & 82.77  & 23.2998 & 12$_{0,12}$-- 11$_{0,11}$   \\
\hline
\end{tabular}
\label{obs_lines}
\end{table*}

\subsection{JCMT observations}
\label{obs_sec}

The frequencies given in the first part of Table~\ref{obs_lines} were
observed for the 7 sources with the JCMT on Mauna Kea, Hawaii from
August 2004 to January 2006. The beam size ($\theta_{\rm beam}$) is
20--21\arcsec\ for observations in the 230~GHz band and 14\arcsec\ in
the 345~GHz band. The spectra were scaled from the observed antenna
temperature scale, $T_A^\ast$, to main beam temperatures, $T_{\rm
MB}$, using main beam efficiencies $\eta_{\rm MB}$ of 0.69 and 0.63 at
230~GHz and 345~GHz, respectively. The front-ends consisted of the
facility receivers A3 and B3; the back-end was the Digital
Autocorrelation Spectrometer (DAS), covering 500~MHz instantaneous
bandwidth with a spectral resolution of 312.5 and 625~kHz
respectively. Pointing was checked every 2~hrs or whenever a new
source was observed and was always within 3\arcsec. To subtract
atmospheric and instrumental backgrounds beam switching with a chop
throw of 180\arcsec\ was used. We aimed for a $T_{\rm rms}$
$\sim$20~mK on the $T_{\rm A}^*$ scale when binned to a
$\sim$1~km~s$^{-1}$ channel, which was obtained with integration times
of $\sim$1~hr at 230~GHz and 2~hrs at 345~GHz dependent on
weather. For NGC~6334~IRS1 the confusion limit was reached after 30
minutes of integration, however, at which point the observation was
terminated. The absolute calibration is accurate to better than 15\%
from comparison to spectral line standards.

The JCMT B3 345~GHz receiver was operated in single side band mode to
minimize line confusion. The A3 230 GHz receiver does not have the
option of image sideband suppression. For line-rich sources like Orion
\citep{sutton1985}, it is difficult to assign a line to the correct
sideband. For our sources the 230~GHz spectra are not confusion
limited except for NGC~6334~IRS1 (see Fig.~\ref{spectra}). For
NGC~6334~IRS1, pairs of spectra with 10~MHz offsets were taken. This
shifts lines in the image sideband by 20~MHz with respect to those
observed in the rest sideband making a unique identification possible.

\subsection{IRAM Observations}
\label{obs_iram}

The frequencies given in the second part of Table~\ref{obs_lines} have
been observed with the IRAM 30~m telescope. The beam size
($\theta_{\rm beam}$ in arcsec) is 29\arcsec\ at 86~GHz, 17\arcsec\ at
140~GHz, and 10.5\arcsec\ at 235~GHz. The front-ends were the B100,
C150, B230, and C270 facility receivers. The VESPA auto-correlator was
used as the back-end, with a bandwidth of 120~MHz, and a channel
spacing of 80~kHz. Pointing was checked every 1.5~hrs and was always
within 2-3\arcsec . Atmospheric and instrumental backgrounds were
subtracted through ``wobbler'' switching where the telescope
alternately observes positions +220\arcsec\ and -220\arcsec\ away from
the source. Integration times differed per source and depended on the
atmospheric conditions, but were generally 30~minutes for AFGL~2591
and G75.78, and 1~hr for NGC~7538~IRS1 and W~3(H$_2$O). All
observations were converted to the main-beam temperature scale using
$F_{\rm eff}/B_{\rm eff}$ of 0.78 at 86~GHz to 0.46 at
260~GHz. Conveniently, all observations were performed in
single-sideband mode.

\section{Data analysis} \label{analysis}

\subsection{Rotation diagrams}\label{rot}
Line assignments were made by comparing the observed frequencies
corrected for source velocity with the JPL\footnote{{\tt
http://spec.jpl.nasa.gov/ftp/pub/catalog/catform.html}},
CDMS\footnote{{\tt http://www.ph1.uni-koeln.de/vorhersagen/}} and
NIST\footnote{{\tt
http://physics.nist.gov/PhysRefData/Micro/Html/contents.html}}
catalogs. Possible assignments had to be within $\sim$1~MHz, unless
the uncertainty of the frequencies in the catalogs was larger. For
NH$_2$CHO, for example, lines were only assigned if multiple
transitions were detected in the same spectrum.

Integrated intensities for the identified lines are given in
Tables~\ref{h2colines}--\ref{ch3och3lines}. These were used to derive
rotation temperatures and column densities through the rotation
diagram method \citep{goldsmith1999} when three or more lines were
detected over a sufficiently large energy range. The integrated
main-beam temperatures are then related to the column density in the
upper energy level by:
\begin{equation}
\frac{N_{\rm u}}{g_{\rm u}} = \frac{3 k \int T_{\rm
MB}dV}{8\pi^3\nu\mu^2S} \label{roteq1}
\end{equation}
where $N_{\rm u}$ is the column density of the upper level 
, $g_{\rm u}$ is the degeneracy in the upper level, $k$ is Boltzmann's
constant, $\nu$ is the transition frequency, $\mu$ is the dipole
moment, and $S$ is the line strength. The total {\it beam-averaged}
column density $N_{\rm T}$ in cm$^{-2}$ can then be computed from:
\begin{equation}
\frac{N_{\rm u}}{g_{\rm u}} = \frac{N_{\rm T}}{Q(T_{\rm rot})}
e^{-E_{\rm u}/T_{\rm rot}}\label{roteq2}
\end{equation}
where $Q(T_{\rm rot})$ is the rotational partition function, and
$E_{\rm u}$ is the upper level energy in K.

The calculated rotation temperatures and column densities can be used
to make predictions for the intensities of other lines for a specific
species. This was used as an independent check on line-assignments. If
lines predicted to have high intensities in the observed frequency
ranges are not seen, a detection is considered questionable. In the
analysis, a correction has been made for the different beam-sizes at
different frequencies. The assumption has been made that the maximum
source size is equal to the smallest observed beam-size of the JCMT
for all sources, i.e. 14\arcsec . The calculations have also been
performed assuming that the emission is extended with respect to the
largest JCMT beam-size of 21\arcsec . In this case, the rotation
temperatures are higher and have larger errors by a factor of 2. In
reality, as will be discussed in Sect.~\ref{beamdilut}, the hot core
gas will be even more severely beam-diluted. Finally, the calibration
uncertainties are taken into account in the rotation diagram analysis
and are shown as error-bars on the rotation diagram plots.

When only very few lines were detected or only a small range in
$E_{\rm u}$ was covered, no rotation temperatures or column densities
could be derived. If the detection of the species was considered real,
the column density was estimated assuming the rotation temperature to
be the average for the same molecule in other sources. If no lines
were detected, a 3$\sigma$ upper limit was determined. For lines from
isotopologues, rotation temperatures equal to the main species were
used to estimate the column densities.

A number of assumptions are implicitly made in the rotation diagram
method. In short, they are that the excitation can be characterized by
a single temperature $T_{\rm rot}$ and that the lines are optically
thin. Even though the first assumption will almost certainly be
incorrect, the rotation diagram method does give the average
excitation temperature of the region from which most of the molecular
emission arises. If the lines are sub-thermally excited because the
density is below the critical density ($n_{\rm cr}$), the rotation
temperature is a lower limit. This could also be partly responsible
for the scatter in the rotation diagrams. Since for most molecules
studied here no collisional rate coefficients are known, a more
sophisticated statistical equilibrium analysis is not possible. The
critical density, $n_{\rm cr}$, for the CH$_3$OH 7$_{6,K}$--6$_{6,K}$
transitions is $\sim$10$^6$~cm$^{-3}$ while for the H$_2$CO
3$_{0,3}$--2$_{0,2}$ transition it is
$\sim$5$\times$10$^6$~cm$^{-3}$. Since $n_{\rm cr}$ is proportional to
$\mu^2$ and $\nu^3$, transitions with high values for either will be
most affected. Most species in this study have dipole moments $\mu$
close to the CH$_3$OH value of 1.7~Debye. H$_2$CO, CH$_3$CHO, HNCO,
NH$_2$CHO, and CH$_3$CN have dipole moments 2-3 times larger and for
these species sub-thermal effects should thus be more
important. Optical depth can significantly influence the results as
well. Column densities derived from optically thick lines will be
severely under- and rotation temperatures overestimated. Lines of
isotopic species can be used to determine the optical
depth. Alternatively, it is possible to predict which lines are
optically thick based on the rotation diagram results and exclude
those from the fit. In this paper both approaches have been used. A
final uncertainty is infrared pumping. It is a well-studied phenomenon
for HNCO \citep{churchwell1986}. Its importance for other species is
unknown, although all molecules have dipole-allowed mid-infrared
transitions. However, no vibrationally excited lines have been
detected in this survey except for CH$_3$OH. With these limitations
the rotation diagram method gives a useful indication of the
temperature region where the abundance of a molecule peaks.

\subsection{Beam dilution correction and hydrogen column density in hot gas}\label{beamdilut}

\begin{figure}
\centering
\includegraphics[width=8cm]{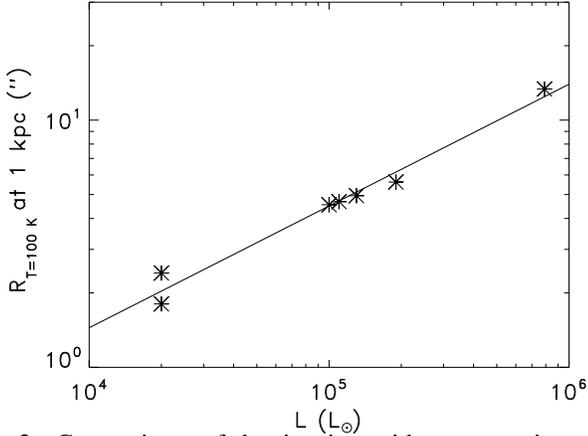}
\caption{Comparison of luminosity with source size where
$T=$100~K. $R_{T=100\rm\ K}$ is given in arcsec assuming all sources
to have a distance of 1~kpc. The solid line represent the best fit to
the data.}\label{lumi}
\end{figure}

If the emission from the complex organic species comes from a warm
region where the temperature is higher than the ice evaporation
temperature of 100~K, it becomes important to properly take the amount
of beam dilution into account, especially for inter-comparison of the
sources. The region where the temperature is above 100~K is expected
to be small ($\sim1$\arcsec) compared to the size of the single-dish
beam, and is related to the luminosity $L$. Also, the beam-dilution
depends on the distance as $d^2$. To correct for the amount of beam
dilution we established a set of models for the dust envelopes of our
sources based on SCUBA 850~$\mu$m data\footnote{The SCUBA 850~$\mu$m
data were retrieved through the JCMT archive as a Guest User at the
Canadian Astronomy Data Center, which is operated by the Dominion
Astrophysical Observatory for the National Research Council of
Canada's Herzberg Institute of Astrophysics}.  For simplification we
fix the inner radii of the envelopes to 500~AU, the outer radii to
100,000~AU ($\approx 0.5$~pc) and adopt a power-law density profile,
$n \propto r^{-1.5}$, typical of a free-falling core. $r_{\rm o}$ is
chosen to be so large that gas extending beyond this radius will not
significantly contribute. We assume that the total luminosity quoted
in Table~\ref{sources} is provided by a single 30,000~K blackbody at
the center of the envelope and fit the observed 850~$\mu$m fluxes of
each source by adjusting the mass of the envelope using the DUSTY dust
radiative transfer code \citep{ivezic1999} as described in
\citet{jorgensen2002}. The results of these models provide a
temperature profile for each source from which the radius where the
temperature increases above 100~K ($R_{T =100\rm\ K}$), and the
beam-dilution can be estimated (see
Table~\ref{profiles}). Figure~\ref{lumi} compares the inferred source
sizes of the warm gas in arcsec with luminosity of the sources with
all distances set to 1~kpc. These correlate very strongly with
luminosity as expected. Our results are consistent with the
independently derived relation from dust modeling of a large range of
sources, $R_{T=100~\rm\ K}\approx 2.3\times 10^{14} (\sqrt{L /
L_\odot})\, {\rm\ cm}$ (Doty private communication). In fact, the
luminosity itself can be used to estimate the linear size of the
region where the temperature is higher than 100~K. We therefore adopt
these radii for the beam dilution correction for molecules that
unambiguously probe warm $\geq100$~K gas estimated from the rotation
diagrams. The derived beam dilution ($\eta_{\rm bf}$),
\begin{equation}
\eta_{\rm bf} = \frac{(R_{T=100\rm\ K}/d)^2}{(\theta_{\rm
beam}/2)^2+(R_{T=100\rm\ K}/d)^2}
\end{equation}
is then used to calculate {\it source-averaged} column densities
($N_{\rm T\geq100\rm\ K}$) of hot molecules (X) in
Sect.~\ref{col_den}:

\begin{equation}
N_{\rm T\geq100\rm\ K}(X) = \frac{N_{\rm T}(X)}{\eta_{\rm bf}}.
\label{beamcol}
\end{equation}

The radiative transfer models used to calibrate the $R_{T=100\rm K}-L$
relation can also be used to calculate the hydrogen mass in the inner
warm regions and thereby provide an H$_2$ column density to determine
molecular abundances for $T\geq100$ K. With the derived density and
temperature profiles listed in Table~\ref{profiles}, the number of
H$_2$ molecules, $Mol$(H$_2$), within $R_{T=100\rm\ K}$, can be
calculated from:
\begin{equation}
Mol({\rm H_2}) = \int_{r_{\rm i}}^{R_{T=100\rm\ K}} 4\pi r^2 n_{\rm i}\left(\frac{r}{r_{\rm i}}\right)^{-\alpha} dr = \frac{4 \pi n_{\rm i} r_{\rm i}^{\alpha}}{3-\alpha}\ (R^{3-\alpha}_{T= 100\rm\ K}-r_{\rm i}^{3-\alpha}), 
\label{N_mol}
\end{equation}
where $n_{\rm i}$ is the density of H$_2$ molecules at the inner
radius, and $\alpha$ the power of the density profiles. The column
density of H$_2$ in gas with $T\geq100{\rm\ K}$, $N_{T\geq100\rm\ K}({\rm\
H_2})$ in cm$^{-2}$ is then simply $Mol({\rm H_2})$ divided by $\pi
R_{T=100\rm\ K}^2$
\begin{equation}
N_{T\geq100\rm\ K}({\rm H_2}) = \frac{4 n_{\rm i} r_{\rm i}^{\alpha}}{(3-\alpha)R^2_{T= 100\rm\ K}}\ (R^{3-\alpha}_{T= 100\rm\ K}-r_{\rm i}^{3-\alpha}). 
\label{h2_col}
\end{equation}
Since both $N_{T\geq100\rm\ K}({\rm H_2})$ and $N_{T\geq100\rm\ K}({\rm X})$
refer to column densities for the same temperature region, they can be
used directly to calculate abundances for a species X in the hot gas through:
\begin{equation}
[{\rm X}] = \frac{N_{T\geq100\rm\ K}({\rm X})}{N_{T\geq100\rm\ K}({\rm
H_2})}.
\label{abcal}
\end{equation}

It is clear from Table~\ref{profiles} that the derived values of
$R_{T=100\rm\ K}$ and $N_{T\geq100\rm\ K}$(H$_2$) are rather similar
for most sources with the exception of AFGL~2591 and G24.78. AFGL~2591
has a significantly smaller and G24.78 a much larger radius and H$_2$
column density. These are related to their low and very high
luminosities. The beam-dilution for G24.78 is thus very similar to
that for the other sources in our sample. This makes a good comparison
between G24.78 and the other sources possible.

Some of our sources have previously been modeled in more detail
\citep[e.g.,][]{vdtak2000, muller2002, hatchell2003}; these models
show some differences, e.g., in the slope of the density distribution
from source to source and even for the same source depending on
method. As an example, \citet{vdtak2000} found that $R_{T=100\rm\ K}$
is 5.3$\times$10$^3$~AU for NGC~6334~IRS1 and 2.8$\times$10$^3$~AU for
W~3(H$_2$O) with a density slope $\alpha$=2. Both values are within
10--15\% of our model result. Together with the good agreement with
the models by Doty (private communication) this implies that the
model-dependent uncertainty has a rather small effect on $R_{T=100\rm\
K}$ and the column densities. We therefore consider our adopted
approach to provide the internally most consistent set of estimates -
even though we caution that a full self-consistent set of models
(explaining, e.g., both single-dish and interferometric continuum
data) is required to estimate abundances to better than a factor 2--3.

Finally, interferometric observations have shown that even on scales
as small as a few thousand AU, there may be strong chemical
differentiation, e.g., oxygen-bearing molecules peaking at a different
position than nitrogen-bearing species
\citep{liu2005,blake1996,blake1987,wyrowski1999,sutton1995}. From
single-dish observations of a single source, it is not possible to
determine whether the emission of various species arises from
different regions. A comparison of the abundances of the species
studied in this paper for a number of sources does, however, show
whether the emission likely comes from the same or a different region
(Sect.~\ref{sec_cor}).

\begin{table}
\caption{Parameters of temperature and density profiles, and
beam-filling factors derived from 850~$\mu$m SCUBA data and source
luminosities.}\label{tab3}
\begin{center}
\begin{tabular}{l|lllll}
\hline
\hline
Sources    & $n_{\rm i}^a$       & $R_{T=100\rm\ K}$      & $\tau^b$   & $N_{T\geq 100\rm\ K}$(H$_2$)  & $\eta_{\rm bf}^c$\\
           & [cm$^{-3}$]       & [AU]                   &            & [cm$^{-2}$]                   &       \\
\hline
AFGL~2591   & 8.5(6) & 1.8(3)   & 0.40     & 7.6(22)            & 6.2(-2)\\     
G24.78      & 1.0(8) & 1.3(4)   & 4.8      & 4.0(23)            & 5.8(-2)\\     
G75.78      & 2.1(7) & 5.6(3)   & 0.98     & 1.2(23)            & 3.7(-2)\\      
NGC~6334~IRS1  & 3.8(7) & 4.7(3)   & 1.8      & 2.4(23)            & 1.3(-1)\\    
NGC~7538~IRS1  & 3.4(7) & 4.9(3)   & 1.6      & 2.1(23)            & 6.0(-2)\\     
W~3(H$_2$O) & 6.2(7) & 2.4(3)   & 2.9      & 1.8(23)            & 2.9(-2)\\
W~33A       & 4.1(7) & 4.5(3)   & 1.9      & 2.6(23)            & 2.6(-2)\\
\hline
\end{tabular}
\end{center}
$^a$ Density at inner radius of 500~AU.

$^b$ The optical depth $\tau$ is given for the dust at 100~$\mu$m.

$^c$ $\eta_{\rm bf}$ is the beam-filling factor.
\label{profiles}
\end{table} 

\section{Results}
\label{results}

\subsection{General comparison between sources}\label{general}
\begin{figure*}
   \centering
\includegraphics[width=17cm]{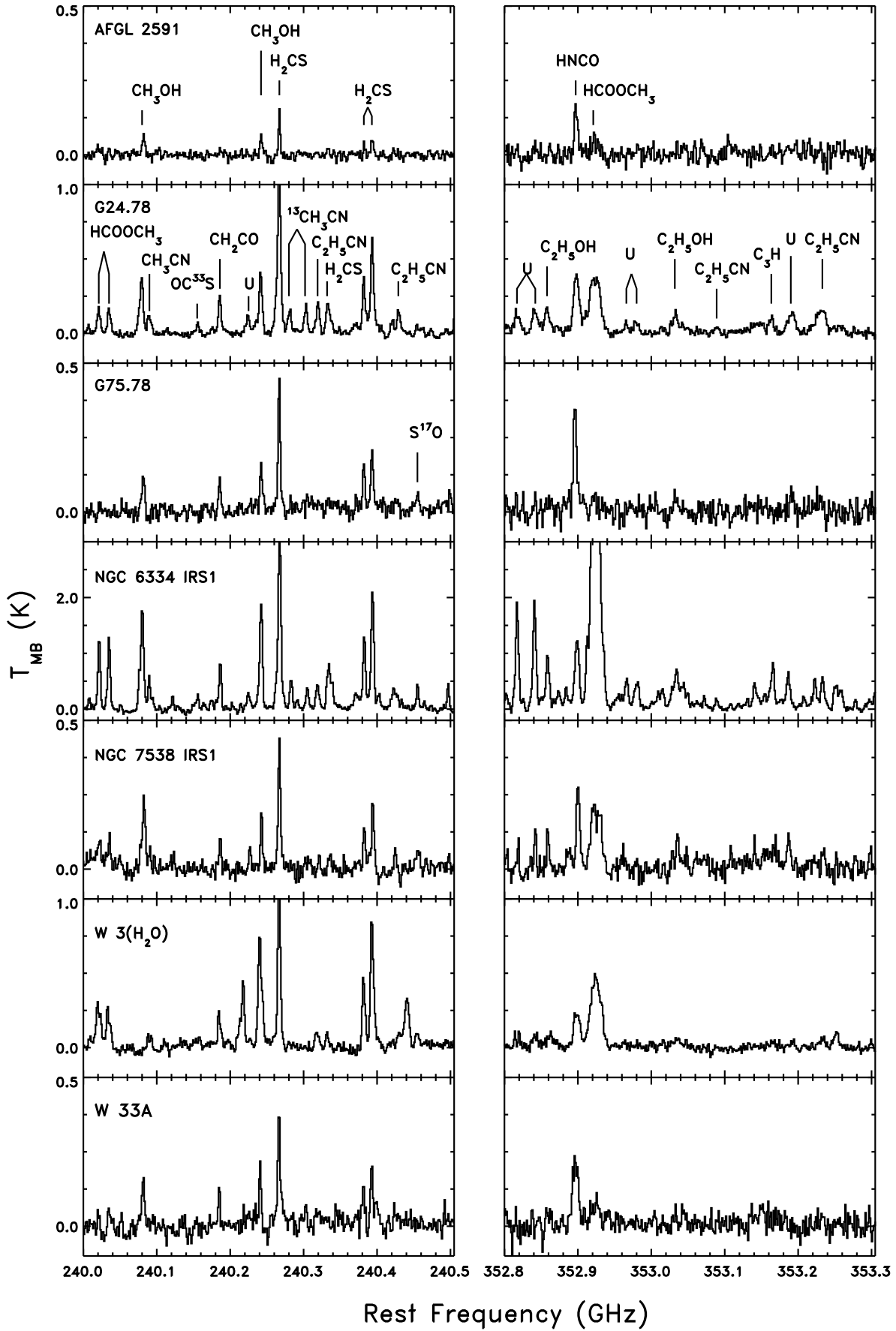}
   \caption{Spectra centered at 240.25 and 353.05~GHz for all
   sources. Note that the $T_{\rm MB}$ scales are different for
   individual sources.}
              \label{spectra}
\end{figure*}

Figure~\ref{spectra} shows the spectra at 240.25 and 353.05~GHz toward
all sources and Tables \ref{h2colines}-\ref{ch3och3lines} contain all
measured line intensities for the molecules discussed in this
paper. There is a clear difference in the ``richness'' of the sources
in molecular lines. The intensities and thus also column densities of
``line-poorer'' sources are lower for most organic species for sources
such as AFGL~2591, G75.78, NGC~7538~IRS1, and W~33A. Some lines, such
as the HCOOCH$_3$ 31$_{4,27}$-- 30$_{4,26}$ at 352.922~GHz, are weak
relative to the HNCO 16$_{1,15}$-- 15$_{1,14}$ at 352.898~GHz for the
``line-poorer'' sources, but strong for the ``line-richer'' sources,
i.e. G24.78, NGC~6334~IRS1, NGC~7538~IRS1, and W~3(H$_2$O). There
therefore seems to be an intrinsic chemical difference between the
``line-rich'' and ``line-poor'' sources.

The two settings shown in Fig.~\ref{spectra} were aimed at detecting
the CH$_2$CO 12$_{1,12}$--11$_{1,11}$ transition at 240.187~GHz and
the C$_2$H$_5$OH 21$_{1,20,2}$--20$_{2,19,2}$ line at
352.858~GHz. Both transitions were detected in most sources,
indicating that the selected settings are suitable for obtaining
detections or otherwise limits. Many lines of other target molecules
have been detected in these settings as well.

\subsection{Optical depth determinations}\label{op_depth}

\begin{table*}
\caption{Column density or line intensity ratios for
H$_2$CO/H$_2^{13}$CO, H$_2^{13}$CO/H$_2$C$^{18}$O, HNCO/HN$^{13}$CO,
and CH$_3$CN/CH$_3^{13}$CN before correction for optical
depth.}\label{isotopes}
\begin{center}
\begin{tabular}{l|lllllll}
\hline
\hline
 & \multicolumn{7}{|c}{Sources}\\
 Species        & AFGL~2591             & G24.78                & G75.78                & NGC~6334~IRS1             & NGC~7538~IRS1             & W~3(H$_2$O)            & W~33A \\
\hline
H$_2$CO/H$_2^{13}$CO$^a$         & \phantom{$<$}37 & \phantom{$<$}9.3 & \phantom{$<$}22 & \phantom{$<$}6.5 & \phantom{$<$}37 & \phantom{$<$}31 & \phantom{$<$}14\\
H$_2^{13}$CO/H$_2$C$^{18}$O$^b$  & $<$2.4 & \phantom{$<$}2.8 & $>$1.7 & \phantom{$<$}2.4 & $<$1.6 & $>$3.0 & $>$3.5\\
HNCO/HN$^{13}$CO$^c$             & $>$2.4 & $>$11 & $>$6.5 & $>$8.6 & $>$2.0 & $>$3.6 & $>$7.3\\
CH$_3$CN/CH$_3^{13}$CN$^d$       & \phantom{$<$}-- & \phantom{$<$}5.8 & $>$5.9 & \phantom{$<$}25 & \phantom{$<$}7 & $>$9.5 & \phantom{$<$}7.6\\
\hline
\end{tabular}
\end{center}
$^a$ Derived from the 3$_{1,2}$--2$_{1,1}$ transition.

$^b$ Derived from the 5$_{2,3}$--4$_{2,2}$ transition.

$^c$ Derived from the 10$_{0,10}$--9$_{0,9}$ transition.

$^d$ Derived from the overall column density determined from all
detected transitions.

\end{table*}

\begin{table*}
\caption{Column densities $N$ for hot core molecules (source-averaged)
and cold molecules (beam-averaged) $^a$ [cm$^{-2}$]. \label{colres}}
\begin{center}
\begin{tabular}{l|lllllll}
\hline
\hline
 & \multicolumn{7}{|c}{Sources}\\
 Species        & AFGL~2591 & G24.78 & G75.78 & NGC~6334~IRS1 & NGC~7538~IRS1 & W~3(H$_2$O) & W~33A \\
\hline
 \multicolumn{8}{c}{Hot molecules}\\
\hline
H$_2$CO$^b$         & \phantom{$<$}1.3(16)     &  \phantom{$<$}6.4(16)  & \phantom{$<$}2.2(16)  &  \phantom{$<$}1.3(17)  & \phantom{$<$}2.5(16)   &  \phantom{$<$}1.8(17)  &  \phantom{$<$}5.4(16)    \\
H$_2^{13}$CO    & \phantom{$<$}2.4(14)     &  \phantom{$<$}1.2(15)  & \phantom{$<$}4.0(14)  &  \phantom{$<$}2.5(15)  & \phantom{$<$}4.8(14)   &  \phantom{$<$}3.3(15)  &  \phantom{$<$}1.0(15)\\
H$_2$C$^{18}$O  & \phantom{$<$}7.4(13) &  \phantom{$<$}4.3(14) &  $<$2.5(14) &  \phantom{$<$}1.1(15) &  \phantom{$<$}1.3(14) &  $<$1.7(15) &  $<$3.0(14)\\
CH$_3$OH$^c$        & \phantom{$<$}4.7(16)      & \phantom{$<$}2.8(17)  & \phantom{$<$}1.1(17) & \phantom{$<$}9.7(17)  & \phantom{$<$}1.2(17)   & \phantom{$<$}1.0(18)  & \phantom{$<$}2.0(17)   \\
$^{13}$CH$_3$OH & $<$1.8(15) & \phantom{$<$}2.9(16) & \phantom{$<$}4.3(15) & \phantom{$<$}8.5(16) & \phantom{$<$}6.7(15) & \phantom{$<$}3.2(16) & \phantom{$<$}6.5(15)\\
C$_2$H$_5$OH    &$<$1.0(15)  &  \phantom{$<$}7.1(15) &$<$2.4(15)&  \phantom{$<$}1.9(16) & \phantom{$<$}5.7(15)  & \phantom{$<$}8.4(15)  &  \phantom{$<$}4.7(15)   \\
HNCO            & \phantom{$<$}8.3(14) &  \phantom{$<$}5.0(15) & \phantom{$<$}3.8(15)  &   \phantom{$<$}4.3(15) & \phantom{$<$}2.3(15)  & \phantom{$<$}4.9(15)  &  \phantom{$<$}6.6(15)   \\
HN$^{13}$CO     & $<$4.7(14) &  $<$9.3(14) & $<$1.2(15) &  $<$3.8(14) &  $<$1.2(14) &  $<$2.7(15) & $<$1.2(15)\\
NH$_2$CHO       & $<$1.3(15) &  \phantom{$<$}7.2(14)&\phantom{$<$}2.0(14) &   \phantom{$<$}7.4(14) & \phantom{$<$}5.7(14) & \phantom{$<$}1.3(15) &  \phantom{$<$}2.1(15)  \\
CH$_3$CN        & $<$3.5(15)  & \phantom{$<$}5.9(16)$^d$   & \phantom{$<$}1.8(15) &  \phantom{$<$}2.9(16)$^d$   &$<$8.2(15)& \phantom{$<$}7.0(15)  &  \phantom{$<$}2.7(16)  \\
CH$_3^{13}$CN   & $<$5.3(14) & \phantom{$<$}1.1(15) &  $<$3.0(14) &  \phantom{$<$}5.6(14) & $<$1.2(15) & $<$7.2(14) &  $<$5.0(14)\\
C$_2$H$_5$CN    & $<$7.5(14) &  \phantom{$<$}4.0(15) & $<$1.2(15) & \phantom{$<$}5.1(15)& $<$9.2(14) & \phantom{$<$}4.5(15) & $<$2.1(15) \\
HCOOCH$_3$      & $<$2.4(16)  & \phantom{$<$}3.1(16)   & \phantom{$<$}7.1(15)  & \phantom{$<$}1.2(17)   & \phantom{$<$}1.4(16)  & \phantom{$<$}5.2(16) &  \phantom{$<$}2.5(16)  \\
CH$_3$OCH$_3$   & $<$7.7(15)   & \phantom{$<$}1.2(17)   & \phantom{$<$}2.3(16)& \phantom{$<$}5.8(17)   &$<$1.6(16) & \phantom{$<$}1.5(17) &  \phantom{$<$}2.7(16)   \\
\hline
 \multicolumn{8}{c}{Cold molecules}\\
\hline
CH$_2$CO        & \phantom{$<$}1.6(13)  &  \phantom{$<$}2.1(14) & \phantom{$<$}6.4(13)  & \phantom{$<$}7.2(14) & \phantom{$<$}9.7(13)  & \phantom{$<$}1.1(14) &  \phantom{$<$}6.3(13) \\
CH$_3$CHO       & \phantom{$<$}3.1(12)   &$<$1.5(13) &$<$2.1(13)& $<$1.2(14) & \phantom{$<$}2.8(13) & \phantom{$<$}3.5(13) & $<$3.0(13) \\
HCOOH           & \phantom{$<$}3.4(13) &  \phantom{$<$}2.1(14) &\phantom{$<$}7.9(13)&   \phantom{$<$}4.9(14) & \phantom{$<$}9.8(13) & \phantom{$<$}1.5(14)  &  \phantom{$<$}1.3(14)  \\
CH$_3$CCH       & \phantom{$<$}7.9(14)     & \phantom{$<$}2.3(15)   & \phantom{$<$}8.6(14)  &  \phantom{$<$}5.2(15) & \phantom{$<$}8.4(14)  & \phantom{$<$}1.5(15)   &  \phantom{$<$}1.3(15)  \\
\hline
\end{tabular}
\end{center}
$^a$ The column densities for the hot core molecules are corrected for
beam-dilution using the values of $\eta_{\rm bf}$ listed in
Table~\ref{lumi}. The cold molecules are given as a beam-averaged
column density for a beam of 14\arcsec .

$^b$ From H$_2^{13}$CO multiplied by 53.

$^c$ From rotation diagram excluding optically thick lines.

$^d$ From CH$_3^{13}$CN multiplied by 53.

\end{table*}

Some species such as CH$_3$OH, H$_2$CO, and CH$_3$CN have previously
been found to be optically thick in their main lines \citep[see
e.g.,][]{helmich1997}. Lines for selected isotopic species were used
to determine whether the lines of the main species are optically thick
and to derive column densities for the main isotopologues (see
Tables~\ref{h2colines}, \ref{ch3ohlines}, \ref{ch3ohlines2},
\ref{hncolines}, and \ref{ch3cnlines}). For H$_2$CO and HNCO, the same
transition could be used for different isotopologues. For CH$_3$CN,
the optical depth was determined by comparing the column density
derived from the rotation diagram analysis for the main species and
that derived from the detected transition for the isotopologues. The
ratios for the beam-averaged column densities are given in
Table~\ref{isotopes}, and the corrected source-averaged column
densities in Table~\ref{colres}. In the remainder of the paper a
$^{12}$C/$^{13}$C ratio value of 53 for the 4~kpc molecular ring has
been used \citep{wilson1994}. Note that this value is uncertain,
however, and known to depend on the distance to the galactic
center. However, for simplicity we have assumed the same ratio for all
sources.

The rotation diagram results by \citet{vdtak2000} were not corrected
for optical depth, because no isotopologues were observed. In our
data, however, lines for H$_2^{13}$CO and H$_2$C$^{18}$O are observed.
The optical depth of H$_2$CO has been determined from the
3$_{1,2}$--2$_{1,1}$ transition. The observed H$_2$CO/H$_2^{13}$CO
ratio ranges from 6.5 for NGC~6334~IRS1 to 37 for AFGL~2591 and
NGC~7538~IRS1, lower than the expected ratio, which implies that the
main isotopologue is optically thick in many of its transitions. The
results for W~3(H$_2$O) are consistent with the optical depth found by
\citet{helmich1997}. The optical depth of H$_2^{13}$CO was estimated
based on the 5$_{2,3}$-- 4$_{2,2}$ transition observed for both
H$_2^{13}$CO and H$_2$C$^{18}$O. The H$_2^{13}$CO/H$_2$C$^{18}$O ratio
is lower than the expected value of 5--8 for most sources, but it is
unclear whether this difference is significant. We assume therefore
for simplicity that H$_2^{13}$CO is optically thin for all sources.

No direct comparison of the same transition is possible for CH$_3$OH
with $^{13}$CH$_3$OH. Many detected lines have strengths that are
lower by more than an order of magnitude compared to the strongest
transitions. It is therefore possible to determine the column density
of the main isotopic species by excluding lines predicted to be
optically thick for the {\it source-averaged} column density from the
rotation diagram. This also circumvents the additional error
introduced by the uncertain isotopic ratios. As seen in
Fig.~\ref{ch3oh}, this excludes mainly lines lying below the fit to
the optically thin lines and results in a much better determined
rotation temperature.

The limits on the HNCO/HN$^{13}$CO ratios are smaller than $\sim$53,
and HNCO could thus be optically thick for all sources.  Note that the
optical depth determination of HNCO is potentially complicated by the
close spacing of the lines of the main species and its isotopologue of
only $\sim$5~MHz. However, since most of our sources have line widths
of only 3-4 km s$^{-1}$, the upper limits for HN$^{13}$CO are
significant. Since HN$^{13}$CO has not been detected for any of our
sources and most sources studied by \citet{zinchenko2000}, HNCO is
assumed to be optically thin throughout the remainder of this paper.

Since it was not one of the target molecules, CH$_3$CN was not
observed for all sources. Its lines and those of CH$_3^{13}$CN were
coincidentally present in the image sidebands for some of the
sources. For G24.78, NGC~6334~IRS1, and W~33A, both the main species
and its isotopologue have been detected and CH$_3$CN is found to be
optically thick in the 12--11 and 13--12 transitions with ratios
ranging from 5.8 for G24.78 to 24 for NGC~6334~IRS1. For G24.78, this
result confirms the findings by \citet{beltran2005}.

Since HCOOCH$_3$ and CH$_3$OCH$_3$ have relatively high column
densities, predictions were made for which lines were optically
thick. All lines detected for both molecules were found to be
optically thin, if we assume that the emission does not come from a
radius smaller than $R_{T=100 \rm K}$.

\subsection{Rotation temperatures}
\label{rot_temp}

In the following subsection we discuss each molecule individually and
compare it with literature; the main results are summarized at the end
of this subsection. The rotation temperatures were derived from single
temperature fits to rotation diagrams (see
Sect.~\ref{rot}). Calibration uncertainties of 15\% are shown as error
bars on the rotation diagrams. All rotation diagrams are shown in
Appendix~\ref{rotdia}, and resulting rotation temperatures in
Table~\ref{tempres}.

\underline{H$_2$CO:} For H$_2$CO, the data are taken from
\citet{vdtak2000} except for G24.78 and G75.78 (see
Fig.~\ref{h2co}). The rotation temperatures for all sources except
NGC~6334~IRS1 and W~3(H$_2$O) are $\sim$80~K. NGC~6334~IRS1 and
W~3(H$_2$O) have temperatures of about 100~K higher. However, as the
optical depths are similar for most sources, this temperature
difference between the sources is real.

\underline{CH$_3$OH:} For CH$_3$OH no clear temperature trend is
seen, but the temperature is generally higher than 100~K. This implies
that the emission is coming from warm gas. The CH$_3$OH rotation
diagram for some sources like W~33A appears to consist of two
temperature components as seen previously by \citet{vdtak2000}
(Fig.~\ref{ch3oh}). However, deducing the presence of two components
from a rotation diagram is non-trivial as this behavior could also
result from sub-thermal excitation and optical depth effects. Indeed,
excluding lines of high optical depth leads to the disappearance of
the two-component structure for most sources other than W~33A. The
fits in Fig.~\ref{ch3oh} are especially sensitive to lines with high
excitation temperatures and are therefore thought to probe the hot
component of the CH$_3$OH gas.

\begin{table*}
\caption{Derived rotation temperatures$^a$ in K from rotation diagram analysis (see Appendix~\ref{rotdia}). \label{tempres}}  
\begin{center}
\begin{tabular}{l|lllllll}
\hline
\hline
 & \multicolumn{7}{|c}{Sources}\\
 Species & AFGL~2591 & G24.78 & G75.78 & NGC~6334~IRS1 & NGC~7538~IRS1 & W~3(H$_2$O) & W~33A \\
\hline
H$_2$CO         &  89$^b$ & 83$\pm$9 &  69$\pm$6  & 193$^b$ &  87$^b$ & 181$^b$ &  88$^b$ \\
CH$_3$OH        &  147$\pm$11    & 211$\pm$13 & 113$\pm$7 & 178$\pm$10 & 156$\pm$10 & 139$\pm$8 & 259$\pm$16\\
C$_2$H$_5$OH    &  [139]  & 104$\pm$7 &[139] & 166$\pm$15 & 164$\pm$17 & 129$\pm$16 & 122$\pm$16 \\
HNCO            &  64$\pm$8  &  96$\pm$11 & 108$\pm$13 & 105$\pm$12 & 278$\pm$88 & 147$\pm$24 &  85$\pm$8 \\
NH$_2$CHO       &  [119]  & 170$\pm$20 & 78$\pm$22 & 166$\pm$29 & 164$\pm$28 &  71$\pm$7 &  40$\pm$13 \\
CH$_3$CN        & [218]   & 261$\pm$20 & 189$\pm$25 & 170$\pm$13 &[218]& 196$\pm$14 & 278$\pm$44 \\
C$_2$H$_5$CN    & [96]   & 99$\pm$5  &[96]&  92$\pm$3 &[96]& 94$\pm$16  &[96]\\
HCOOCH$_3$      & [119]   & 121$\pm$6 &  87$\pm$13 & 144$\pm$7 & 134$\pm$8 & 109$\pm$7 & 112$\pm$15 \\
CH$_3$OCH$_3$   & [130]   & 134$\pm$12 &[130]& 241$\pm$35 &[130]&  94$\pm$5 &  43$\pm$7 \\
\hline
CH$_2$CO        &  [47] &  39$\pm$6 &  40$\pm$6 &  67$\pm$20 & 37$\pm$4  &  54$\pm$8 &  45$\pm$7 \\
CH$_3$CHO       &  [20] & [20] & [37.5] & [37.5] & 18$\pm$5 & 16$\pm$3 & [37.5] \\
HCOOH           &  [55] &  42$\pm$7 &[55]& 75$\pm$15 &  73$\pm$12 &  189$\pm$108 &  38$\pm$6 \\
CH$_3$CCH       &     45$\pm$12  &  59$\pm$6 &  59$\pm$14 &  63$\pm$12 &  58$\pm$10 &  82$\pm$15 &  40$\pm$6 \\
\hline
\end{tabular}
\end{center}
$^a$ The bracket notation indicates assumed rotation temperature, when
no temperature could be derived from a rotation diagram.

$^b$ H$_2$CO data from \citet{vdtak2000}.
\end{table*}

\underline{CH$_2$CO:} CH$_2$CO has temperatures that are very low,
generally only around 40~K. Slightly higher temperatures are found for
NGC~6334~IRS1 and W~3(H$_2$O). These temperatures suggest that this
molecule is not present in hot regions, but rather in the cold
envelope. The IRAM and JCMT detections for NGC~7538~IRS1 and
W~3(H$_2$O) are consistent. \citet{macdonald1996} find very high
values of 273~K for G34.3 based on three lines with $E_{\rm
u}>$100~K. However the integrated intensity for the CH$_2$CO
17$_{1,16}$--16$_{1,15}$ line at 346.6~GHz with $E_{\rm u}>$150~K in
this survey, also detected by \citet{macdonald1996}, is consistent
with the low rotation temperature found in our survey.

\underline{CH$_3$CHO:} Due to unfavorable weather conditions, IRAM
observations of CH$_3$CHO have only been performed for AFGL~2591,
G75.78, NGC~7538~IRS1, and W~3(H$_2$O). Almost all targeted lines were
detected toward the sources NGC~7538~IRS1 and W~3(H$_2$O). Both
sources have rotation temperatures $<$20 K, consistent with the
results found by \citet{ikeda2001,ikeda2002}. Although its relatively
high dipole moment of 2.69 D implies that sub-thermal excitation can
play a role, it is unlikely that the temperature is above 100~K,
especially since molecules with similarly high dipole moments such as
HNCO and CH$_3$CN have much higher rotation temperatures.

\underline{C$_2$H$_5$OH:} C$_2$H$_5$OH consistently has temperatures
of more than 100~K. This is somewhat higher than was found by
\citet{ikeda2001}, which could be due to both the detection of more
high excitation lines in this study. For the lower energy transitions,
a steeper slope is observed, which indicates a lower temperature.

\underline{HCOOH:} HCOOH has relatively low temperatures ranging from
$\sim$40~K for G24.78, and W~33A to $\sim$70~K for NGC~6334~IRS1 and
NGC~7538~IRS1. For NGC~7538~IRS1 and W~3(H$_2$O), it was possible to
combine results from both IRAM and the JCMT. For NGC~7538~IRS1, low
rotation temperatures are confirmed, whereas for W~3(H$_2$O) a scatter
of more than one order of magnitude is present. As the W~3(OH) region
is close to W~3(H$_2$O) it is possible that emission from this region
is picked up in one of the settings. Alternatively, HCOOH could be
present in both hot and cold gas with only the cold component present
for NGC~7538~IRS1. Due to the larger beam size of the JCMT, the 230
GHz observations are more sensitive to extended emission, whereas the
IRAM 270 GHz observations probe a more compact region. Since most
sources have low rotation temperatures, we conclude that it is mostly
located in a cold environment. Low rotation temperatures have
previously been inferred for HCOOH in other high mass star forming
regions \citep{ikeda2001}.

\underline{HNCO:} The rotation temperature of HNCO varies between the
different sources. This is not unexpected since the HNCO excitation is
known to be dominated by radiation and not collisions
\citep{churchwell1986}. It is therefore not likely to provide much
information about the actual kinetic temperature of the
species. However it does need to be close to an infrared source to be
excited, and thus must be present in hot gas. Finally, it is important
to note that excluding the line with $E_{\rm u}>$300~K (see
Fig.~\ref{hnco}) strongly reduces the inferred rotation temperature
for NGC~6334~IRS1 from 180~K to 106~K.

\underline{NH$_2$CHO:} NH$_2$CHO has a variety of rotation
temperatures. The scatter in the derived rotation temperatures appear
to be relatively large, with especially NGC~7538~IRS1 being an outlier
(Fig.~\ref{nh2cho}). There is also some indication that there may be
two temperature components - most clearly seen for the sources where
many lines with $E_{\rm u}>$200~K are detected. Since NH$_2$CHO has
lines very close to the detection limit, this component may also be
present for the other sources, but simply below the detection limit.

\underline{CH$_3$CN:} Rotation temperatures of CH$_3$CN range from
170--280~K. It is thus always present in hot gas (see
Fig.~\ref{ch3cn}). The particularly high temperatures for G24.78 and
W~33A are upper limits due to high optical
depths. \citet{pankonin2001} find lower rotation temperatures of 89~K
for G24.78 which confirm this picture. For NGC~6334~IRS1 only three
lines with $E_{\rm u} \geq$~500~K have been detected, making the
resulting rotation temperature uncertain. However, it seems clear that
the rotation temperature for CH$_3$CN is high, and the species must be
present in hot gas.

\underline{C$_2$H$_5$CN:} The other cyanide, C$_2$H$_5$CN, only has
enough detected lines for three of the sources to allow a
determination of the rotation temperature (see Fig.~\ref{c2h5cn}). For
all three sources temperatures of $\sim$90~K are found. Fewer lines
have been detected for W~3(H$_2$O), but the fact that the results for
this source are rather similar gives additional credibility to the
derived temperatures.

\underline{CH$_3$CCH:} CH$_3$CCH generally has rotation temperatures
below $\sim$80~K - except for G24.78, NGC~6334~IRS1, and
W~3(H$_2$O). The higher temperatures for these three sources are,
however, questionable. Removing the point with $E_{\rm u}>$300~K gives
significantly lower temperatures for G24.78 and NGC~6334~IRS1 (see
Fig.~\ref{ch3cch}). The W~3(H$_2$O) generally has a higher uncertainty
and scatter on the data, which may in fact cause the higher
temperature found for this source.

\underline{HCOOCH$_3$:} HCOOCH$_3$ is detected in almost all of
the sources, but with a relatively large scatter on the data. The
temperature generally seems to be between 100--150 K, with somewhat
lower values when less lines have been detected. For G75.78 the
results are inconclusive as the upper limits on high energy
transitions are not significant. Somewhat lower temperatures were
found by \citet{ikeda2001}, which could be due to a difference in the
number of detected lines.

\underline{CH$_3$OCH$_3$:} CH$_3$OCH$_3$ also appears to have a
relatively large spread in rotation temperatures (see
Fig.~\ref{ch3och3}). For W~33A, only lines with $E_{\rm u}<$100~K were
detected, which result in low rotation temperatures. The upper limits
on lines with $E_{\rm u}>$150 K are not significant, however, and do
not exclude temperatures above 100 K. Rotation temperatures above
100~K are commonly seen for CH$_3$OCH$_3$ \citep{ikeda2001}.

\underline{In summary}, the temperatures for most molecules are rather
consistent between all sources. The molecules can clearly be divided
into ``hot'' ($T>100$~K) and ``cold'' ($T<$100~K) where H$_2$CO,
CH$_3$OH, C$_2$H$_5$OH, HNCO, NH$_2$CHO, CH$_3$CN, C$_2$H$_5$CN,
HCOOCH$_3$, and CH$_3$OCH$_3$ belong to the first and CH$_2$CO,
CH$_3$CHO, HCOOH and CH$_3$CCH to the second category.

\subsection{Column densities}
\label{col_den}
\begin{figure*}
\includegraphics[width=17cm]{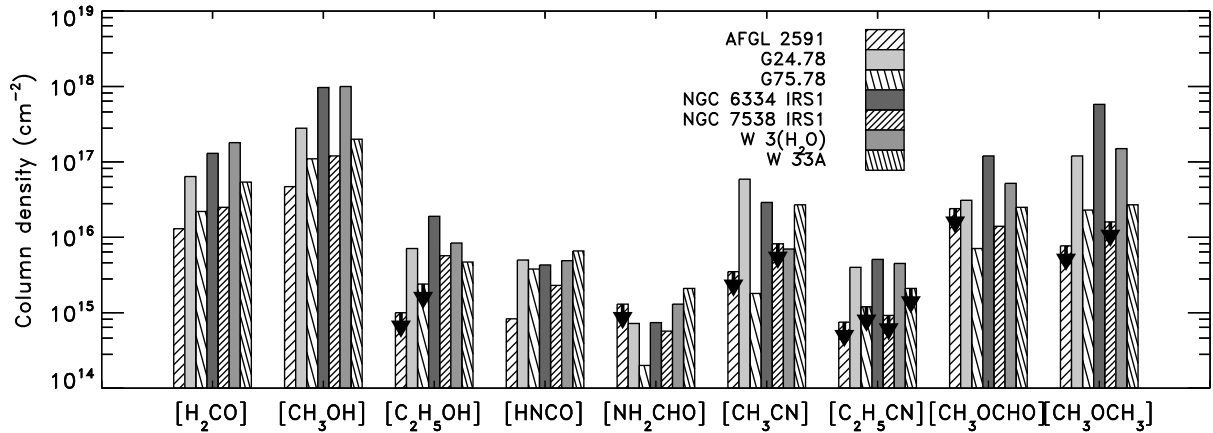}
\caption{Histogram of molecular column densities of hot molecules
corrected for beam dilution for all sources for a source size defined
by $T>$100~K.}\label{col_plot}
\end{figure*}

\begin{figure}
\includegraphics[width=8cm]{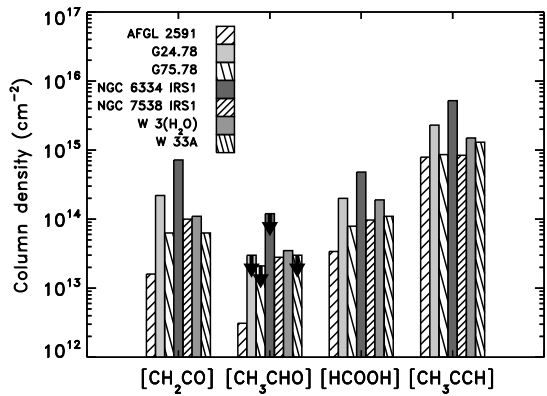}
\caption{Histogram of molecular beam-averaged column densities of the
cold molecules for all sources.}\label{col_plot2}
\end{figure}

Table~\ref{colres} and Figure~\ref{col_plot} present the {\it
source-averaged} column densities, i.e., corrected for beam dilution
(Sect.~\ref{beamdilut}), for the hot molecules per source. Absolute
column densities are compared to determine the variability between
different sources. Since we have assumed the beam dilution to be the
same for all molecules, the trends for the {\it beam-averaged} column
densities are similar. Molecules with rotation temperatures below
100~K are present in a different, cold region, presumably the envelope
surrounding the hot region, and beam-averaged column densities are
therefore given for those. Thus, the column densities for the hot and
cold molecules in Fig.~\ref{col_plot} and \ref{col_plot2} cannot be
compared directly.

Clearly, most molecules have column densities that vary by about an
order of magnitude in the hot gas between the different sources. The
column density variations do not scale with the luminosity and thus
total mass in the hot region. CH$_3$OH, H$_2$CO, but also HCOOCH$_3$,
and CH$_3$OCH$_3$ - if detected - have column densities that are
orders of magnitude higher than those of C$_2$H$_5$OH, HNCO, and
NH$_2$CHO. As will be discussed in Sect.~\ref{sec_cor} abundances show
a similar trend. The advantage of comparing column densities is that
they come directly from the observations and no assumption on the
H$_2$ column in the hot gas enters the analysis.

Most hot molecules have source-averaged column densities that are
highest for the line-rich sources NGC~6334~IRS1, G24.78, and
W~3(H$_2$O). For the oxygen-bearing molecules C$_2$H$_5$OH,
CH$_3$OCH$_3$, HCOOCH$_3$, G24.78 has a higher column density than
W~3(H$_2$O). W~3(H$_2$O) has a higher column density of CH$_3$OH and
H$_2$CO, however. Thus H$_2$CO and CH$_3$OH have variable column
densities compared to the other oxygen-bearing molecules. Nonetheless
the column density trends between the oxygen-bearing species are
similar.

The HNCO and NH$_2$CHO column densities peak for W~33A and G24.78, and
have very similar column density trends with respect to each other.
W~33A has the highest and G75.78 the lowest column
densities. NGC~7538~IRS1 has a relatively high column density for
these species compared with the oxygen-bearing species. The column
densities of the nitrogen-bearing species do, therefore, behave
different from the oxygen-bearing species.

CH$_3$CN has a relatively high column density for G24.78 and W~33A,
but the uncertainties make it more difficult to compare its column
density trends with other molecules (see Sect.~\ref{rot_temp}).  The
general CH$_3$CN column density trends appear more similar to those of
HNCO and NH$_2$CHO than to the O-bearing species. However, if we
consider only the two sources for which both the main species and its
isotopologue CH$_3^{13}$CN were detected with $E_{\rm u}<$500~K, the
relative column densities appear more similar to those of the
oxygen-bearing species.

Figure~\ref{col_plot2} shows the {\it beam-averaged} column densities
for the cold molecules, CH$_2$CO, HCOOH, and CH$_3$CCH. Clearly all
three have very similar trends. NGC~6334~IRS1 has the highest column
density and AFGL~2591 the lowest. For CH$_3$CHO, the number of sources
with detections is limited and it was thus not possible to compare its
column density trends with the other species.

In summary, the molecular abundances vary about an order of magnitude
between the sources. Furthermore, all oxygen-bearing species have
similar abundance trends, but the abundances of HNCO and NH$_2$CHO
behave differently.

\section{Column density and abundance correlations} \label{sec_cor}
Molecular column densities and abundances can be compared between
sources in different ways, two of which will be presented
here. Abundances can be estimated relative to column densities of (i)
H$_2$ calculated on the basis of CO, (ii) H$_2$ derived from dust
continuum observations (see Sect.~\ref{beamdilut}) or (iii) one of the
observed molecules, typically CH$_3$OH. The disadvantage of method (i)
is that CO gas is likely extended over much larger scales than the hot
molecules due its lower desorption temperature. Observations of high
excitation lines of CO isotopologues which probe only hot gas, will
make this method possible in the future. In this paper, we use (ii)
and (iii), and adopt (i) only for the cold molecules.

Abundances with respect to H$_2$ column densities from modeling of
dust continuum observations (Sect.~\ref{beamdilut}) and to CH$_3$OH
column densities (Sect.~\ref{ch3oh_cor}) have the advantage that they
allow comparisons to ``only material at high temperatures'' or
``species probing the regions solely of interest''. Both approaches
also have disadvantages, however. Since H$_2$ is not directly
observed, one has to rely on other tracers, in our case
dust. Extrapolations of density and temperature profiles are made to
scales smaller than actually probed by the observations. CH$_3$OH is
directly observed, but this method implicitly assumes that the
abundance of CH$_3$OH is constant from source to source. If this is
not the case incorrect correlations of abundances with respect to
CH$_3$OH can be found, as we will return to in Sect.~\ref{ch3oh_cor}.

\subsection{Abundances in $T\geq$100~K gas}
\label{res_100}

Table~\ref{ab_var100} lists the molecular abundances within the gas
with $T\geq$100~K for each source in our sample calculated through
Eq.~(\ref{abcal}). The abundances have the same trends between the
different sources as the column densities. Most molecules have
abundances that vary one order of magnitude from source to source, as
was also seen for CH$_3$OH by \citet{vdtak2000}. The standard
deviation on the average abundances is larger than our estimated
uncertainty of a factor 2-3 (Sect.~\ref{analysis}). Thus the observed
variations are real. Furthermore, the overall abundances are
relatively high compared to values in the literature. This is because
we have determined the abundances with respect to the H$_2$ column
density within $R_{T=100\rm\ K}$, whereas in e.g., \citet{ikeda2001}
they are determined based on CO transitions, which trace gas that is
cooler and extended with respect to the hot core.

\begin{table*}
\caption{Molecular abundances ($N$(X)/$N$(H$_2$)). }\label{ab_var100}
\begin{center}
\begin{tabular}{l|lllllllll}
\hline
\hline
            & AFGL~2591           & G24.78             & G75.78             & NGC~6334~IRS1           & NGC~7538~IRS1          & W~3(H$_2$O)         & W~33A                & Average & $\sigma$$^a$\\
\hline
 \multicolumn{10}{c}{Hot molecules$^b$}\\
\hline
H$_2$CO     & \phantom{$<$}1.7(-7) & \phantom{$<$}1.6(-7) & \phantom{$<$}1.8(-7) & \phantom{$<$}5.4(-7)  & \phantom{$<$}1.2(-7) & \phantom{$<$}1.0(-6) & \phantom{$<$}2.1(-7) & \phantom{$<$}3.4(-7) & 3.2\\
CH$_3$OH    & \phantom{$<$}7.0(-7) & \phantom{$<$}6.5(-7) & \phantom{$<$}9.2(-7) & \phantom{$<$}4.0(-6) & \phantom{$<$}5.7(-7) & \phantom{$<$}5.4(-6) & \phantom{$<$}7.3(-7) & \phantom{$<$}1.9(-6) & 2.0\\
C$_2$H$_5$OH & $<$1.4(-8) & \phantom{$<$}1.9(-8) & \phantom{$<$}2.0(-8) & \phantom{$<$}7.5(-8) & \phantom{$<$}2.5(-8) & \phantom{$<$}4.4(-8) & \phantom{$<$}1.8(-8) & \phantom{$<$}3.6(-8) & 2.3\\
HNCO        & \phantom{$<$}7.0(-9) & \phantom{$<$}1.3(-8) & \phantom{$<$}3.2(-9) & \phantom{$<$}1.8(-8) & \phantom{$<$}1.1(-8) & \phantom{$<$}2.7(-8) & \phantom{$<$}2.5(-8) & \phantom{$<$}1.6(-8) & 0.91\\
NH$_2$CHO & $<$1.7(-8) & \phantom{$<$}1.8(-9) & \phantom{$<$}1.7(-9) & \phantom{$<$}3.1(-9) & \phantom{$<$}2.6(-9) & \phantom{$<$}5.6(-9) & \phantom{$<$}8.5(-9) & \phantom{$<$}3.9(-9) & 2.7\\
CH$_3$CN & $<$4.6(-8) & \phantom{$<$}1.5(-7) & \phantom{$<$}1.5(-8) & \phantom{$<$}1.2(-7) & $<$3.9(-8) & \phantom{$<$}3.9(-8) & \phantom{$<$}1.0(-7) & \phantom{$<$}8.5(-8) & 5.6\\
C$_2$H$_5$CN & $<$9.9(-9) & \phantom{$<$}1.0(-8) & $<$1.0(-8) & \phantom{$<$}2.1(-8) & $<$4.4(-9) & \phantom{$<$}2.5(-8) & $<$8.1(-9) & \phantom{$<$}1.7(-8) & 0.77\\
HCOOCH$_3$ & $<$3.2(-7) & \phantom{$<$}7.3(-8) & \phantom{$<$}5.9(-8) & \phantom{$<$}4.6(-7) & \phantom{$<$}6.7(-8) & \phantom{$<$}2.9(-7) & \phantom{$<$}9.6(-8) & \phantom{$<$}1.7(-7) & 1.6\\
CH$_3$OCH$_3$ & $<$1.0(-7) & \phantom{$<$}3.0(-7) & \phantom{$<$}1.9(-7) & \phantom{$<$}2.4(-6) & $<$7.6(-8) & \phantom{$<$}8.3(-7) & \phantom{$<$}1.0(-7) & \phantom{$<$}7.7(-7) & 9.6\\
\hline
  \multicolumn{10}{c}{Cold molecules$^c$}\\
\hline 
CH$_2$CO  & \phantom{$<$}3.1(-11) & \phantom{$<$}8.8(-10) & \phantom{$<$}8.5(-11)    & \phantom{$<$}2.8(-10)    & \phantom{$<$}4.1(-10) & \phantom{$<$}1.3(-10) & \phantom{$<$}4.1(-10) &  \phantom{$<$}3.2(-10) &  2.9\\
CH$_3$CHO & \phantom{$<$}6.1(-12) & $<$6.3(-11)           & $<$2.8(-11)    & $<$4.7(-11)              & \phantom{$<$}1.2(-10) & \phantom{$<$}4.1(-11) & $<$1.9(-10) &  \phantom{$<$}5.6(-11) & 5.8\\
HCOOH     & \phantom{$<$}6.7(-11) & \phantom{$<$}8.8(-10) & \phantom{$<$}1.0(-10)    & \phantom{$<$}1.9(-10)    & \phantom{$<$}4.2(-10) & \phantom{$<$}1.8(-10) & \phantom{$<$}8.4(-10) &  \phantom{$<$}3.8(-10) & 3.4\\
CH$_3$CCH & \phantom{$<$}1.5(-9)  & \phantom{$<$}9.6(-9)  & \phantom{$<$}1.1(-9)    & \phantom{$<$}2.1(-9)     & \phantom{$<$}3.6(-9)  & \phantom{$<$}1.8(-9)  & \phantom{$<$}8.4(-9) &  \phantom{$<$}4.0(-9) & 3.5\\
\hline
\end{tabular}
\end{center}
$^a$ $\sigma$ refers to the same order of magnitude as the average.

$^b$ The abundances are derived with respect to the column density of H$_2$ within $R_{T=100\rm\ K}$ in Sect.~\ref{beamdilut} and Table~\ref{tab3}.

$^c$ The abundances are derived using the CO column densities from
\citet{vdtak2000}, except G24.78 and G75.78 for which the data comes
from \citet{cesaroni2003} and \citet{hatchell1998} respectively.

$^c$ $\sigma$ refers to the same order of magnitude as the average.
\end{table*}

\begin{figure}
   \centering
\includegraphics[width=8cm]{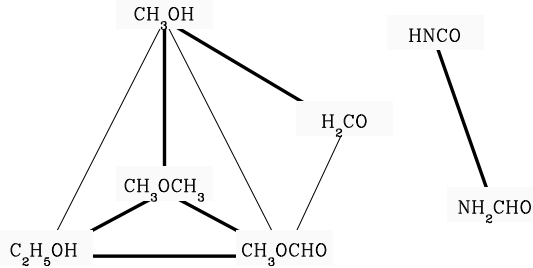}
  \caption{Empirical correlations between the different molecules
  based on the Pearson correlation coefficients. All solid lines
  indicate correlations coefficients $>$0.8 and the thick lines
  $>$0.9. When two species correlate, it does not mean that they are
  directly chemically related. Two species can also correlate if they
  are both formed from the same mother species.}
 \label{scheme}
\end{figure}

\begin{figure*}
   \centering
\includegraphics[width=17cm]{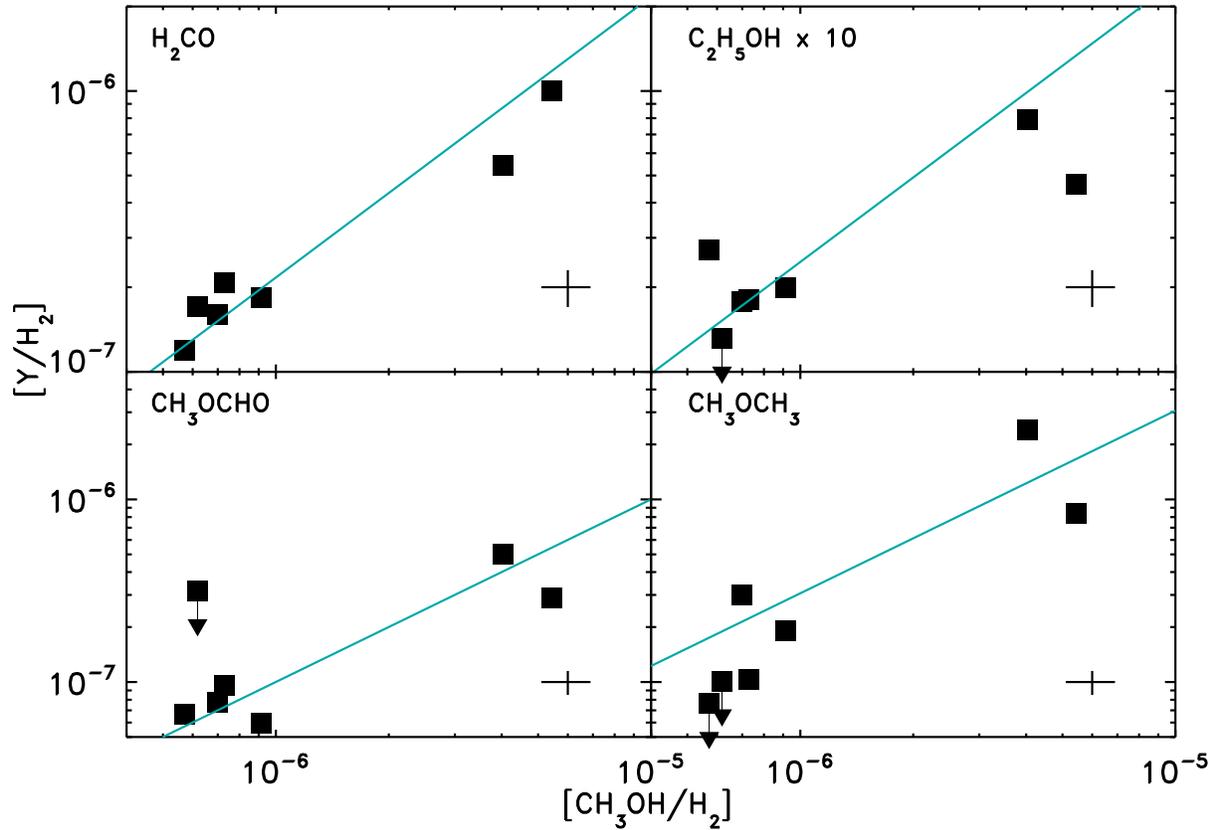}
   \caption{Comparison of molecular abundances with respect to H$_2$
   within $R_{T=100 \rm\ K}$ for the oxygen-bearing molecules. The
   gray lines indicate the linear fit to the data. An error-bar
   representing the calibration uncertainties is depicted in the lower
   right corner.}
              \label{ratio100}
    \end{figure*}

\begin{table*}
\caption{Abundance correlations with respect to H$_2$ for the
abundances within $R_{T=100 \rm\ K}$.\label{cor_100} for hot core
molecules.}
\begin{centering}
\begin{tabular}{l|lllllllll}
\hline
\hline
             & H$_2$CO     & CH$_3$OH & C$_2$H$_5$OH & HNCO     & NH$_2$CHO  & CH$_3$CN & HCOOCH$_3$ & CH$_3$OCH$_3$ & HCOOH\\
\hline
H$_2$CO      & \phantom{-}1.00        & \phantom{-}0.98     & \phantom{-}0.76         & \phantom{-}0.46     & \phantom{-}0.53       & -0.07               & \phantom{-}0.88       & \phantom{-}0.75    & \phantom{-}0.76\\
CH$_3$OH     & \phantom{-}0.98        & \phantom{-}1.00     & \phantom{-}0.86         & \phantom{-}0.38     & \phantom{-}0.39       & -0.07               & \phantom{-}0.92       & \phantom{-}0.86    & \phantom{-}0.70 \\
C$_2$H$_5$OH & \phantom{-}0.76        & \phantom{-}0.86     & \phantom{-}1.00         & \phantom{-}0.02     & \phantom{-}0.22       & \phantom{-}0.12     & \phantom{-}0.92       & \phantom{-}0.95    & \phantom{-}0.30 \\
HNCO         & \phantom{-}0.46        & \phantom{-}0.38     & \phantom{-}0.02         & \phantom{-}1.00     & \phantom{-}0.38       & \phantom{-}0.81     & \phantom{-}0.16       & -0.27              & \phantom{-}0.98\\
NH$_2$CHO    & \phantom{-}0.53        & \phantom{-}0.39     & \phantom{-}0.22         & \phantom{-}0.38     & \phantom{-}1.00       & \phantom{-}0.17     & \phantom{-}0.42       & -0.02              & \phantom{-}0.65\\
CH$_3$CN     & -0.07                  & -0.07               & \phantom{-}0.12         & -0.81               & \phantom{-}0.17       & \phantom{-}1.00     & \phantom{-}0.29       & \phantom{-}0.20    & -0.51\\
HCOOCH$_3$   & \phantom{-}0.88        & \phantom{-}0.92     & \phantom{-}0.92         & \phantom{-}0.16     & \phantom{-}0.42       & \phantom{-}0.29     & \phantom{-}1.00       & \phantom{-}0.90    & \phantom{-}0.49\\
CH$_3$OCH$_3$& \phantom{-}0.75        & \phantom{-}0.86     & \phantom{-}0.95         & -0.27               & -0.02                 & \phantom{-}0.20     & \phantom{-}0.90       & \phantom{-}1.00    & \phantom{-}0.09\\
HCOOH$^a$        & \phantom{-}0.76        & \phantom{-}0.70     & \phantom{-}0.30         & \phantom{-}0.98     & \phantom{-}0.65       & -0.51               & \phantom{-}0.49       & \phantom{-}0.09    & \phantom{-}1.00\\
\hline
\end{tabular}
\end{centering}

$^a$ Assuming that the HCOOH emission comes from within $R_{T=100\rm\
K}$.
\end{table*}

The abundances of the species detected in our sources with $T>100$~K
were compared by estimating the Pearson correlation coefficients
between log(X/H$_2$) vs log(Y/H$_2$) (see
Table~\ref{cor_100}). C$_2$H$_5$CN is not compared with the other
species due to the low number of sources with detections. Although the
number of sources for the other molecules are limited and correlations
therefore only suggestive, a number of interesting trends are
seen. These result from groups of molecules with very similar
abundance trends.

\begin{figure}
   \centering
\includegraphics[width=8cm]{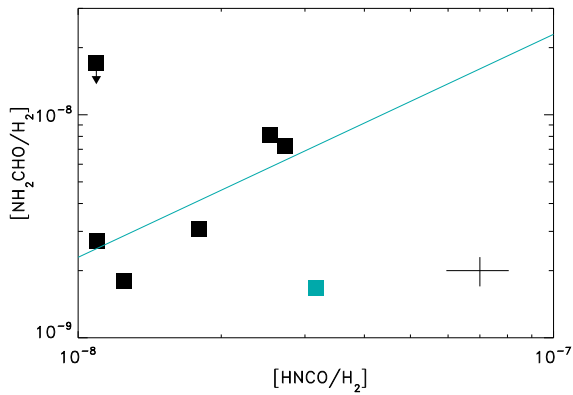}
   \caption{Comparison of molecular abundances with respect to H$_2$ for
   HNCO with NH$_2$CHO within $R_{T=100 \rm K}$. The source indicated in gray is G75.78 and the limit is AFGL~2591. An error-bar
   representing the calibration uncertainties is depicted in the lower
   right corner.}
              \label{n_cor}
    \end{figure}

First, the abundances of the nitrogen-bearing species do not correlate
with those of oxygen-bearing species. Strong correlations are,
however, found between the oxygen-bearing species H$_2$CO, CH$_3$OH,
C$_2$H$_5$OH, HCOOCH$_3$, and CH$_3$OCH$_3$. Figure~\ref{ratio100}
shows the abundances of all hot oxygen-bearing species with respect to
H$_2$ compared to CH$_3$OH/H$_2$. The nitrogen-bearing species HNCO and
NH$_2$CHO also appear to be well correlated, which follows from the
column density trends (Sect.~\ref{results} and shown in
Fig.~\ref{n_cor}). The NH$_2$CHO abundance for G75.78 is uncertain, as
only few lines were detected. The correlation coefficient is 0.38 with
this source included and 0.92 without.

\begin{table*}
\caption{Average abundance ratio for our sources and other well-known
high mass YSOs with respect to CH$_3$OH for the oxygen-bearing species
and with respect to HNCO for NH$_2$CHO. $\sigma$ is the standard
deviation between the sources in our sample.}\label{branch}
\begin{center}
\begin{tabular}{l|llllll}
\hline
\hline
Molecule      & abundance ratio($\sigma$) & Orion Compact Ridge$^a$ & G34.3$^b$ & Sgr B2(N)$^c$ & G327.3$^d$ & GC clouds$^e$\\
\hline
\multicolumn{5}{c}{with respect to CH$_3$OH}\\
\hline
H$_2$CO         & \phantom{$>$}0.22$\pm$0.05    & \phantom{$>$}0.10   & \phantom{$>$}0.009   & $>$0.0025           & \phantom{$>$}-- & $\sim$0.01\\
C$_2$H$_5$OH    & \phantom{$>$}0.025$\pm$0.013  &\phantom{$>$}0.004 & \phantom{$>$}0.19    & \phantom{$>$}0.005  & \phantom{$>$}0.05 & $\sim$0.04\\
HCOOCH$_3$      & \phantom{$>$}0.10$\pm$0.03   & \phantom{$>$}0.08  & \phantom{$>$}0.86    & \phantom{$>$}0.005  & \phantom{$>$}0.96 & $\sim$0.04\\
CH$_3$OCH$_3$   & \phantom{$>$}0.31$\pm$0.20    & \phantom{$>$}0.05  & \phantom{$>$}--       & \phantom{$>$}0.015  & \phantom{$>$}7 & $\sim$0.04\\
\hline
\multicolumn{5}{c}{with respect to HNCO}\\
\hline
NH$_2$CHO       & \phantom{$>$}0.36$\pm$0.08    & \phantom{$>$}--      & \phantom{$>$}--       & \phantom{$>$}0.23 & \phantom{$>$}0.07\\
\hline
\end{tabular}
\end{center}
$^a$ \citet{sutton1995}

$^b$ \citet{macdonald1996}

$^c$ \citet{nummelin2000}.

$^d$ \citet{gibb2000a}.

$^e$ \citet{requena2006}.
\end{table*}

If the HCOOH emission is assumed to arise in hot instead of cold gas,
its abundance correlates with very few molecular species. Its
strongest correlation is with HNCO with a coefficient of 0.98. This
result is surprising and due to the low rotation temperatures for
HCOOH questionable.

The abundances of the hot molecules presented in Table~\ref{ab_var100}
and Fig.~\ref{ratio100} assume the same beam dilution factor
corresponding to $R_{\rm 100~K}$. As Table~\ref{tempres} shows, the
rotation diagram temperatures of several species are significantly
higher than 100~K. The $T=$200~K radius is typically a factor of
$\sim$3 smaller than that for 100~K so that the beam dilution and
abundances can be a factor $\sim$9 larger. If all emission of hot
molecules would arise from a smaller region, the correlations would
remain the same. If some species are present in 100~K gas and others
in 200~K gas, the correlations would disappear, however. The
differences in rotation temperatures in Table~\ref{tempres} are
assumed to arise largely from other effects such as optical depth,
sub-thermal excitation, or the detection of very high or low lying
energy levels affecting the overall fit. Future interferometric
observations of multiple high mass sources could determine whether the
assumption of a single source size for all molecules is correct.

For the molecules that have abundances that do correlate, the
abundance ratios were calculated with respect to either CH$_3$OH/H$_2$
or HNCO/H$_2$ and are shown in Table~\ref{branch}. Figure~\ref{scheme}
shows the empirical relations for molecules with Pearson coefficients
of 0.8 and higher. Per set of species the ratios vary no more than a
factor of three, the previously found uncertainty in the
abundances. H$_2$CO, HCOOCH$_3$, and CH$_3$OCH$_3$ have 5-10 times
lower abundances than CH$_3$OH. NH$_2$CHO is a factor 3 times lower
than HNCO. Possibly, coefficients between 0.8--0.9 are present between
species that are not directly chemically related, but that are both
strongly correlated with a third species.

\subsection{Abundances with respect to CH$_3$OH} \label{ch3oh_cor}

The Pearson correlation coefficients of log(X/CH$_3$OH)
vs. log(Y/CH$_3$OH) differ significantly from the correlations in
Sect.~\ref{res_100}. First, strong correlations are not present
between the same species. This is due to the variability of the
CH$_3$OH abundance between the different sources, and to the
dependence of the correlations on three (X, Y, and CH$_3$OH) rather
than two (X and Y) species. The abundances of two species with respect
to CH$_3$OH will correlate when their abundances with respect to H$_2$
correlate but also when the abundance of one of the species correlates
with that of CH$_3$OH. This makes it difficult to use correlations of
abundance ratios with respect to CH$_3$OH to chemically link
molecules. Instead it is better to compare X/CH$_3$OH with the
CH$_3$OH column density or abundance, i.e. N(CH$_3$OH) or
CH$_3$OH/H$_2$ as has been done by e.g., \citet{requena2006}.

\subsection{Comparison to other well known sources} \label{comp}

The abundance ratios with respect to CH$_3$OH are compared in
Table~\ref{branch} to other well-studied high mass star forming
regions, specifically the Orion Compact Ridge \citep{sutton1995},
G34.3 \citep{macdonald1996}, Sgr B2(N) \citep{nummelin2000}, and
G327.3 \citep{gibb2000a}. The ratios for the Orion Compact Ridge all
agree within the error bars for H$_2$CO and HCOOCH$_3$. CH$_3$OCH$_3$
has a relatively low abundance with respect to CH$_3$OH. Only three
lines have been detected for this species in the Compact Ridge and it
has a rather low excitation temperature of 75~K implying that it is
perhaps present in a colder region.

In contrast, the relative abundances for G34.3 deviate more from those
in our work. The H$_2$CO/CH$_3$OH ratio by \citet{macdonald1996} is
probably a lower limit as only one line was detected for G34.3 and the
column density of H$_2^{13}$CO was calculated to be higher than the
main H$_2$CO isotopologue based on their respective line
intensities. Both C$_2$H$_5$OH and HCOOCH$_3$ have relatively high
abundances with respect to CH$_3$OH. This could potentially be due to
high optical depth in the lines of CH$_3$OH. \citet{macdonald1996}
assumed the emission to be optically thin, and thus underestimate the
CH$_3$OH column density. The relative abundances of C$_2$H$_5$OH and
HCOOCH$_3$ are similar between G34.3 and the sources studied in this
paper.

The abundance ratios with respect to CH$_3$OH for Sgr B2(N) from
\citet{nummelin2000} appear to be significantly lower than for the
sources in our sample. This is due to the CH$_3$OH emission being
decomposed into a core and a halo component for Sgr B2(N), whereas
that of the other molecules is not. For this source it is more
relevant to compare the relative ratios between the different
molecules. The C$_2$H$_5$OH abundance is high compared to HCOOCH$_3$
and CH$_3$OCH$_3$. However, the HCOOCH$_3$/CH$_3$OCH$_3$ and the
NH$_2$CHO/HNCO ratios of 0.33 are similar to those for our sources.

G327.3 in the study of \citet{gibb2000a} has a C$_2$H$_5$OH/CH$_3$OH
abundance ratio a factor of 2 higher compared to the average for our
sample. Small deviations could be caused by their relatively low
rotation temperature for C$_2$H$_5$OH of 66 K. HCOOCH$_3$ and
CH$_3$OCH$_3$ have relatively high abundances with respect to
CH$_3$OH, but HCOOCH$_3$/CH$_3$OCH$_3$ is within a factor two of the
ratio derived for our sample. Again NH$_2$CHO/HNCO is the same within
error to our sample.

Molecular abundance ratios can also be compared to those found in
shocked gas in the galactic center by \citet{requena2006}. Their
sources have relatively constant ratios of complex molecules with
respect to CH$_3$OH of 0.04. The H$_2$CO/CH$_3$OH and
CH$_3$OCH$_3$/CH$_3$OH ratios are significantly higher for our hot
cores. Small differences may be due to the fact that
\citet{requena2006} observed rotational transitions with $E_{\rm
u}<$40--50~K, whereas lines with excitation energies that range from
20--900~K have been observed in this study.

In summary, the overall emerging picture is that of relatively
constant abundance ratios between various molecules comparable to the
sources for our sample. Distinct differences for individual species
can often be ascribed to specific observation or analysis problems.

\subsection{Comparison with ice abundances}\label{ice}

\begin{table*}
\caption{Solid state abundances with respect to H$_2$, assuming H$_2$O
ice to have an abundance of 10$^{-4}$.}\label{w33a}
\begin{center}
\begin{tabular}{l|llllll}
\hline
\hline
 Species     & W~33A$^a$     & NGC~7538~IRS1$^b$ & AFGL~2591$^b$     & HH~46$^c$ (low mass) & CK~2$^c$ (field star) &  gas phase$^d$\\
             & 10$^{-5}$     & 10$^{-5}$      & 10$^{-5}$         & 10$^{-5}$  & 10$^{-5}$ & \\
\hline 
H$_2$CO      & 0.65          & \phantom{$>$}--              & \phantom{$>$}--                 & \phantom{$>$}--           & \phantom{$>$}--          & 3.2(-7)\\
CH$_3$OH     & 1.4-1.7       & $<$0.4         & \phantom{$>$}1.4               & \phantom{$>$}1.2         & $<$0.11    & 2.0(-6)\\
HCOOH        & 0.37-0.71     & \phantom{$>$}2.2            & $<$1.7            & $<$1.5      & \phantom{$>$}0.11       & 3.8(-10)\\
CH$_3$CHO    & 0.98          & \phantom{$>$}1.2            & $<$0.66           & \phantom{$>$}--           & \phantom{$>$}--          & 5.6(-11)\\
OCN$^-$      & 0.63          & $<$0.05        & \phantom{$>$}--                 & $\lesssim$0.12  & \phantom{$>$}--          & 1.6(-8)$^e$\\
\hline
\end{tabular}
\end{center}
$^a$ From \citet{gibb2000b,dartois1999,schutte1999,whittet2001,demyk1998}.

$^b$ From \citet{gibb2004}.

$^c$ From \citet{knez2005} and references therein.

$^d$ Average abundances from this paper.

$^e$ Assuming that OCN$^-$ evaporates as HNCO.
\end{table*}

It is interesting to relate the gas-phase abundances derived above to
the ice abundances from mid-infrared observations. Of the sources in
our sample, AFGL 2591, NGC~7538 IRS1 and W~33A were also observed at
mid-infrared wavelengths with ISO \citep{gibb2004}. Table~\ref{w33a}
lists the abundances of the ice species from that paper assuming an
H$_2$O abundance of 10$^{-4}$ with respect to H$_2$. W~33A is
particularly rich in organics in the solid state whereas the other two
sources mainly have upper limits. The solid state abundances of
CH$_3$OH are at least an order of magnitude higher than the comparable
source-averaged gas phase abundances. The H$_2$CO abundance of
$\sim$2$\times$10$^{-6}$, although more uncertain, is also larger by
an order of magnitude \citep{keane2001a}. For W~33A, however, the
solid H$_2$CO/CH$_3$OH abundance ratio is consistent with the gas
phase abundance ratio.

Where detected toward high mass YSOs, solid HCOOH and CH$_3$CHO have
abundances of 10$^{-6}$--10$^{-5}$ and $\sim$10$^{-5}$ respectively
\citep{schutte1999,keane2001a,gibb2000b,gibb2004}. Similar abundances
for solid HCOOH are found toward background stars by
\citet{knez2005}. These absolute solid state abundances are four to
five orders of magnitude higher than the gas phase abundances for the
same sources.

The XCN band is detected in many objects and usually assigned to
OCN$^-$
\citep{grim1987,demyk1998,pendleton1999,gibb2000b,broekhuizen2005a}. OCN$^-$
is thought to convert to and evaporate off grains as HNCO
(Sect.~\ref{models}). Abundances of solid OCN$^-$ vary and range from
1$\times$10$^{-7}$--4$\times$10$^{-6}$
\citep{broekhuizen2005a,thi2006}. Gas phase abundances of HNCO are
factors of 100--1000 times lower.

Many of the species are expected to have abundances in the ice below
the detection limit. \citet{boudin1998} determined upper limits for
the column densities of a number of more complex organic molecules,
including C$_2$H$_6$ and C$_2$H$_5$OH toward NGC~7538 IRS9. The upper
limit on the C$_2$H$_5$OH abundance is 1.2$\times10^{-6}$ with
C$_2$H$_5$OH/CH$_3$OH $<$0.28. This is consistent with the
C$_2$H$_5$OH/CH$_3$OH abundance ratio of 0.009 in the gas phase for
our sample of sources.

In summary, the absolute ice abundances of observed ``first
generation'' species, CH$_3$OH and OCN$^-$, are 10--10$^2$ times
higher compared to the gas phase values, and as much as a factor of
10$^4$--10$^5$ for HCOOH and CH$_3$CHO.

\section{Discussion and comparison to astrochemical models}
\label{discussion}

\subsection{General model considerations}\label{models}

Pure gas phase chemistry models such as presented in e.g.,
\citet{lee1996}, are unable to explain the high abundances of many
complex organics, especially of highly saturated molecules in hot gas
phase environments. CH$_3$OH, for example, is estimated to have
abundances of $\sim$10$^{-12}$ through pure gas phase chemistry,
whereas abundances of 10$^{-6}$ are found. For CH$_3$CN abundances of
$\sim$10$^{-10}$ are modeled \citep{lee1996} whereas abundances of
$\sim$10$^{-7}$ are detected. Furthermore, they are not able to
explain the presence of complex organics in ices by simple freeze-out
of gas phase species \citep[see][ for a review]{herbst2005}. For
highly saturated molecules, solid state chemistry provides a likely
alternative. Here one can distinguish basic grain surface chemistry
and chemistry induced by energetic processing (UV irradiation,
energetic particle bombardment and/or thermal heating) which can
proceed also inside the ice.

Surface chemistry models have been developed by several groups
\citep[e.g.,][]{hasegawa1993,charnley2001,keane2001} (see Fig.~1). The
results are generally very sensitive to the atomic C, O, and N
abundances in the gas. Moreover, the dust temperature ($T_{\rm dust}$)
plays a crucial role in the mobility of the atoms and radicals as well
as the availability of CO as a reaction partner on the grain, which
evaporates at $T_{\rm dust}>$20~K
\citep{tielens1982,caselli1993}. However, if CO is located in an ice
environment that is dominated by H$_2$O it will be available for
reactions in the ice up to much higher temperatures
\citep{sandford1993,collings2003b}. 

If grain surface chemistry is the explanation for our observed
correlations, the constant abundance ratios for different sources must
imply very similar conditions during ice formation, at least of the
ice layer that contains the ``hot'' molecules. This is also confirmed
by the similarity between the composition of ices observed toward
background stars compared to those toward low and high mass star
forming regions \citep{knez2005}. Perhaps the simplest explanation is
that the bulk of the ice forms at similarly low temperature conditions
in which the bulk of the oxygen and nitrogen are in atomic form and
C/CO is at a fairly constant ratio, around 0.01 as computed for dense
clouds \citep{gredel1989}. When these ``first~generation'' species
evaporate together and have similar destruction rates, a constant
abundance ratio would result even in regions where the temperatures of
the gas and UV-flux are very different.

In contrast, ``second~generation'' species are expected to peak at
different times dependent on their formation and destruction
processes. While this argument has been used to explain the
anti-coincidences between the nitrogen and oxygen-bearing species
\citep{caselli1993,rodgers2003}, it is inconsistent with our observed
correlations of supposedly ``second generation'' species HCOOCH$_3$
and CH$_3$OCH$_3$ with other ``first generation'' oxygen-bearing
molecules. Indeed, in models by \citet{garrod2006} where HCOOCH$_3$ is
formed in the solid state, many of the oxygen-bearing species have
similar gas phase abundance variations throughout the chemical
evolution and have constant abundance ratios. Solid state formation is
therefore more likely for the supposedly ``second generation''
oxygen-bearing species. However, the abundances of oxygen-bearing and
nitrogen-bearing species do not correlate. This can be due to a
different time-dependence of formation and destruction or a variable
N/C abundance ratio.

Laboratory experiments show that beyond a certain minimum dose of
irradiation the reaction products often do reach a constant plateau
\citep{gerakines1996,moore2000,broekhuizen2004,hagen1979}. Oxygen-bearing
species, such as H$_2$CO, and nitrogen-bearing species, e.g.,
OCN$^{-}$ and NH$_2$CHO, can be formed this way
\citep{grim1989}. However, there is not enough information in the
literature on the formation of the more complex molecules observed
here to make a quantitative test of this scheme. It is plausible,
though, that an equilibrium between formation and destruction is
reached in interstellar ices and could be responsible for the constant
abundance ratios between the oxygen-bearing species or HNCO/NH$_2$CHO.

Recent work by \citet{garrod2006} combines both grain surface
chemistry and UV-induced chemistry with thermal evolution of the ices
during the protostellar phase. In particular, radicals created in the
ices by ultraviolet radiation become mobile at the higher dust
temperatures during the warm-up phase and lead to the formation of
many complex organic species. This suggests that the ice composition
changes with temperature. Potentially only the ices on grains that are
close to the evaporation temperature have similar relative abundances
to those observed in the gas phase. Since the detected ice features
are observed in a column, they will contain both cold and warmer ices,
which complicates the comparison of the ice abundances with the gas
phase, even within one source.

Species such as HCOOH and CH$_3$CHO are detected with much lower
column densities in the gas phase than in the ice (see
Section~\ref{ice}) and have low rotation temperatures. Possibly, these
species are formed in the solid state at low temperatures and
destroyed due to reactions at high temperatures. Since their
desorption temperatures are typically $\sim$100~K
\citep{viti2004,bisschop2007b}, thermal desorption cannot account for
their low temperatures. However, non-thermal mechanisms induced by,
for example, cosmic rays \citep{leger1985}, can explain the low gas
phase abundances if they cause a small fraction (10$^{-4}$) of the ice
to evaporate (see Sect.~\ref{indi}). Alternatively, cold gas phase
ion-molecule reactions are able to reproduce the abundances of e.g.,
CH$_3$CHO and CH$_2$CO \citep[][ see also Sect.~\ref{indi}]{lee1996}.

Ices are expected to be layered because different molecules such as
H$_2$O, CO, and N$_2$ form or condense at different densities. Since
nitrogen is transformed into N$_2$ deepest in the cloud it is expected
to freeze-out last. To form species such as HNCO or NH$_2$CHO a more
reactive form of nitrogen, i.e. N-atoms are needed. N-atoms could
potentially be present in a non-hydrogen bonding layer dominated by CO
or a layer dominated by H$_2$O. If they were present in the
non-hydrogen bonding layer a small fraction of non-thermal desorption
would result in cold gas phase N-bearing species. The lack of these
cold N-bearing molecules and high rotation temperatures of HNCO and
NH$_2$CHO suggests that they desorb at higher temperatures and must
thus be present in a different, more tightly bound ice layer.

In summary, pure gas-phase reactions cannot reproduce the abundances
of most species depicted in Fig.~\ref{grainchem}, and second
generation gas phase chemistry is inconsistent with the observed
correlations for the oxygen-bearing species. Constant abundance ratios
after ice evaporation can be explained either by similar initial
conditions for ice formation or an equilibrium between formation and
destruction due to energetic processes. The low abundances of cold gas
phase species could be due to gas phase ion-molecule reactions or
non-thermal desorption of a small fraction of the ices e.g., by cosmic
rays.

\subsection{Individual molecules}\label{indi}

This subsection elaborates on the general conclusions stated in
Sect.~\ref{models} for individual molecules, grouped according to the
different branches of Fig.~\ref{grainchem}.

\underline{H$_2$CO and CH$_3$OH:} CH$_3$OH and H$_2$CO are molecules
 commonly expected to originate from grain-surface chemistry. They are
 formed through successive hydrogenation of CO on grains, a scheme
 that has been tested experimentally by \citet{hiraoka2002},
 \citet{watanabe2004}, and Fuchs et al. (in prep.). Both H$_2$CO and
 CH$_3$OH have been detected in these experiments with different
 abundance ratios dependent on the exact conditions. H$_2$CO can also
 be formed by UV-irradiation of CH$_3$OH containing ices
 \citep{allamandola1988,bernstein1995}. However, the high ice
 abundances of CH$_3$OH \citep[e.g.,][]{gibb2004} and the constant
 abundance ratio between H$_2$CO/CH$_3$OH in this paper are
 inconsistent with this CH$_3$OH destruction mechanism. Our data imply
 that hydrogenation reactions are more feasible formation mechanisms.

Furthermore as discussed in Sect.~\ref{ice}, the solid state abundance
of CH$_3$OH appears to be at least an order of magnitude higher than
the comparable source-averaged gas phase abundances. This could be due
to a number of effects: i) CH$_3$OH is partly destroyed upon
evaporation, ii) intrinsic CH$_3$OH ice variations from source to
source \citep{brooke1999,dartois1999,pontoppidan2004}, iii) an
underestimate of the beam-filling factor of the CH$_3$OH emitting
gas. Possibly, all three factors contribute to the relatively low gas
phase abundance.

\underline{CH$_2$CO, CH$_3$CHO, and C$_2$H$_5$OH:} In
Figure~\ref{grainchem} CH$_2$CO, CH$_3$CHO, and C$_2$H$_5$OH are
thought to be interrelated through successive hydrogenation as well as
to H$_2$CO and CH$_3$OH by successive H- and C-atom addition. All
three species are expected to evaporate from the ice together with
H$_2$O \citep{viti2004} and have similar rotation temperatures. This
is not the case, however. Indeed, interferometric observations of
CH$_3$CHO in high mass star forming regions show that CH$_3$CHO
emission is extended with respect to hot cores species such as
HCOOCH$_3$ and CH$_3$OCH$_3$ \citep{liu2005}. This suggest that
CH$_2$CO, CH$_3$CHO, and C$_2$H$_5$OH gas are not present in the same
region. Possibly, H-atom addition to CH$_2$CO and CH$_3$CHO on grains
is very efficient and completely converts them to C$_2$H$_5$OH at high
temperatures, so that only C$_2$H$_5$OH evaporates from the ice and is
detected in the warm inner regions. Non-thermal desorption of a small
fraction due to e.g., cosmic ray spot heating of CH$_2$CO and
CH$_3$CHO ice in cold regions could explain the low gas phase
abundances and rotation temperatures, as well as the high ice
abundances of CH$_3$CHO. Alternatively, C$_2$H$_5$OH could be formed
through another surface reaction related to CH$_3$OH and H$_2$CO
formation, explaining its constant abundance ratio.

For CH$_2$CO and CH$_3$CHO there are also potential gas phase
formation mechanisms. \citet{millar1985} argue that CH$_2$CO could
result from radiative associative reactions between CH$_3^+$ and
CO. Assuming that CH$_3$CO$^+$ dissociative recombination results in
CH$_2$CO for 50\% of the cases, this would allow for CH$_2$CO
abundances of $\sim$10$^{-10}$, comparable to our observed abundances
and also to those by \citet{turner1999}. Gas phase models by
\citet{lee1996} predict fractional abundances of 5$\times$10$^{-12}$
for CH$_3$CHO and its isomers. This is lower than the abundances of
\citet{ikeda2001} and \citet{charnley2004} but close to the abundances
for some of our sources. \citet{herbst1989} proposed that H$_3$O$^+$
and C$_2$H$_2$ react to form CH$_3$CHO after dissociative
recombination, which results in higher abundances of up to
$\sim$10$^{-10}$ in dark clouds. Many of these conclusions depend on
high branching ratios toward these products in dissociative
recombination. If three body break-up prevails as found for some
species, pure gas phase production becomes untenable
\citep{geppert2005}.

\underline{HCOOH:} High abundances of HCOOH in the ice, low gas phase
rotation temperatures, and orders of magnitude lower abundances in the
gas phase compared to the solid state can be due to two reasons: i)
HCOOH is quickly destroyed in the gas phase upon evaporation, and ii)
HCOOH is abundant in low temperature ice, but destroyed at higher
temperatures due to reactions in the ice. Gas phase destruction is a
possibility, but it is unclear why the destruction rate of HCOOH would
be orders of magnitude larger compared to other species. Even if this
would be the case, most high mass star-forming regions are expected to
have an evaporation front that moves outward through the envelope as
the luminosity increases. HCOOH should thus continuously freshly
evaporate from icy grains. The second possibility of HCOOH destruction
in the ice at high temperatures is much more likely and can be tested
in the laboratory. The lower gas phase abundance can then be explained
by a small fraction of the ice ($\sim$10$^{-4}$) that evaporates due
to non-thermal desorption mechanisms. The low rotation temperatures of
$\sim$40--70~K for most sources in our sample are also consistent with
this picture. Finally, it can explain the high HCOOH abundances in
shocked regions by e.g. \citet{requena2006}, where the ice is warmed
up and evaporated over much shorter timescales.

The precise formation of HCOOH through surface chemistry still remains
a puzzle, however. In Figure~\ref{grainchem} the formation mechanisms
in the solid state is either through CO or CO$_2$. Laboratory
experiments indicate that it only evaporates at very high temperatures
($\sim100$~K) similar to H$_2$O \citep{collings2004}. This is even the
case for HCOOH diluted in CO or CO$_2$ ices (Bisschop et al. in
prep.). \citet{keane2001} predicts the formation of HCOOH through
successive atom addition to CO and finds that the HCOOH ice
concentration is proportional to the O/O$_2$ ratio. Alternatively,
\citet{allen1977} proposed the reaction HCO~+~OH in the solid state
leading to HCOOH. New models by \citet{garrod2006} have HCOOH gas
phase abundances that are much higher than found for our sources with
most of the HCOOH actually formed in the gas. If this occurs at
intermediate temperatures, a fraction of the HCOOH may freeze out on
grains. Further laboratory experiments are needed to elucidate the
exact formation mechanism for HCOOH.

\underline{HNCO and NH$_2$CHO:} HNCO formation has been explained by
surface chemistry formation and gas phase reactions
\citep{turner1999,keane2001}. Both \citet{turner1999} and
\citet{keane2001} argue that NH$_2$CHO is formed on grain-surfaces due
to its high level of hydrogen saturation. The strong correlation
between the two species in Sect.~\ref{res_100} suggests they are
related. Alternatively, both OCN$^-$ and NH$_2$CHO can form in ices
due to energetic processes such as UV-irradiation
\citep{broekhuizen2004,munoz2003,allamandola1999}. Van Broekhuizen et
al.(2004) conclude that thermal formation of OCN$^{-}$ can, however,
also not be ruled out. Concluding, solid state processes appear to be
the most likely formation mechanism for both species, whether they
involve energetic processing or not.

\underline{HCOOCH$_3$ and CH$_3$OCH$_3$:} HCOOCH$_3$ and CH$_3$OCH$_3$
were previously thought to form from protonated methanol in the gas
phase \citep{millar1991}. \citet{horn2004} recently found the barrier
for the gas phase formation of HCOOCH$_3$ to be higher than previously
anticipated. Moreover, the gas phase routes assume that
H$_2$COOCH$_3^+$ and (CH$_3$)$_2$OH$^+$ recombine to HCOOCH$_3$ and
CH$_3$OCH$_3$ respectively.  Measurements by
\citet{geppert2004,geppert2005} have shown that dissociative
recombination of CH$_3$OH$_2^+$ mostly results in a three-body
break-up instead of its deprotonated counterpart. If this would also
be the case for species such as HCOOCH$_3$ and CH$_3$OCH$_3$, they are
more likely to be at least partially formed on grain surfaces. The
constant abundance ratios between both species and CH$_3$OH also
confirm this scenario. Indeed \citet{garrod2006} explain surface
formation of HCOOCH$_3$ and CH$_3$OCH$_3$ through reactions of CH$_3$O
and HCO radicals, first proposed by \citet{allen1977}.

\underline{CH$_3$CN, C$_2$H$_5$CN, and CH$_3$CCH:} CH$_3$CN can be
formed from the fast association reaction CH$_3^+$ + HCN and the
radiative association reaction of CH$_3$ + CN
\citep{charnley1992}. The abundances in the gas phase for Orion are,
however, $\sim$10$^{-10}$ \citep{charnley1992} compared to 10$^{-8}$
in our sources. The relatively high abundances plus their high degree
of hydrogenation for C$_2$H$_5$CN and CH$_3$CN imply that they are
possibly formed on grain-surfaces. CH$_3$CCH usually has very low
rotation temperatures, and its formation is explained in dark and
translucent clouds through gas phase ion-molecule or neutral-neutral
reactions. 

\underline{In summary}, most species, such as H$_2$CO, CH$_3$OH,
C$_2$H$_5$OH, HCOOCH$_3$, and CH$_3$OCH$_3$ probably result from
``first generation'' chemistry in ices, likely dominated by surface
chemistry. This is possibly also true for CH$_3$CN and
C$_2$H$_5$CN. CH$_2$CO, CH$_3$CHO, HCOOH and CH$_3$CCH have all been
detected in the gas with low rotation temperatures and, for some
species, abundances that are orders of magnitude lower than those in
ices. Non-thermal desorption of a small fraction of the ice
($\sim$10$^{-4}$) is sufficient to explain the abundances in the cold
gas phase. For CH$_2$CO, CH$_3$CHO, and HCOOH formation plus
destruction in the ice at higher temperature are plausible, but gas
phase chemistry cannot be excluded. Models of gas phase reactions are
able to accurately reproduce the observed CH$_3$CCH abundances.

\section{Summary and Conclusions}
\label{conclusions}

We have performed a survey of selected frequency settings for 7 high
mass YSOs. The survey was targeted at detecting species thought to be
produced by the grain surface chemistry depicted in
Fig.~\ref{grainchem}. Other species such as CH$_3$CN, C$_2$H$_5$CN,
CH$_3$CCH, HCOOCH$_3$ and CH$_3$OCH$_3$ have also been detected. To
compare abundances between sources, the source luminosity and
850~$\mu$m dust emission was used to constrain the radius within which
$T\geq100$~K and the column density out to this radius,
$N_{T\geq100\rm K}$. The main conclusion derived in this work are:
\begin{itemize}

\item Most detected molecules have similar rotation temperatures for
all sources. The following molecules can be classified as cold
($T<$100~K): CH$_2$CO, CH$_3$CHO, HCOOH, and CH$_3$CCH; and the
following as hot ($T\geq$100~K): H$_2$CO, CH$_3$OH, C$_2$H$_5$OH,
HNCO, NH$_2$CHO, CH$_3$CN, C$_2$H$_5$CN, HCOOCH$_3$, and
CH$_3$OCH$_3$.

\item Groups of molecules have their highest column densities for the
same sources. This is the case for the oxygen-bearing species H$_2$CO,
CH$_3$OH, C$_2$H$_5$OH, HCOOCH$_3$ and CH$_3$OCH$_3$. HNCO and
NH$_2$CHO have their highest column in other sources than the
oxygen-bearing species, but both are highest for the same sources. Due
to optical depth effects and a limited number of detections it is
difficult to establish the trends for CH$_3$CN and C$_2$H$_5$CN.

\item Absolute abundances in the hot regions vary by an order of
magnitude. Relative abundances of the oxygen-bearing species and the
two nitrogen-bearing species are, however, very constant. The absolute
abundances are much higher than can be produced by gas phase
chemistry. The constant abundance ratios plus the similar rotation
temperatures imply that the oxygen-bearing species, H$_2$CO, CH$_3$OH,
C$_2$H$_5$OH, HCOOCH$_3$ and CH$_3$OCH$_3$, as well as the two
nitrogen-bearing species, HNCO and NH$_2$CHO, share a common solid
state formation scheme. This can be the result of very similar
conditions during ice formation or a balance between formation and
destruction in the ice. The oxygen-bearing species and the two
nitrogen-bearing species are not related, however. The abundances of
two cyanides, CH$_3$CN and C$_2$H$_5$CN, are likely due to ``first
generation'' ice chemistry as well.

\item CH$_2$CO, CH$_3$CHO, CH$_3$CCH and HCOOH have very low rotation
temperatures. The formation of CH$_2$CO, CH$_3$CHO, and HCOOH can be
explained by solid state or gas phase reactions. For HCOOH and
CH$_3$CHO gas phase abundances are a factor of 10$^4$ lower than those
observed in the solid state. Solid state formation for these species
is thus more likely. A low fraction of non-thermal desorption could be
responsible for the low gas phase abundances. Models of ion-molecule
reactions in the cold gas phase can, however, also reproduce the
abundances of CH$_2$CO and CH$_3$CHO. CH$_3$CCH on the other hand is
probably created by cold gas phase chemistry, as its abundance is
similar to that found in dark clouds. Interestingly, no cold N-bearing
molecules are found.

\end{itemize}

The abundance correlations and branching ratios determined in this
paper are key to understanding chemical relationships between these
complex organic molecules and addressing their origins. Future
spatially resolved observations with interferometers such as ALMA will
make it possible to confirm molecular relations and address the
chemical variations within each source. Additional laboratory
experiments can explore the pathways to the formation of the most
abundant species given these observational constraints of constant
abundance ratios. Taken together such studies will be important to
fully understand the chemistry of complex organic molecules in star
forming regions.

\begin{acknowledgements}
We thank Peter Schilke for supplying the XCLASS program used to
calculate line intensities, and Remo Tilanus and Alain Castets for the
support for the observations at the JCMT and IRAM respectively. We are
also grateful for useful discussions with Steve Doty, Eric Herbst,
Miguel Requena-Torres, Floris van der Tak, Xander Tielens, as well as
Malcolm Walmsley and an anonymous referee for comments on the
paper. Funding was provided by NOVA, the Netherlands Research School
for Astronomy and a NWO Spinoza grant. The research of JKJ was
supported by NASA Origins Grant NAG5-13050.
\end{acknowledgements}

\clearpage
\begin{appendix}
\section{Rotation diagrams per molecule for all sources}
\label{rotdia}

\begin{figure*}
   \centering \includegraphics[width=16cm]{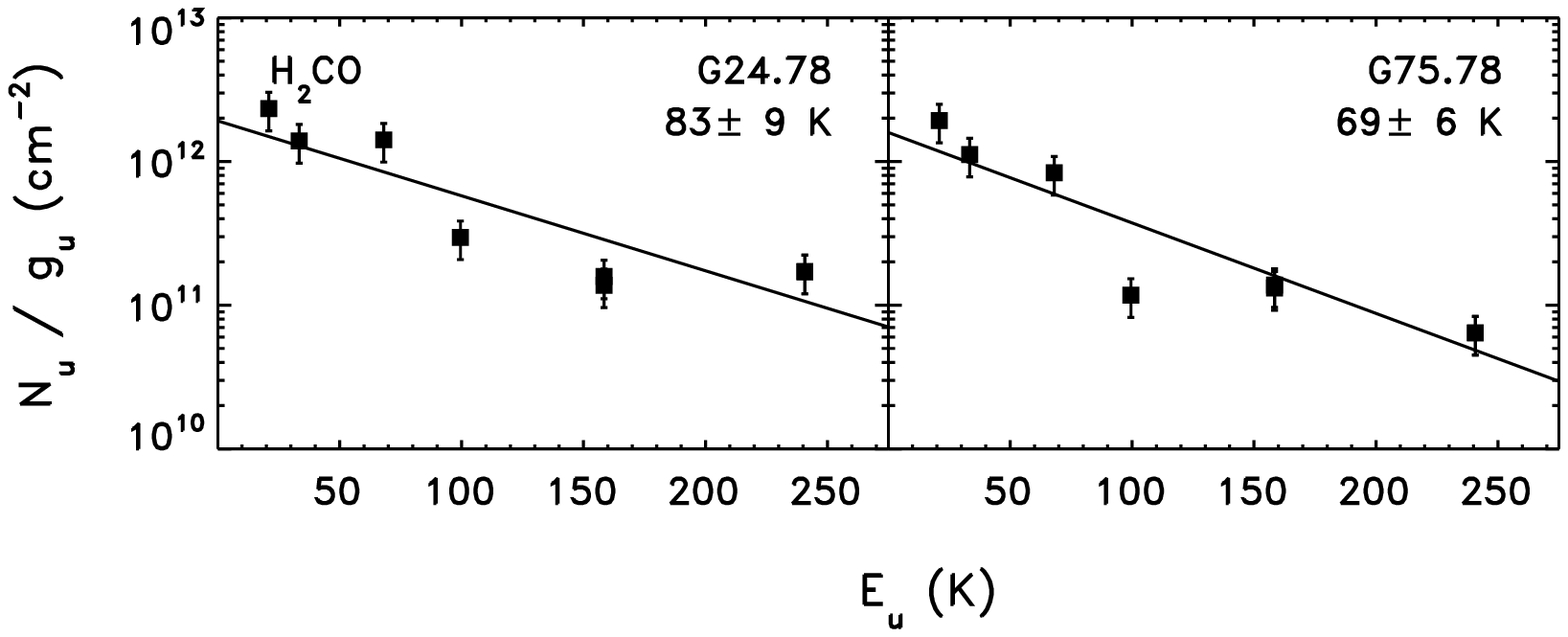}
   \caption{Rotation diagrams for H$_2$CO for G24.78, and G75.78.}
              \label{h2co}
    \end{figure*}

\begin{figure*}
   \centering
\includegraphics{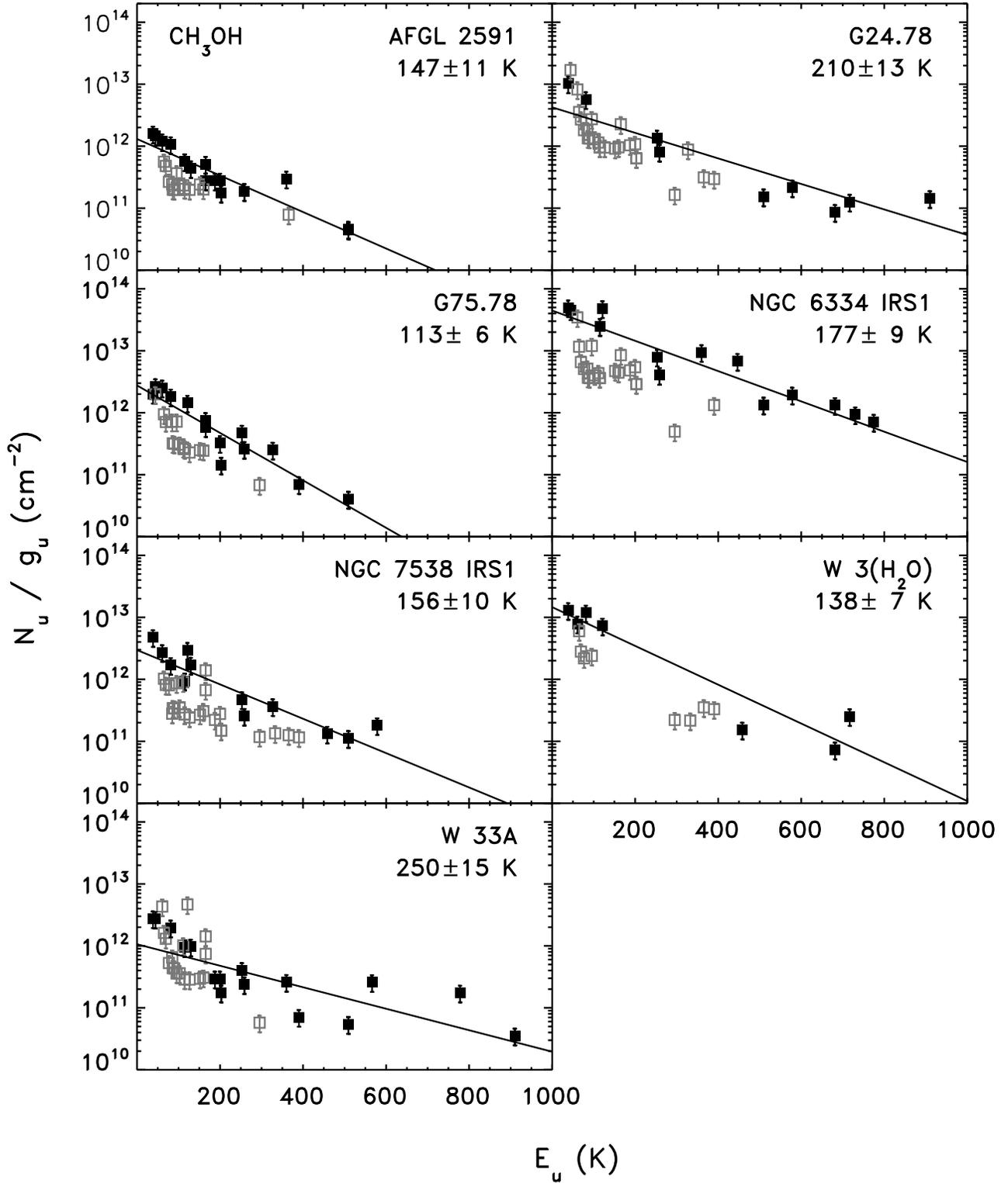}
   \caption{Rotation diagrams for CH$_3$OH for respectively
   AFGL2591, G24.78, G75.78, NGC~6334IRS1, NGC~7538IRS1, W3(H$_2$O), and
   W33A. The filled black squares are optically thin lines, the open
   gray squares optically thick lines, based on the estimated column
   densities. The rotation temperatures are fits to the optically
   thin lines only.}
              \label{ch3oh}
    \end{figure*}

\begin{figure*}
   \centering
\includegraphics[width=16cm]{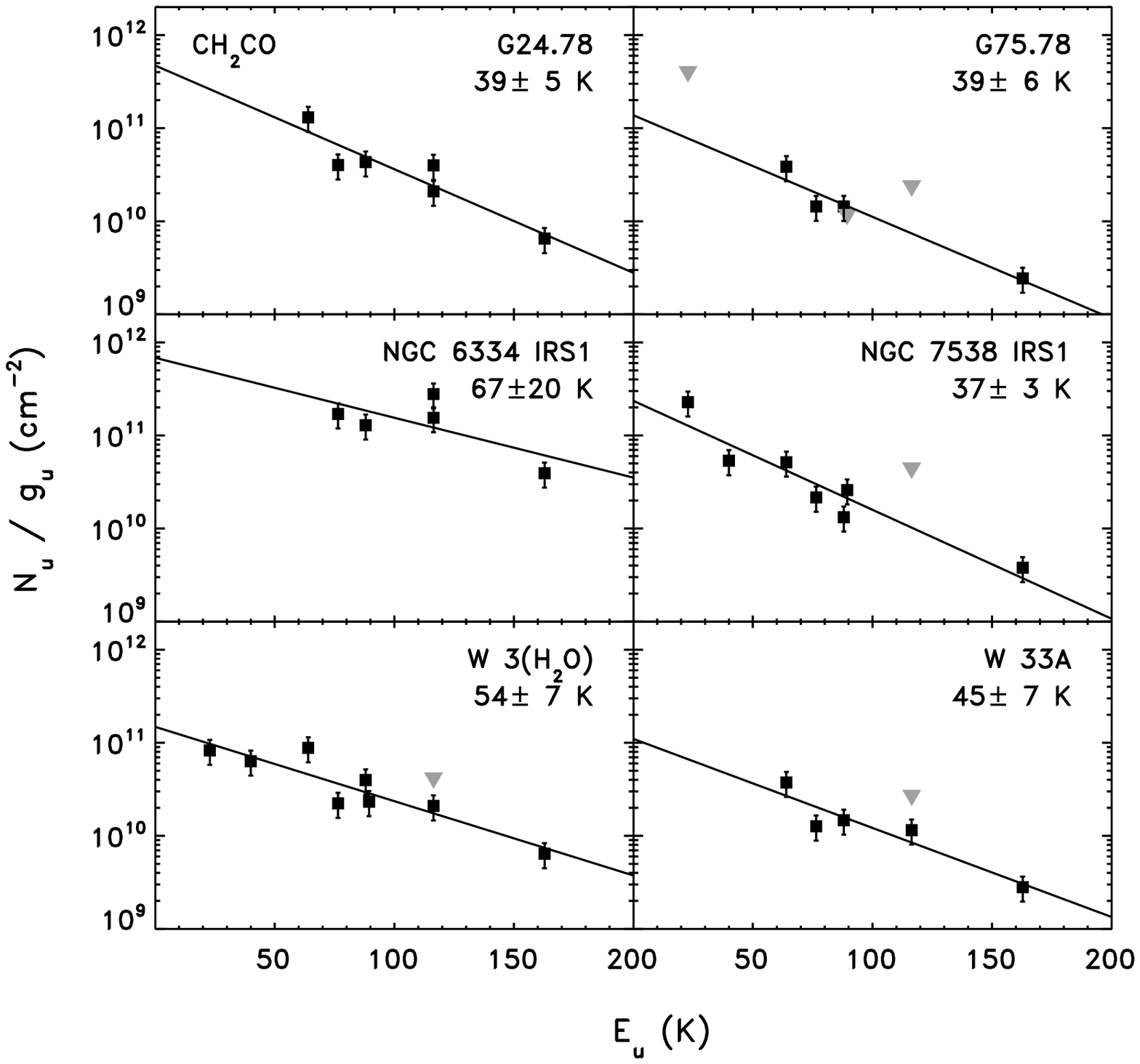}
   \caption{Rotation diagrams for CH$_2$CO for respectively G24.78,
   G75.78, NGC~6334IRS1, NGC~7538, W3(H$_2$O), and W33A. The gray
   triangles indicate upper limits.}
              \label{ch2co}
    \end{figure*}

\begin{figure*}
   \centering
\includegraphics[width=16cm]{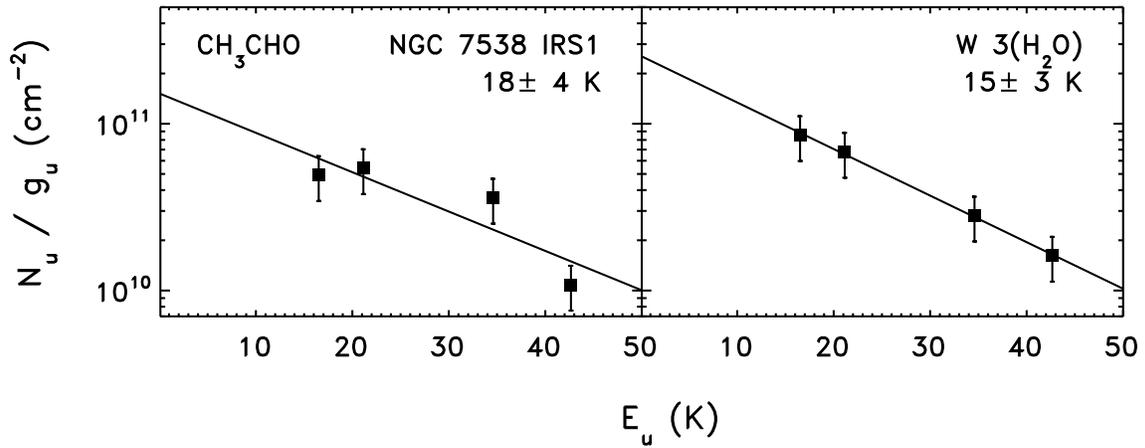}
   \caption{Rotation diagrams for CH$_3$CHO for respectively
   NGC~7538~IRS1, W3(H$_2$O), and W33A.}
              \label{ch3cho}
    \end{figure*}

\begin{figure*}
   \centering
\includegraphics[width=16cm]{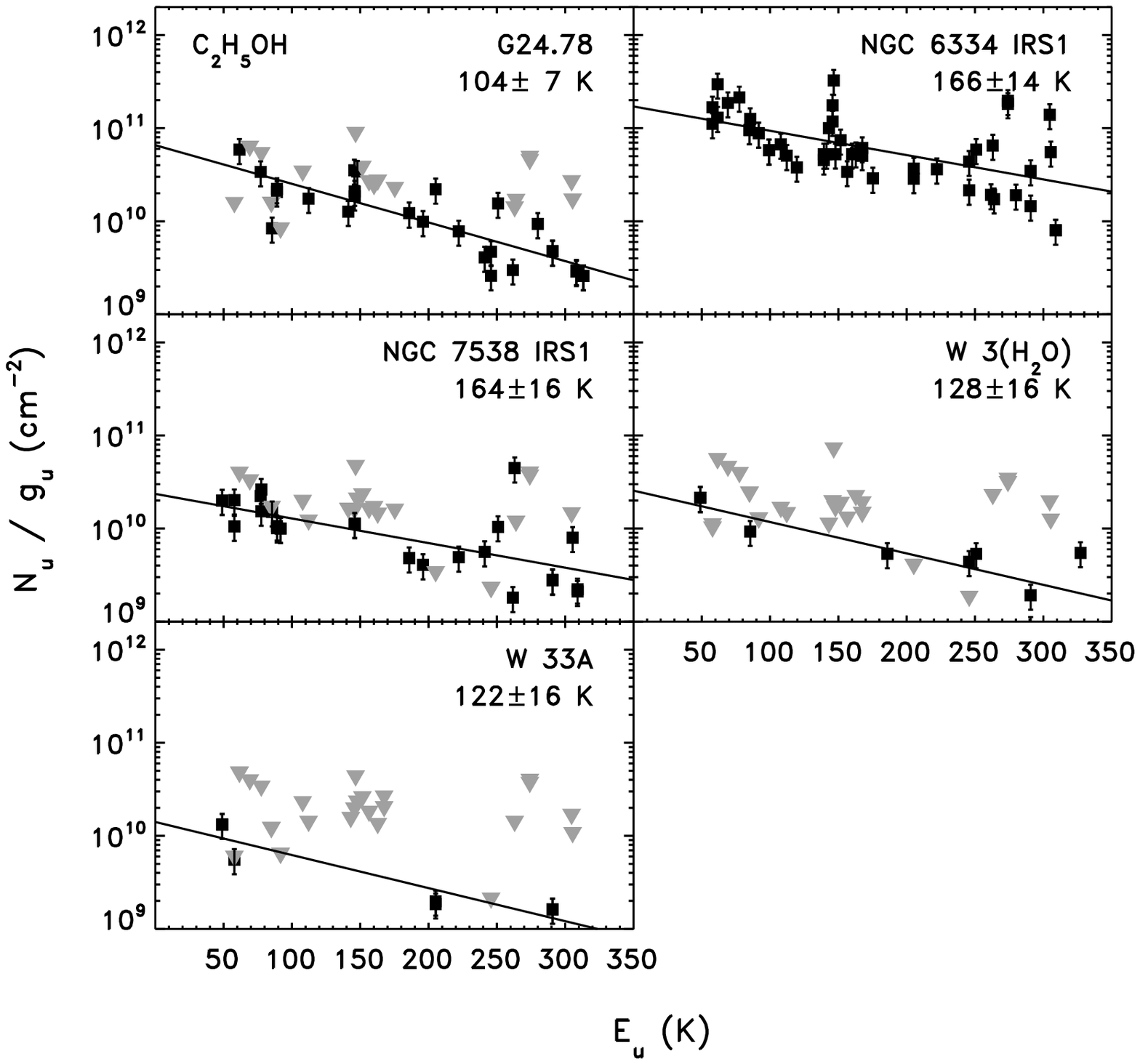}
   \caption{Rotation diagrams for C$_2$H$_5$OH for respectively G24,
   NGC~6334IRS1, NGC~7538IRS1, W3(H$_2$O), and W33A. The gray
   triangles indicate upper limits.}
              \label{c2h5oh}
    \end{figure*}

\begin{figure*}
   \centering
\includegraphics[width=16cm]{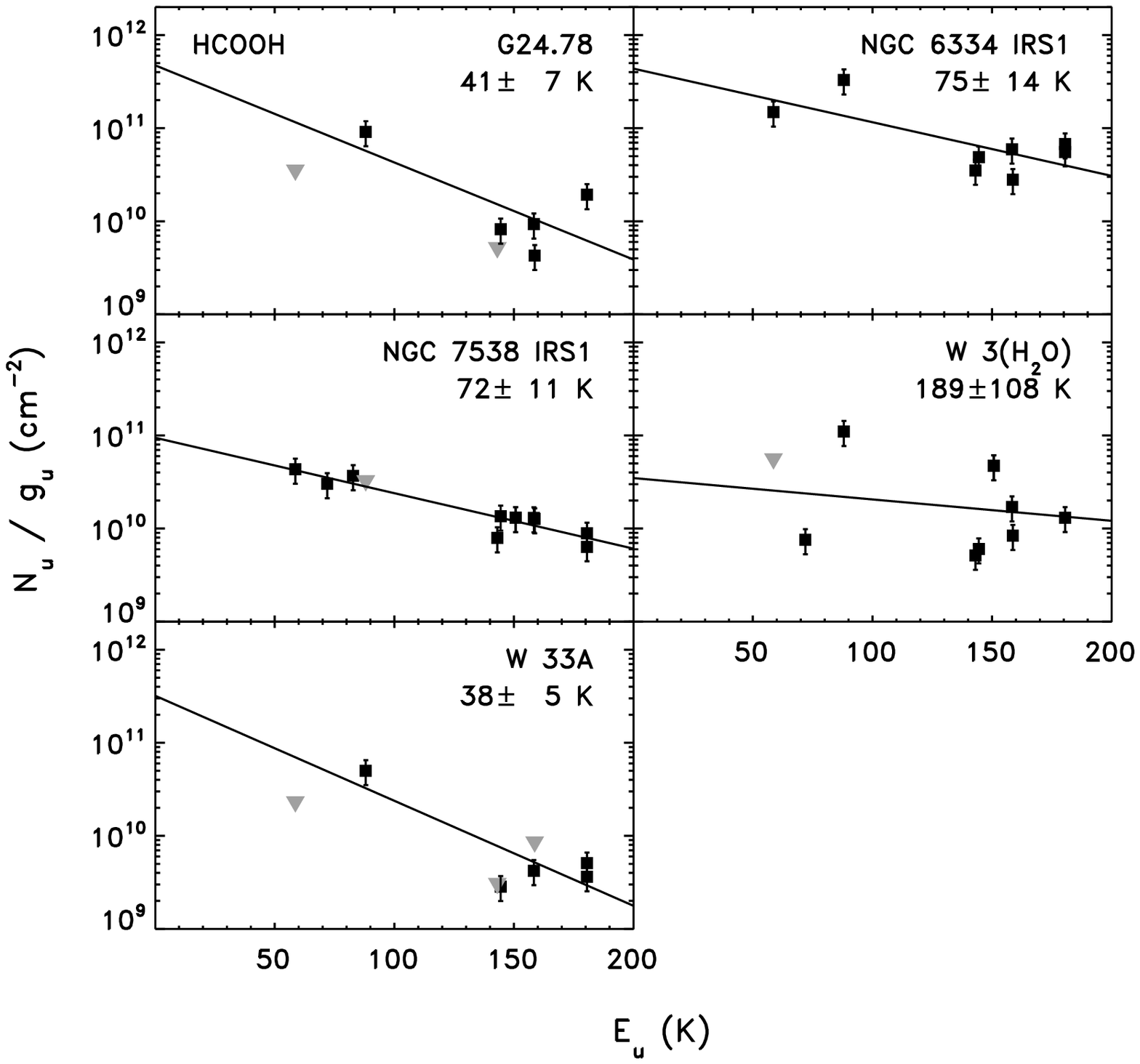}
   \caption{Rotation diagrams for HCOOH for respectively G24.78,
   NGC~6334IRS1, NGC~7538IRS1, W3(H$_2$O), and W33A. The gray
   triangles indicate upper limits.}
              \label{hcooh}
    \end{figure*}

\begin{figure*}
   \centering
\includegraphics[width=16cm]{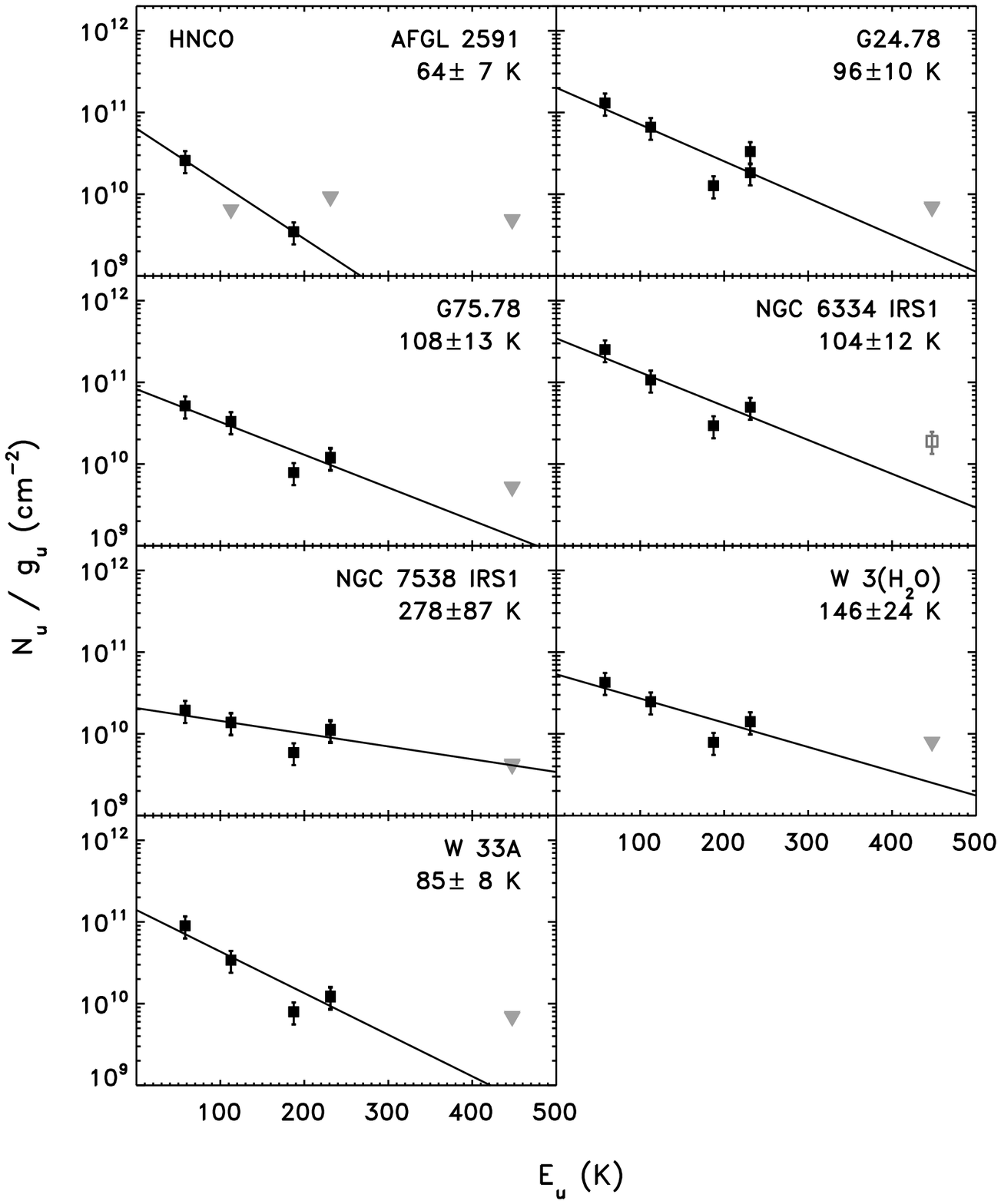}
   \caption{Rotation diagrams for HNCO for respectively G24.78,
   G75.78, NGC~6334IRS1, NGC~7538IRS1, W3(H$_2$O), and W33A. The
   filled black squares are the points included and the gray squares
   excluded from the fit. The gray triangles indicate upper limits.}
              \label{hnco}
    \end{figure*}

\begin{figure*}
   \centering
\includegraphics[width=16cm]{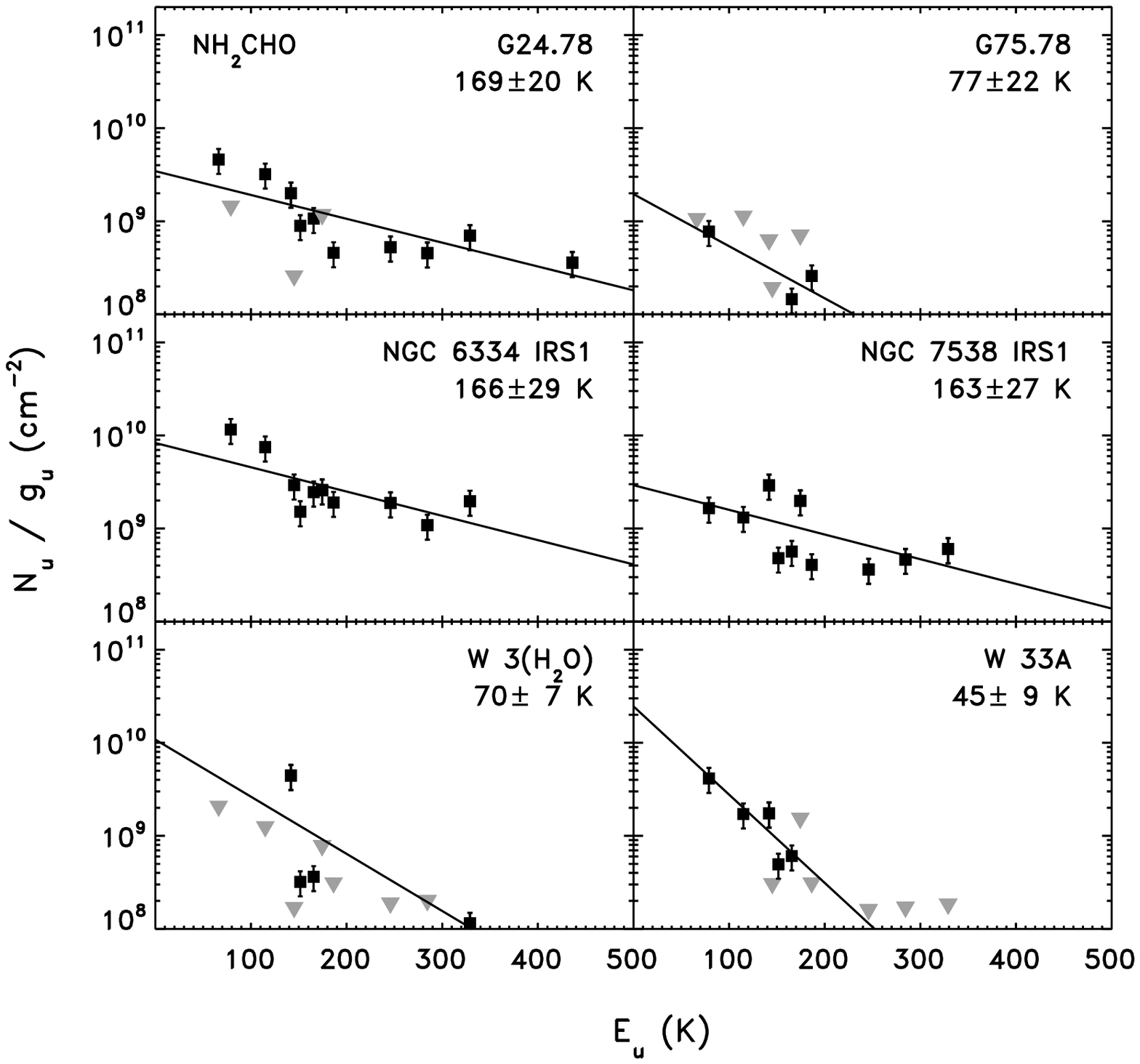}
   \caption{Rotation diagrams for NH$_2$CHO for respectively G24.78,
   G75.78, NGC~6334IRS1, NGC~7538IRS1, W3(H$_2$O), and W33A. The gray
   triangles indicate upper limits.}
              \label{nh2cho}
    \end{figure*}

\begin{figure*}
   \centering
\includegraphics[width=16cm]{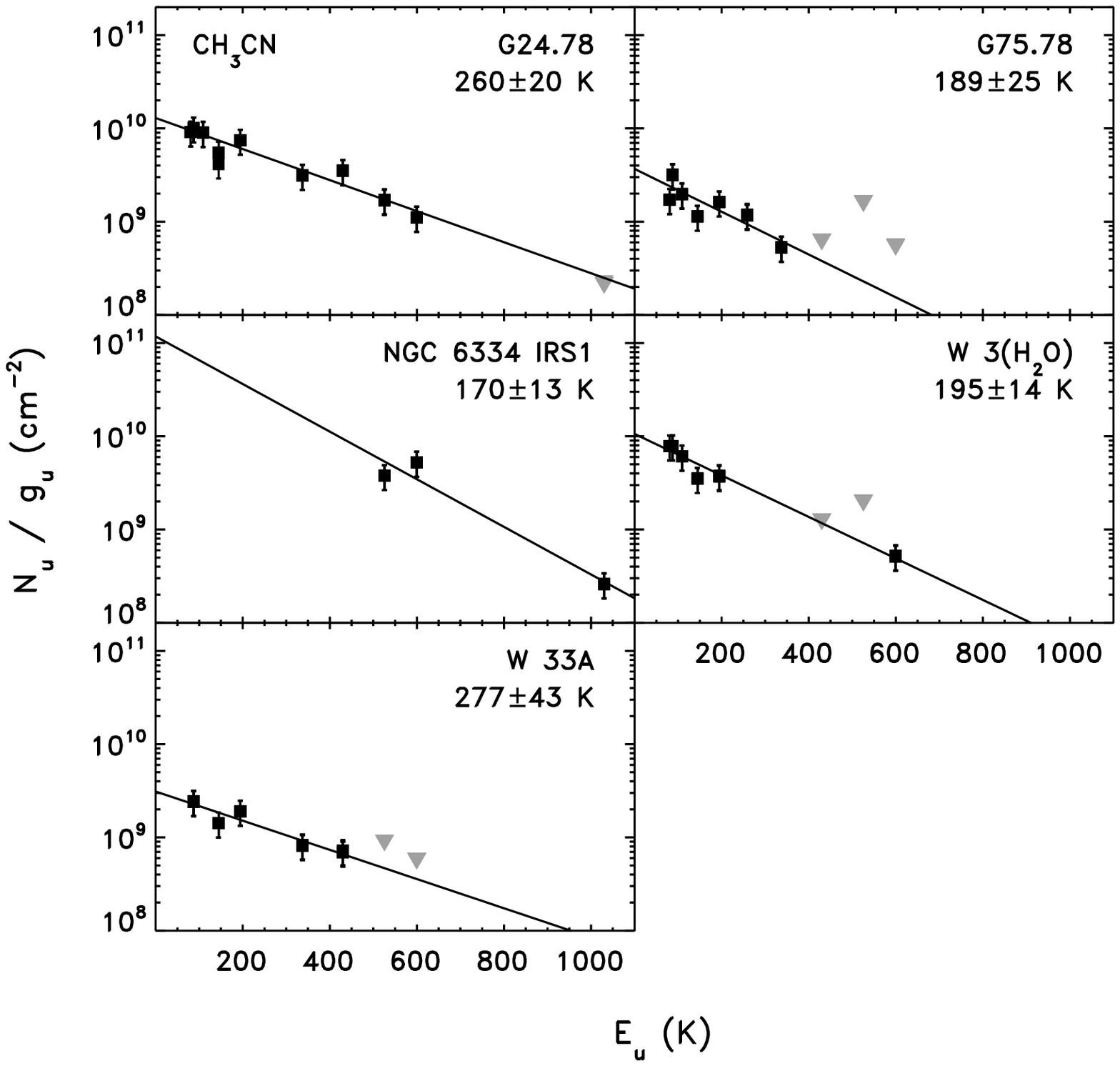}
   \caption{Rotation diagrams for CH$_3$CN for respectively G24.78,
   G75.78, NGC~6334IRS1, W3(H$_2$O), and W33A. The gray triangles
   indicate upper limits.}
              \label{ch3cn}
    \end{figure*}

\begin{figure*}
   \centering
\includegraphics[width=16cm]{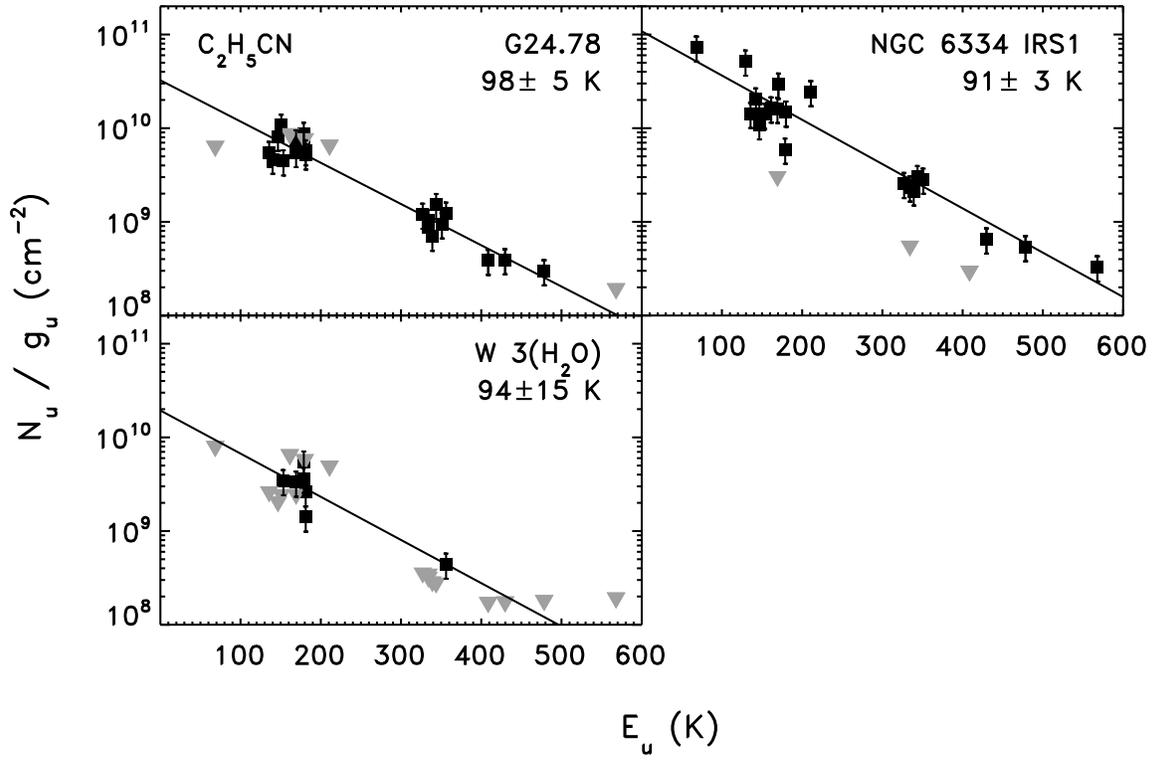}
   \caption{Rotation diagrams for C$_2$H$_5$CN for respectively
   G24.78, and NGC~6334IRS1. The gray triangles indicate upper
   limits.}
              \label{c2h5cn}
    \end{figure*}

\begin{figure*}
   \centering
\includegraphics[width=16cm]{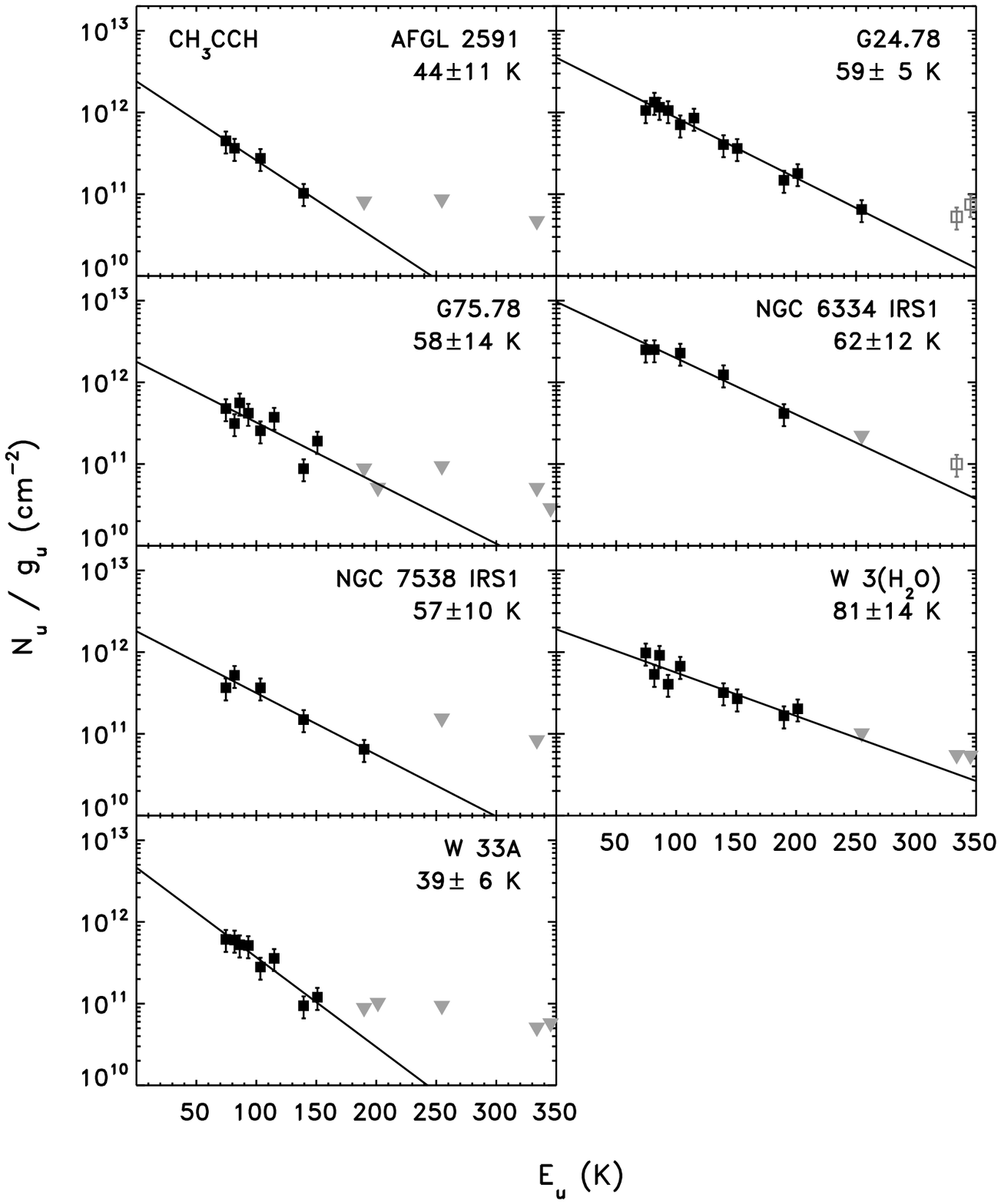}
   \caption{Rotation diagrams for CH$_3$CCH for respectively AFGL2591,
   G24.78, G75.78, NGC~6334IRS1, NGC~7538IRS1, W3(H$_2$O), and
   W33A. The filled black squares are the points included, the gray
   squares excluded from the fit and the gray triangles indicate upper
   limits.}
              \label{ch3cch}
    \end{figure*}

\begin{figure*}
   \centering
\includegraphics[width=16cm]{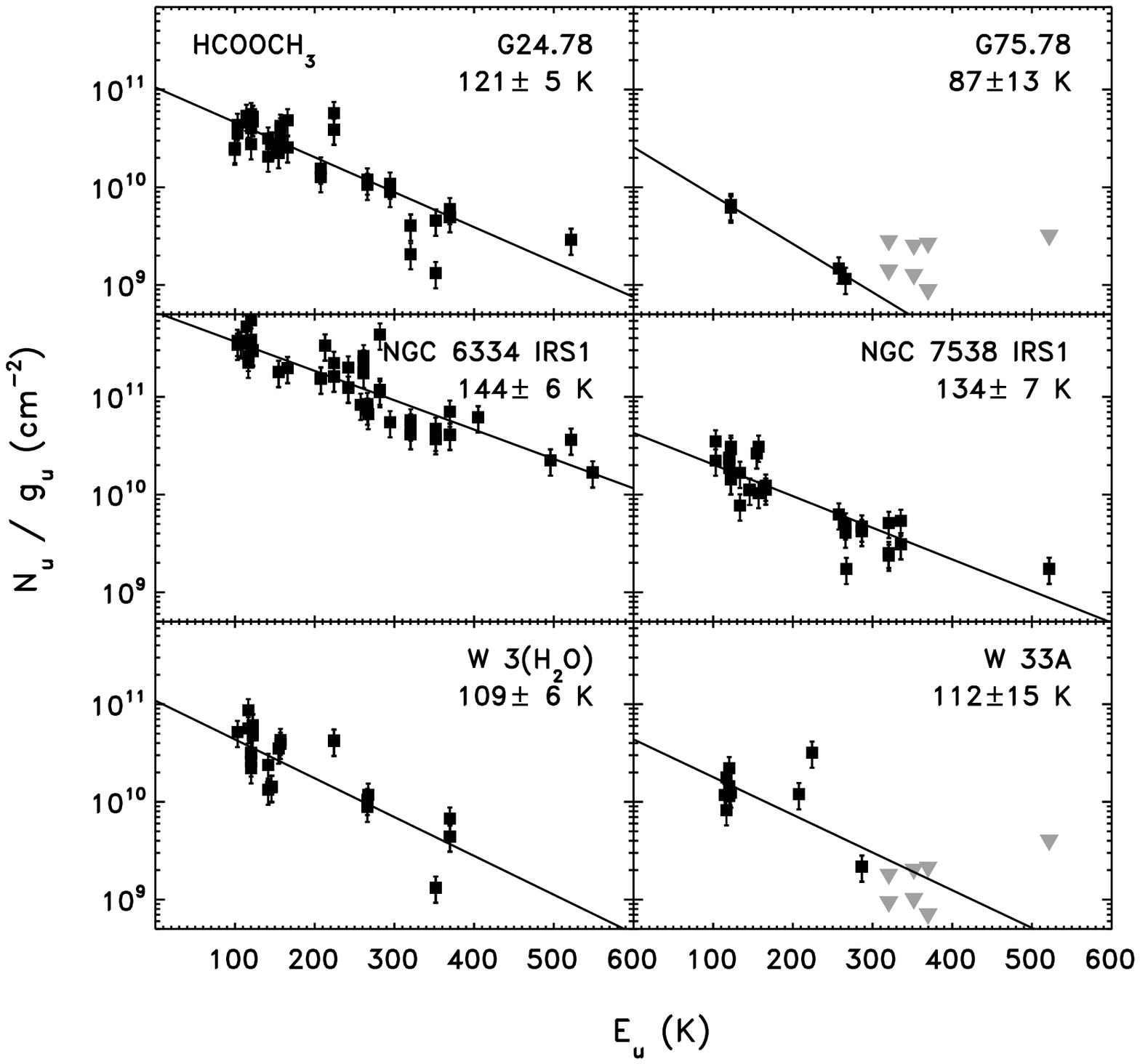}
   \caption{Rotation diagrams for HCOOCH$_3$ for respectively G24.78,
   G75.78, NGC~6334IRS1, NGC~7538IRS1, W3(H$_2$O), and W33A. The
   filled black squares are the points included and the gray triangles
   indicate upper limits.}
              \label{ch3ocho}
    \end{figure*}

\begin{figure*}
   \centering
\includegraphics[width=16cm]{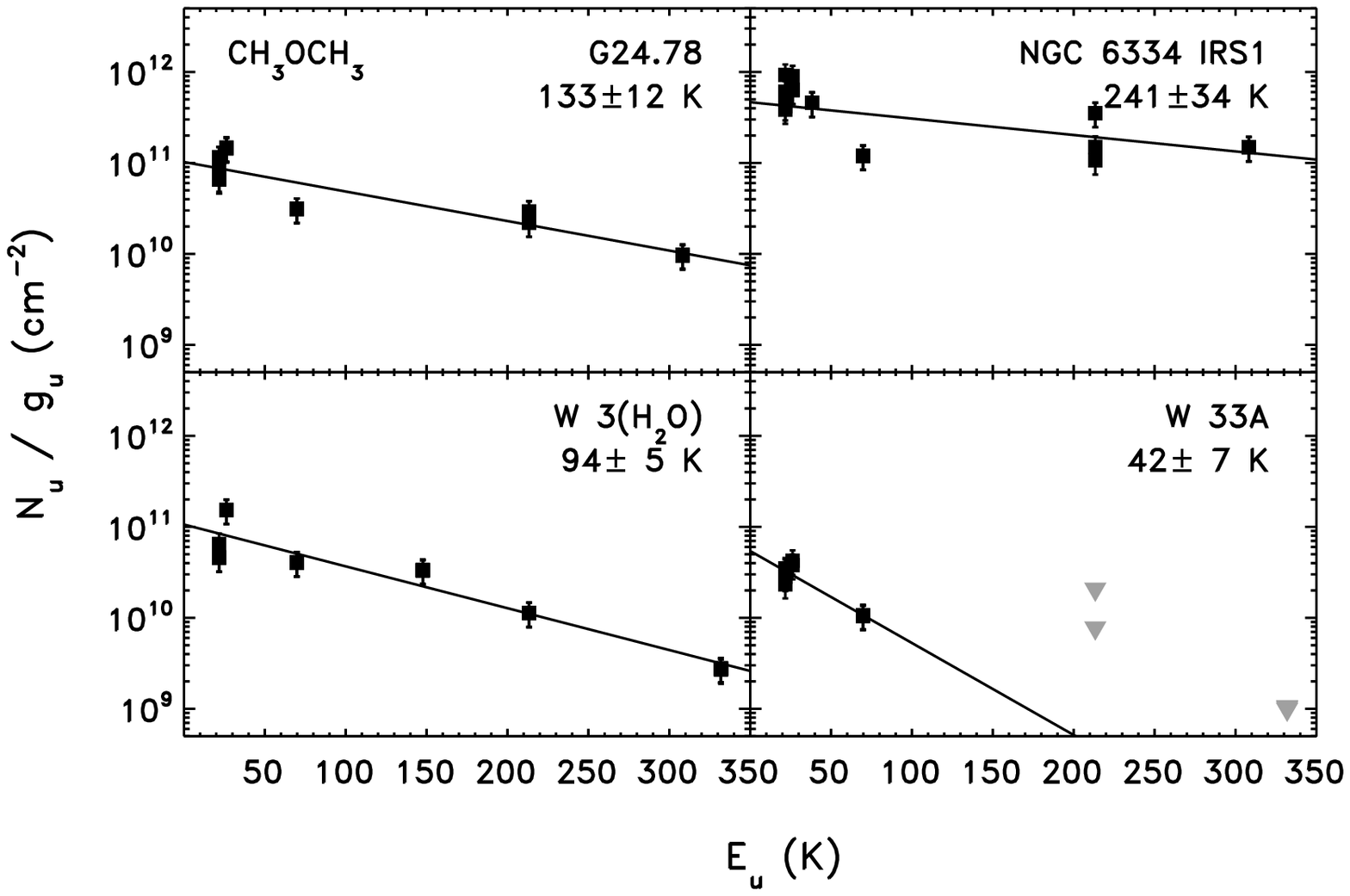}
   \caption{Rotation diagrams for CH$_3$OCH$_3$ for respectively G24.78,
   NGC~6334IRS1, W3(H$_2$O), and W33A. The filled black squares are
   the points included and the gray triangles indicate upper limits.}
              \label{ch3och3}
    \end{figure*}

\clearpage

\section{Detected lines per species for all sources}
\label{detected}

\begin{table*}
\caption{Observed line fluxes $\int T_{\rm MB}dV$ (K km s$^{-1}$) for
H$_2$CO and its isotopic species. }  
\begin{tabular}{l|l|lllllll}
\hline
\hline
Frequency & Transition & \multicolumn{7}{|c}{Sources}\\
{[GHz]}     &            & AFGL~2591 & G24.78  & G75.78 & NGC~6334~IRS1 & NGC~7538~IRS1 & W~3(H$_2$O) & W~33A  \\
\hline
\multicolumn{9}{c}{H$_2$CO}\\
\hline
218.222 & 3$_{0,3}$-- 2$_{0,2}$        & \phantom{$<$}--         & \phantom{$<$}17    & \phantom{$<$}14   &  \phantom{$<$}--        &   \phantom{$<$}--       &  \phantom{$<$}--         &  \phantom{$<$}--    \\
218.476 & 3$_{2,2}$-- 2$_{2,1}$        & \phantom{$<$}--         & \phantom{$<$}5.6     & \phantom{$<$}3.3    &  \phantom{$<$}--        &   \phantom{$<$}--       &  \phantom{$<$}--         &  \phantom{$<$}--    \\
225.698 & 3$_{1,2}$-- 2$_{1,1}$        & \phantom{$<$}15      & \phantom{$<$}27    & \phantom{$<$}22   &  \phantom{$<$}57     &   \phantom{$<$}31    &  \phantom{$<$}34      &  \phantom{$<$}22 \\
363.946 & 5$_{2,4}$-- 4$_{2,3}$        & \phantom{$<$}--         & \phantom{$<$}15    & \phantom{$<$}5.9    &  \phantom{$<$}--        &   \phantom{$<$}--       &  \phantom{$<$}--         &  \phantom{$<$}--    \\
364.103 & 5$_{4,1/2}$-- 4$_{4,0/1}$    & \phantom{$<$}--         & \phantom{$<$}7.3     & \phantom{$<$}2.7    &  \phantom{$<$}--        &   \phantom{$<$}--       &  \phantom{$<$}--         &  \phantom{$<$}--    \\
\hline
\end{tabular}

-- means frequency not observed.
\label{h2colines}
\end{table*}

\begin{table*}
\caption{Observed line fluxes $\int T_{\rm MB}dV$ (K km s$^{-1}$) for CH$_3$OH in the 230~GHz window.\label{ch3ohlines}}
\begin{tabular}{l|l|lllllll}
\hline\hline
Frequency & Transition & \multicolumn{7}{|c}{Sources}\\
 {[GHz]}    &                              & AFGL~2591 & G24.78  & G75.78 & NGC~6334~IRS1 & NGC~7538~IRS1 & W3(H$_2$O) & W33A    \\
\hline
\multicolumn{9}{c}{CH$_3$OH}\\
\hline
218.440 & 4$_{3,1,0}$-- 3$_{2,1,0}$           & \phantom{$<$}--         & \phantom{$<$}12.5    & \phantom{$<$}3.0    & \phantom{$<$}--         & \phantom{$<$}--         &    \phantom{$<$}--       & \phantom{$<$}--       \\
219.994 & 23$_{14,9,0}$-- 22$_{14,8,0}$       & $<$0.36   & $<$0.49 & $<$0.26& \phantom{$<$}2.0       & $<$0.32   &    $<$0.47 & $<$0.41 \\
220.079 & 8$_{4,4,0}$-- 7$_{4,3,0}$           & \phantom{$<$}0.55      & \phantom{$<$}4.1     & \phantom{$<$}1.1    & \phantom{$<$}18      & \phantom{$<$}1.3       &    \phantom{$<$}3.6     & \phantom{$<$}0.54    \\
227.815 & 16$_{9,8,0,1}$-- 15$_{9,7,0,1}$     & \phantom{$<$}--         & \phantom{$<$}2.1     & \phantom{$<$}0.60   & \phantom{$<$}--         & \phantom{$<$}0.87      &    \phantom{$<$}1.8     & \phantom{$<$}--       \\
229.939 & 19$_{12,7,0,2}$-- 20$_{12,8,0,2}$   & $<$0.31   & \phantom{$<$}0.56    & $<$0.32& \phantom{$<$}5.1       & \phantom{$<$}0.47      &    $<$0.38 & $<$0.25 \\
\hline
\end{tabular}

-- means frequency not observed.

$^a$ Blend with HCOOCH$_3$.

$^b$ Blend with CH$_3$CN.

$^c$ Blend with 5$_{2,3,2}$-- 4$_{2,3,2}$.

$^d$ Blend with HCOOCH$_3$.
\end{table*}

\begin{table*}
\caption{Observed line fluxes $\int T_{\rm MB}dV$ (K km s$^{-1}$) for CH$_3$OH in the 345~GHz window and its isotopic species. \label{ch3ohlines2}}
\begin{tabular}{l|l|lllllll}
\hline\hline
Frequency & Transition & \multicolumn{7}{|c}{Sources}\\
 {[GHz]}    &                              & AFGL~2591 & G24.78  & G75.78 & NGC~6334~IRS1 & NGC~7538~IRS1 & W3(H$_2$O) & W33A    \\
\hline
\multicolumn{9}{c}{CH$_3$OH continued}\\
\hline
337.969 & 7$_{4,3,1,2}$-- 6$_{4,3,1,2}$       & \phantom{$<$}0.48      & \phantom{$<$}3.2     & \phantom{$<$}0.76   & \phantom{$<$}14      & \phantom{$<$}1.3       &    \phantom{$<$}3.6     & \phantom{$<$}0.77    \\
338.125 & 7$_{4,4,0}$-- 6$_{3,3,0}$           & \phantom{$<$}2.9       & \phantom{$<$}20    & \phantom{$<$}4.7    & \phantom{$<$}56      & \phantom{$<$}5.9       &    \phantom{$<$}24    & \phantom{$<$}5.8     \\
338.345 & 7$_{3,4,0}$-- 6$_{3,4,0}$           & \phantom{$<$}3.9       & \phantom{$<$}35    & \phantom{$<$}7.7    & \phantom{$<$}71      & \phantom{$<$}8.1       &    \phantom{$<$}30    & \phantom{$<$}11    \\
338.409 & 7$_{4,4,0,1}$-- 6$_{3,3,0,1}$       & \phantom{$<$}3.9       & \phantom{$<$}39    & \phantom{$<$}7.9    & \phantom{$<$}59      & \phantom{$<$}--         &    \phantom{$<$}66    & \phantom{$<$}11    \\
338.431 & 7$_{1,7,0}$-- 6$_{0,6,0}$           & \phantom{$<$}--         & \phantom{$<$}4.0     & \phantom{$<$}0.84   & \phantom{$<$}--         & \phantom{$<$}--         &    \phantom{$<$}--       &  \phantom{$<$}--      \\
\hline
\end{tabular}

-- means frequency not observed.

$^a$ Blend with 7$_{6,2,0,1/2}$-- 6$_{5,1,0,1/2}$.

\end{table*}

\begin{table*}
\caption{Observed line fluxes $\int T_{\rm MB}dV$ (K km s$^{-1}$) for
CH$_2$CO. }
\begin{tabular}{l|l|lllllll}
\hline
\hline
Frequency & Transition  & \multicolumn{7}{|c}{Sources}\\
 {[GHz]}    &                          & AFGL~2591 & G24.78  & G75.78 & NGC~6334~IRS1 & NGC~7538~IRS1 & W3(H$_2$O) & W33A   \\
\hline
81.586     & 4$_{1,3}$-- 3$_{1,2}$     & \phantom{$<$}0.32   & \phantom{$<$}--       &  $<$0.75                   & \phantom{$<$}--         & \phantom{$<$}0.45          & \phantom{$<$}0.16           & \phantom{$<$}--\\
140.127    & 7$_{1,7}$-- 6$_{1,6}$     &  \phantom{$<$}--   & \phantom{$<$}--       &  \phantom{$<$}--            & \phantom{$<$}--         & \phantom{$<$}0.72          & \phantom{$<$}0.85           & \phantom{$<$}--\\
244.712    & 12$_{1,11}$-- 11$_{1,10}$ & $<$0.39             & \phantom{$<$}--       &  $<$0.78                   & \phantom{$<$}--         & \phantom{$<$}1.8           & \phantom{$<$}1.6           & \phantom{$<$}--\\
262.619    & 13$_{2,12}$-- 12$_{2,11}$ & $<$0.33             & \phantom{$<$}--       &  $<$0.47                   & \phantom{$<$}--         & $<$1.7                     & \phantom{$<$}1.5           & \phantom{$<$}--\\
\hline
220.178    & 11$_{1,11}$-- 10$_{1,10}$ & $<$0.19   & \phantom{$<$}1.2     & \phantom{$<$}0.41   & \phantom{$<$}5.0       & \phantom{$<$}0.63      & \phantom{$<$}0.65       & \phantom{$<$}0.37   \\
222.199$^a$& 11$_{0,11}$-- 10$_{0,10}$ & $<$0.31   & \phantom{$<$}1.3     & \phantom{$<$}0.38   & \phantom{$<$}5.9       & \phantom{$<$}0.51      & \phantom{$<$}0.87       & \phantom{$<$}0.37   \\
\hline

\end{tabular}

-- means frequency not observed.

$^a$ Blend of lines 11$_{0 11}$ -- 10$_{0 10}$, 11$_{3 9}$ -- 10$_{3 8}$, and  11$_{3 8}$-- 10$_{3 7}$.
\label{ch2colines}
\end{table*}

\begin{table*}
\caption[]{Observed line fluxes $\int T_{\rm MB}dV$ (K km s$^{-1}$) for CH$_3$CHO.}\label{ch3cholines}
\begin{tabular}{l|l|lllllll}
\hline
Frequency & Transition & \multicolumn{7}{|c}{Sources}\\
 {[GHz]}    &                                  & AFGL 2591 & G24.78  & G75.78  & NGC~6334~IRS1 & NGC~7538~IRS1 & W3(H$_2$O) & W33A   \\
\hline
98.901     & 5$_{1,4}$-- 4$_{1,3}$            & \phantom{$<$}0.29 & \phantom{$<$}--       & $<$0.14        & \phantom{$<$}--           & \phantom{$<$}0.22           & \phantom{$<$}0.38           & \phantom{$<$}-- \\
112.249    & 6$_{1,6}$-- 5$_{1,5}$            & \phantom{$<$}--    & \phantom{$<$}--       & \phantom{$<$}-- & \phantom{$<$}--           & \phantom{$<$}0.40           & \phantom{$<$}0.50           & \phantom{$<$}-- \\
149.507    & 8$_{1,8}$-- 7$_{1,7}$            & $<$0.15           & \phantom{$<$}--       & $<$0.36        & \phantom{$<$}--           & \phantom{$<$}0.69           & \phantom{$<$}0.54           & \phantom{$<$}-- \\
168.093    & 9$_{1,9}$-- 8$_{1,8}$            & $<$0.21           & \phantom{$<$}--       & $<$0.30        & \phantom{$<$}--           & \phantom{$<$}0.30           & \phantom{$<$}0.48                     & \phantom{$<$}-- \\
\hline
\end{tabular}

-- means frequency not observed.
\end{table*}

\begin{table*}
\caption[]{Observed line fluxes $\int T_{\rm MB}dV$ (K km s$^{-1}$) for C$_2$H$_5$OH.}\label{c2h5ohlines}
\begin{tabular}{l|l|lllllll}
\hline
\hline
Frequency & Transition & \multicolumn{7}{|c}{Sources}\\
 {[GHz]}    &                                  & AFGL 2591 & G24.78  & G75.78  & NGC~6334~IRS1 & NGC~7538~IRS1 & W3(H$_2$O) & W33A   \\
\hline
220.155 & 24$_{3,22,2}$-- 24$_{2,23,2}$        & $<$0.24   & $<$0.38 & $<$0.36 & \phantom{$<$}1.1       & \phantom{$<$}0.78      & $<$0.39    & $<$0.24\\
222.138 & 19$_{7,13/12,0}$-- 18$_{7,11/12,1}$  & $<$0.39   & $<$0.43 & $<$0.32 & \phantom{$<$}1.6       & $<$0.32   & $<$0.27    & $<$0.32\\ 
225.279 & 13$_{5,9,1}$-- 12$_{5,8,1}$          & $<$0.41   & $<$0.43 & $<$0.24 & \phantom{$<$}0.57      & $<$0.26   & $<$0.21    & $<$0.29\\ 
225.283 & 13$_{5,8,1}$-- 12$_{5,7,1}$          & $<$0.41   & $<$0.43 & $<$0.24 & \phantom{$<$}0.9       & $<$0.26   & $<$0.21    & $<$0.29\\ 
225.457 & 13$_{3,11,1}$-- 12$_{3,10,1}$        & $<$0.41   & \phantom{$<$}0.43    & $<$0.24 & \phantom{$<$}0.66      & $<$0.26   & $<$0.21    & $<$0.29\\ 
225.546 & 13$_{4,10,1}$-- 12$_{4,9,1}$         & $<$0.41   & $<$0.43 & $<$0.24 & \phantom{$<$}0.56      & $<$0.26   & $<$0.21    & $<$0.29\\
\hline
\end{tabular}

-- means frequency not observed.

$^a$ Blend with C$_2$H$_5$CN.
\end{table*}

\begin{table*}
\caption{Observed line fluxes $\int T_{\rm MB}dV$ (K km s$^{-1}$) for
HCOOH.}  
\begin{tabular}{l|l|lllllll}
\hline
\hline
Frequency & Transition  & \multicolumn{7}{|c}{Sources}\\
 {[GHz]}    &                             & AFGL~2591 & G24.78  & G75.78 & NGC~6334~IRS1 & NGC~7538~IRS1 & W3(H$_2$O) & W33A   \\
\hline
223.916    & 10$_{2,9}$-- 9$_{1,8}$      & $<$0.21            & \phantom{$<$}--          & $<$0.17        & \phantom{$<$}--        & \phantom{$<$}0.76           & \phantom{$<$}0.19          & \phantom{$<$}--     \\
247.514    & 11$_{5,6/7}$-- 10$_{5,5/6}$ & \phantom{$<$}--     & \phantom{$<$}--          & \phantom{$<$}-- & \phantom{$<$}--        & \phantom{$<$}0.66           & \phantom{$<$}2.4           & \phantom{$<$}--     \\
257.975    & 12$_{1,12}$-- 11$_{1,11}$   & $<$0.38            & \phantom{$<$}--          & $<$1.9         & \phantom{$<$}--        & $<$0.89                     & $<$0.95                    & \phantom{$<$}--     \\
262.103    & 12$_{0,12}$-- 11$_{0,11}$   &  \phantom{$<$}--    & \phantom{$<$}--          & \phantom{$<$}-- & \phantom{$<$}--        & $<$0.36                     & $<$1.3                     & \phantom{$<$}--     \\
\hline
220.038    & 10$_{0,10}$-- 9$_{0,9}$      & $<$0.19   & $<$0.29 & $<$0.29& \phantom{$<$}1.3       & \phantom{$<$}0.37      & $<$0.46    & $<$0.19\\
225.513    & 10$_{3,7}$-- 9$_{3,6}$       & $<$0.32   & \phantom{$<$}0.73    & $<$0.19& \phantom{$<$}2.6       & $<$0.25   & \phantom{$<$}0.88       & \phantom{$<$}0.40   \\
\hline
\end{tabular}

-- means frequency not observed.

$^a$ Blend with C$_2$H$_5$CN.
\label{hcoohlines}
\end{table*}

\begin{table*}
\caption{Observed line fluxes $\int T_{\rm MB}dV$ (K km s$^{-1}$) for
HNCO.}  
\begin{tabular}{l|l|lllllll}
\hline
\hline
Frequency & Transition & \multicolumn{7}{|c}{Sources}\\
 {[GHz]}    &                            & AFGL 2591 & G24.78  & G75.78 & NGC~6334~IRS1 & NGC~7538~IRS1 & W3(H$_2$O) & W33A   \\
\hline
\multicolumn{9}{c}{HNCO}\\
\hline
219.656    & 10$_{3,7/8}$-- 9$_{3,6/7}$  & $<$0.29   & $<$0.41 & $<$0.31& \phantom{$<$}1.2      & $<$0.25   & $<$0.47    & $<$0.41\\ 
219.734    & 10$_{2,9}$-- 9$_{2,8}$      & $<$0.29   & \phantom{$<$}0.59    & \phantom{$<$}0.77   & \phantom{$<$}3.2       & \phantom{$<$}0.72      & \phantom{$<$}0.90       & \phantom{$<$}0.78   \\  
219.737    & 10$_{2,8}$-- 9$_{2,7}$      & $<$0.29   & \phantom{$<$}1.1     & \phantom{$<$}$^a$   & \phantom{$<$}$^a$      & \phantom{$<$}$^a$      & \phantom{$<$}$^a$       & \phantom{$<$}$^a$   \\  
219.798    & 10$_{0,10}$--9$_{0,9}$      & \phantom{$<$}0.88      & \phantom{$<$}4.4     & \phantom{$<$}1.7    & \phantom{$<$}8.4       & \phantom{$<$}0.65      & \phantom{$<$}1.4        & \phantom{$<$}3.0    \\  
240.876    & 11$_{1,11}$-- 10$_{1,10}$   & $<$0.25   & \phantom{$<$}2.7     & \phantom{$<$}1.3    & \phantom{$<$}4.3       & \phantom{$<$}0.55      & \phantom{$<$}0.99       & \phantom{$<$}1.4    \\  
352.898    & 16$_{1,15}$-- 15$_{1,14}$   & \phantom{$<$}0.89      & \phantom{$<$}3.3     & \phantom{$<$}2.0    & \phantom{$<$}7.7       & \phantom{$<$}1.5       & \phantom{$<$}2.0        & \phantom{$<$}2.1    \\  
\hline
\multicolumn{9}{c}{HN$^{13}$CO}\\
\hline
219.803    & 10$_{2,9}$-- 9$_{2,8}$       & $<$0.36   & $<$0.41 & $<$0.26 & $<$0.98 & $<$0.32 & $<$0.39 & $<$0.41\\
240.881    &11$_{1,11}$-- 10$_{1,10}$    & $<$0.25   & $<$0.46 & $<$0.36& $<$0.41   & $<$0.28   & $<$0.54    & $<$0.26\\ 
\hline
\end{tabular}

-- means frequency not observed.

$^a$ Blend with HNCO line at 219734 MHz.
\label{hncolines}
\end{table*}

\begin{table*}
\caption{Observed line fluxes $\int T_{\rm MB}dV$ (K km s$^{-1}$) for
NH$_2$CHO.}  
\begin{tabular}{l|l|lllllll}
\hline
\hline
Frequency & Transition & \multicolumn{7}{|c}{Sources}\\
 {[GHz]}    &                              & AFGL 2591 & G24.78  & G75.78 & NGC~6334~IRS1 & NGC~7538~IRS1 & W3(H$_2$O) & W33A   \\
\hline
227.606    & 11$_{0,11}$-- 10$_{0,10}$     & \phantom{$<$}--         & \phantom{$<$}0.90    & $<$0.20& \phantom{$<$}--         & \phantom{$<$}$^a$      & $<$0.39    & \phantom{$<$}--      \\
232.274    & 11$_{2,10}$-- 10$_{2,9}$      & $<$0.5    & $<$0.27 & \phantom{$<$}0.15   & \phantom{$<$}2.2      & \phantom{$<$}0.32      & \phantom{$<$}--          & $<$0.60\\
233.529    & 11$_{6,5/6}$-- 10$_{6,4/5}$   & $<$0.15   & $<$0.32 & $<$0.19& \phantom{$<$}0.73      & \phantom{$<$}0.56      & $<$0.21    & $<$0.42\\
233.595    & 11$_{5,6/7}$-- 10$_{5,5/6}$   & $<$0.15   & \phantom{$<$}0.64    & $<$0.19& \phantom{$<$}$^b$      & \phantom{$<$}0.93      & \phantom{$<$}1.4        & \phantom{$<$}0.56   \\
233.736    & 11$_{4,8}$-- 10$_{4,7}$       & $<$0.15   & \phantom{$<$}0.56    & $<$0.19& \phantom{$<$}1.3       & \phantom{$<$}0.23      & $<$0.21    & \phantom{$<$}0.30   \\
339.716    & 16$_{8,8/9}$-- 15$_{8,7/8}$   & $<$0.3    & \phantom{$<$}1.3     & $<$0.17& \phantom{$<$}3.8       & \phantom{$<$}1.2       & \phantom{$<$}0.22       & $<$0.34\\
\hline
\end{tabular}

-- means frequency not observed.

$^a$ Blend with C$_2$H$_5$OH.

$^b$ Blend with C$_2$H$_5$OH at 225404 MHz in other sideband.

\label{nh2cholines}
\end{table*}

\begin{table*}
\caption{Observed line fluxes $\int T_{\rm MB}dV$ (K km s$^{-1}$) for
CH$_3$CN.}  
\begin{tabular}{l|l|lllllll}
\hline
\hline
Frequency & Transition & \multicolumn{7}{|c}{Sources}\\
 {[GHz]}    &                                & AFGL 2591 & G24.78  & G75.78 & NGC~6334~IRS1 & NGC~7538~IRS1 & W3(H$_2$O) & W33A   \\
\hline
\multicolumn{9}{c}{CH$_3$CN}\\
\hline
220.476    & 12$_{8.0,\rm K}$-- 11$_{8,0,\rm L}$         & $<$0.24   & \phantom{$<$}0.45    & $<$0.43& \phantom{$<$}1.0       & $<$0.54   & $<$0.53    & $<$0.24\\
238.913    & 13$_{7,0,\rm K}$-- 12$_{7,0,\rm L}$                 & \phantom{$<$}--         & \phantom{$<$}1.4     & $<$0.25& \phantom{$<$}--         & \phantom{$<$}--         & $<$0.50    & \phantom{$<$}0.38   \\ 
238.972    & 13$_{6,0,\rm K}$-- 12$_{6,0,\rm K}$                 & \phantom{$<$}--         & \phantom{$<$}2.8     & \phantom{$<$}0.47   & \phantom{$<$}--         & \phantom{$<$}--         & \phantom{$<$}1.8        & \phantom{$<$}0.73   \\ 
239.064    & 13$_{4,0,\rm K}$-- 12$_{4,0,\rm L}$                 & \phantom{$<$}--         & \phantom{$<$}3.8     & \phantom{$<$}0.83   & \phantom{$<$}--         & \phantom{$<$}--         & \phantom{$<$}1.9        & \phantom{$<$}0.97   \\ 
239.097    & 13$_{3,0,\rm K}$-- 12$_{3,0,\rm L}$                 & \phantom{$<$}--         & \phantom{$<$}5.9     & \phantom{$<$}1.2   & \phantom{$<$}--         & \phantom{$<$}--         & \phantom{$<$}3.8        & \phantom{$<$}1.5    \\
\hline
\end{tabular}

-- means frequency not observed.

K refers to main transition of the upper level i.e. 12, 13, or 19
$+$1, $+0$, or $-$1

L refers to the main transition of the lower level i.e. 11, 12 , or 18
$+$1, $+0$, or $-$1.

$^a$ Blend with CH$_3$OH.
\label{ch3cnlines}
\end{table*}

\begin{table*}
\caption{Observed line fluxes $\int T_{\rm MB}dV$ (K km s$^{-1}$) for
C$_2$H$_5$CN.}\label{c2h5cnlines} 
\begin{tabular}{l|l|lllllll}
\hline
\hline
Frequency & Transition & \multicolumn{7}{|c}{Sources}\\
 {[GHz]}    &                              & AFGL~2591 & G24.78  & G75.78 & NGC~6334~IRS1 & NGC~7538~IRS1 & W3(H$_2$O) & W33A   \\
\hline
218.390 & 24$_{3,21}$-- 23$_{3,20}$        & \phantom{$<$}--         & \phantom{$<$}0.71    & $<$0.41& \phantom{$<$}--         & \phantom{$<$}--         & \phantom{$<$}--          & \phantom{$<$}--       \\
219.506 & 24$_{2,22}$-- 23$_{2,21}$        & $<$0.36   & \phantom{$<$}0.85    & $<$0.26& \phantom{$<$}2.2       & $<$0.32   & $<$0.39    & $<$0.41 \\
222.918 & 25$_{1,24}$-- 24$_{1,23}$        & $<$0.32   & \phantom{$<$}--       & \phantom{$<$}--      & \phantom{$<$}3.4       & $<$0.36   & \phantom{$<$}--          & \phantom{$<$}--       \\
223.385 & 26$_{1,26}$-- 25$_{1,25}$        & $<$0.32   & \phantom{$<$}--       & \phantom{$<$}--      & \phantom{$<$}1.9       & $<$0.36   & \phantom{$<$}--          & \phantom{$<$}--       \\
227.781 & 25$_{3,22}$-- 24$_{3,21}$        & \phantom{$<$}--         & \phantom{$<$}1.8     & $<$0.26& \phantom{$<$}--         & $<$0.32   & $<$0.39    & \phantom{$<$}--       \\
228.483 & 25$_{2,23}$-- 24$_{2,22}$        & $<$0.24   & \phantom{$<$}1.4     & $<$0.36& \phantom{$<$}2.4       & $<$0.30   & $<$0.33    & $<$0.24 \\
\hline
\end{tabular}

-- means frequency not observed.

$^a$ Blend with NH$_2$CHO. 
\end{table*}

\begin{table*}
\caption{Observed line fluxes $\int T_{\rm MB}dV$ (K km s$^{-1}$) for
CH$_3$CCH.}  
\begin{tabular}{l|l|lllllll}
\hline
\hline
Frequency & Transition & \multicolumn{7}{|c}{Sources}\\
 {[GHz]}    &                       & AFGL 2591 & G24.78  & G75.78 & NGC~6334~IRS1 & NGC~7538~IRS1 & W3(H$_2$O) & W33A   \\
\hline
222.014    & 13$_6$-- 12$_6$         & $<$0.23   & $<$0.27 & $<$0.25& \phantom{$<$}0.51      & $<$0.41   & $<$0.27    & $<$0.25\\
222.061    & 13$_5$-- 12$_5$         & $<$0.23   & \phantom{$<$}0.18    & $<$0.25& $<$0.59   & $<$0.41   & $<$0.27    & $<$0.25\\
222.099    & 13$_4$-- 12$_4$         & $<$0.23   & \phantom{$<$}0.44    & $<$0.25& \phantom{$<$}1.2       & \phantom{$<$}0.19      & \phantom{$<$}0.49       & $<$0.25\\ 
222.129    & 13$_3$-- 12$_3$         & \phantom{$<$}0.63      & \phantom{$<$}2.5     & \phantom{$<$}0.54   & \phantom{$<$}7.6       & \phantom{$<$}0.92      & \phantom{$<$}2.0        & \phantom{$<$}0.58   \\ 
222.150    & 13$_2$-- 12$_2$         & \phantom{$<$}0.87      & \phantom{$<$}2.2     & \phantom{$<$}0.81   & \phantom{$<$}7.2       & \phantom{$<$}1.2       & \phantom{$<$}2.1        & \phantom{$<$}0.89   \\  
222.163    & 13$_1$-- 12$_1$         & \phantom{$<$}1.2       & \phantom{$<$}4.3     & \phantom{$<$}1.0    & \phantom{$<$}8.1       & \phantom{$<$}1.7       & \phantom{$<$}1.7        & \phantom{$<$}1.9    \\
\hline
\end{tabular}

-- means frequency not observed.

$^a$ Blend with CH$_3$CN.

$^b$ Blend with OCS at 231062 MHz in the main sideband.
\label{ch3cchlines}
\end{table*}

\begin{table*}
\caption{Observed line fluxes $\int T_{\rm MB}dV$ (K km s$^{-1}$) for HCOOCH$_3$ for the 230 GHz band.}\label{ch3ocholines}
\begin{tabular}{l|l|lllllll} 
\hline\hline
Frequency & Transition & \multicolumn{7}{|c}{Sources}\\
 {[GHz]}    &                              & AFGL 2591 & G24.78  & G75.78 & NGC~6334~IRS1 & NGC~7538~IRS1 & W3(H$_2$O) & W33A   \\
\hline
211.537 & 17$_{6,11}$-- 16$_{6,10}$ E     & $<$0.36   & \phantom{$<$}--       & \phantom{$<$}--      & \phantom{$<$}8.6       & \phantom{$<$}--        & \phantom{$<$}--          & $<$0.14\\
211.575 & 17$_{6,11}$-- 16$_{6,10}$ A     & $<$0.36   & \phantom{$<$}--       & \phantom{$<$}--      & \phantom{$<$}6.3       & \phantom{$<$}--        & \phantom{$<$}--          & \phantom{$<$}0.19   \\ 
211.771 & 17$_{5,13}$-- 16$_{5,12}$ E     & $<$0.36   & \phantom{$<$}--       & \phantom{$<$}--      & \phantom{$<$}6.2       & \phantom{$<$}--        & \phantom{$<$}--          & $<$0.14\\ 
211.785 & 17$_{5,13}$-- 16$_{5,12}$ A     & $<$0.36   & \phantom{$<$}--       & \phantom{$<$}--      & \phantom{$<$}6.2       & \phantom{$<$}--        & \phantom{$<$}--          & $<$0.14\\ 
218.281 & 17$_{1,16}$-- 16$_{1,15}$ E     & \phantom{$<$}--         & \phantom{$<$}0.48    & $<$0.41& \phantom{$<$}--         & \phantom{$<$}--        & \phantom{$<$}--          & \phantom{$<$}--      \\
218.297 & 17$_{3,14}$-- 16$_{3,13}$ A     & \phantom{$<$}--         & \phantom{$<$}0.46    & $<$0.41& \phantom{$<$}--         & \phantom{$<$}--        & \phantom{$<$}--          & \phantom{$<$}--      \\
\hline
\end{tabular}

-- means frequency not observed.

$^a$ Blend with 21$_{2,20}$-- 20$_{2,19}$ E,21$_{1,21}$-- 20$_{1,20}$ A, and 21$_{0,21}$-- 20$_{0,20}$ A.

$^b$ Blend with line at 233397 19$_{14,5}$-- 18$_{14,4}$ E. 

$^c$ Blend with 19$_{13,6}$-- 18$_{13,5}$ A, and 19$_{13,7}$-- 18$_{13,6}$ A.

$^d$ Blend with C$_2$H$_5$CN.

$^e$ Blend with 23$_{1,23}$-- 22$_{1 22}$ E, 23$_{2,22}$-- 22$_{2,21}$ E, 23$_{1,23}$-- 22$_{2,21}$ E,
\end{table*}

\begin{table*}
\caption{Observed line fluxes $\int T_{\rm MB}dV$ (K km s$^{-1}$) for HCOOCH$_3$ in the 345 GHz band.}\label{ch3ocholines2}
\begin{tabular}{l|l|lllllll} 
\hline\hline
Frequency & Transition & \multicolumn{7}{|c}{Sources}\\
 {[GHz]}    &                              & AFGL 2591 & G24.78  & G75.78 & NGC~6334~IRS1 & NGC~7538~IRS1 & W3(H$_2$O) & W33A   \\
\hline
338.356 & 27$_{8,19}$-- 26$_{8,18}$ E     & $<$0.30   & $<$0.30 & $<$0.20& \phantom{$<$}8.8       & \phantom{$<$}0.23      & \phantom{$<$}1.6        & $<$0.29\\  
338.397 & 27$_{7,21}$-- 26$_{7,20}$ E     & $<$0.30   & $<$0.30 & \phantom{$<$}0.13   & \phantom{$<$}11     & \phantom{$<$}0.85      & $<$0.30    & $<$0.29\\  
340.221 & 31$_{13,19}$-- 31$_{12,19}$ E$^a$ & $<$0.36   & $<$0.35 & $<$0.26& \phantom{$<$}2.7       & $<$0.41   & $<$0.36    & $<$0.23\\  
345.069 & 28$_{14,14}$-- 27$_{14,13}$ E$^b$ & $<$0.29   & \phantom{$<$}1.7     & $<$0.30& \phantom{$<$}14      & $<$0.33   & \phantom{$<$}1.5        & $<$0.24\\  
345.090 & 28$_{14,15}$-- 27$_{14,14}$ E   & $<$0.29   & \phantom{$<$}0.69    & $<$0.30& \phantom{$<$}8.2       & $<$0.33   & \phantom{$<$}0.78       & $<$0.24\\  
345.281 & 18$_{13,5/6}$-- 18$_{12,6/7}$ A & $<$0.29   & $<$0.27 & $<$0.30& \phantom{$<$}3.5       & $<$0.33   & $<$0.21    & $<$0.24\\
\hline
\end{tabular}

-- means frequency not observed.

$^a$ Blend with 31$_{13,18}$-- 31$_{12,19}$ A, and 31$_{13,19}$-- 31$_{12,20}$ A.

$^b$ Blend with 28$_{14,14}$-- 27$_{14,13}$ A, and 28$_{14,15}$-- 27$_{14,14}$ A.
\end{table*}			     

\begin{table*}
\caption{Observed line fluxes $\int T_{mb}dV$ (K km s$^{-1}$) for CH$_3$OCH$_3$.\label{ch3och3lines}} 
\begin{tabular}{l|l|ccccccc}
\hline
\hline
Frequency & Transition & \multicolumn{7}{|c}{Sources}\\
 {[GHz]}    &                                  & AFGL 2591 & G24.78  & G75.78 & NGC~6334~IRS1 & NGC~7538~IRS1 & W3(H$_2$O) & W33A   \\
\hline
222.029    & 21$_{2,20,2/3}$-- 21$_{1,21,2/3}$ & $<$0.39   & \phantom{$<$}0.26    & $<$0.32& \phantom{$<$}3.1       & $<$0.41   & \phantom{$<$}0.42       & $<$0.25\\
222.035    & 21$_{2,20,0/1}$-- 21$_{1,21,0/1}$ & $<$0.39   & \phantom{$<$}0.72    & $<$0.32& \phantom{$<$}2.5       & $<$0.41   & \phantom{$<$}0.37       & $<$0.25\\  
222.239    & 4$_{3,2,2}$--  3$_{2,1,2}$        & $<$0.39   & \phantom{$<$}0.16    & $<$0.32& \phantom{$<$}1.6       & $<$0.41   & $<$0.27    & $<$0.25\\  
222.248    & 4$_{3,2,1/3}$--  3$_{2,1,1/3}$    & $<$0.39   & \phantom{$<$}0.86    & $<$0.32& \phantom{$<$}5.3       & $<$0.41   & \phantom{$<$}0.37       & \phantom{$<$}0.37   \\  
222.255    & 4$_{3,2,0}$--  3$_{2,1,0}$        & $<$0.39   & \phantom{$<$}0.53    & $<$0.32& \phantom{$<$}4.7       & $<$0.41   & \phantom{$<$}0.50       & \phantom{$<$}0.18   \\ 
222.259    & 4$_{3,1,1/2}$--  3$_{2,1,1/2}$    & $<$0.39   & \phantom{$<$}0.60    & $<$0.32& \phantom{$<$}2.6       & $<$0.41   & \phantom{$<$}0.48       & \phantom{$<$}0.18   \\
\hline	
\end{tabular}

-- means frequency not observed.

K is either 0,1,2, or 3.

$^a$ 5$_{3,3,1}$--  4$_{2,2,1}$ transition is blended with 5$_{3,3,2}$-- 4$_{2,2,2}$ and 5$_{3,3,0}$--  4$_{2,2,0}$.
\end{table*}
\end{appendix}
				   
\end{document}